\documentclass[prl,twocolumn,superscriptaddress,showpacs]{revtex4}
\usepackage{graphicx}% Include figure files
\usepackage{epsfig}% Include figure files
\usepackage{amssymb,amsmath,amsfonts,hyperref}
\usepackage{wasysym}
\usepackage{latexsym}
\usepackage{verbatim}
\usepackage[normalem]{ulem}
\usepackage{color}

\newcommand{\be}{\begin{eqnarray}}
\newcommand{\ee}{\end{eqnarray}}

\begin{document}
%\preprint{APS/123-QED}
%draft
%\twocolumn[\hsize\textwidth\columnwidth\hsize\csname @twocolumnfalse\endcsname
\title{Vacancy-induced low-energy states in undoped graphene}
\author{Sambuddha Sanyal}
\affiliation{\small{International Center for Theoretical Sciences, Tata Institute of Fundamental Research, Bengaluru 560089, India}}
\author{Kedar Damle}
\affiliation{\small{Department of Theoretical Physics, Tata Institute of Fundamental Research, Mumbai 400005, India}}
\author{Olexei I. Motrunich}
\affiliation{\small{Department of Physics, California Institute of Technology, Pasadena, California 91125, USA}}
\begin{abstract}
We demonstrate that a nonzero concentration $n_v$ of static, randomly-placed vacancies in graphene leads
to a density $w$ of zero-energy quasiparticle states at the band-center $\epsilon=0$ within a tight-binding description with nearest-neighbour hopping $t$ on the honeycomb lattice. We show that $w$ remains generically nonzero in the compensated case (exactly equal number of vacancies on the two sublattices) even in the presence of hopping disorder, and depends sensitively on $n_v$ and correlations between vacancy positions. For low, {\em but not-too-low}
$|\epsilon|/t$ in this compensated case, we show that the density of states (DOS) $\rho(\epsilon)$ 
exhibits a strong divergence of the form $\rho_{\rm 1D}(\epsilon) \sim |\epsilon|^{-1}/ [\log(t/|\epsilon|)]^{(y+1)} $, which crosses over to the universal low-energy asymptotic form expected on symmetry grounds $\rho_{\rm GW}(\epsilon)  \sim |\epsilon|^{-1}e^{-b[\log(t/|\epsilon|)]^{2/3} }$ below a crossover scale $\epsilon_c \ll t$. $\epsilon_c$ is found to decrease rapidly with decreasing $n_v$, while $y$ decreases much more slowly.
\end{abstract}


\maketitle

Static impurities, which give rise to random time-independent terms in the single-particle Hamiltonian for quasiparticle excitations of a condensed matter system, can lead
to the phenomenon of Anderson localization, whereby quasiparticle wavefunctions lose their plane-wave
character and become localized~\cite{Lee_Ramakrishnan}. Such localization transitions
and universal low-energy properties of the localized phase have been successfully described in many cases using effective field-theories~\cite{Altland_Simons_Zirnbauer,Evers_Mirlin} whose form depends on symmetry properties
of the quasiparticle Hamiltonian in the presence of impurities. In some cases~\cite{Gade,Gade_Wegner}, it has also been possible to refine these field theoretical predictions using real-space strong-disorder renormalization group ideas~\cite{Motrunich_Damle_Huse}.

In this Letter, we study the effects of a nonzero concentration $n_v$ of static, randomly-located vacancies
in graphene. We use a tight-binding description for electronic states of graphene, with hopping amplitude $t$ between nearest-neighbour sites on a honeycomb lattice, and model vacancies by the deletion
of the corresponding site in this tight-binding model~\cite{Araujo_Terrones_Dresselhaus,Forte_etal,Pereira_dosSantos_CastroNeto,Pereira_Guinea_dosSantos_Peres_CastroNeto,Wehling_Yuan_Lichtenstein_Geim_Katsnelson}. We focus on the compensated case, {\em i.e.}, exactly equal numbers of vacancies
on the two sublattices of the honeycomb lattice, and demonstrate that vacancies generically lead
to a nonuniversal density $w$ of zero-energy quasiparticle states at the band-center $\epsilon=0$ even in this compensated case, including in the presence of hopping disorder. For low, {\em but not-too-low}
$|\epsilon|/t$ in this compensated case,
the density of states (DOS) $\rho(\epsilon)$ 
exhibits a strong divergence of the form:
\begin{equation}
\rho_{\rm 1D}(\epsilon) \sim |\epsilon|^{-1}/ [\log(t/|\epsilon|)]^{(y+1)} \; ,
\label{1Dform}
\end{equation}
familiar in the context of various random-hopping problems in one dimension~\cite{Dyson,Theodorou_Cohen,Eggarter_Riedenger,Motrunich_Damle_Huse_PRB0,Motrunich_Damle_Huse_PRB1,Gruzberg_Read_Vishweshwara,Brouwer_Furusaki_Gruzberg_Mudry,Brouwer_Mudry_Furusaki,Titov_Brouwer_Furusaki_Mudry}.
At still lower energies, below a crossover scale $\epsilon_c$ that is
several orders of magnitude smaller than $t$ even for moderately small values of $n_v$ ($0.05$--$0.1$), we show that the DOS crosses over to the low-energy asymptotic behaviour~\cite{Gade,Gade_Wegner,Motrunich_Damle_Huse,Mudry_Ryu_Furusaki} of the chiral orthogonal universality class (to which our tight-binding model belongs on symmetry grounds):
\begin{equation}
\rho_{\rm GW}(\epsilon)  \sim |\epsilon|^{-1}e^{-b[\log(t/|\epsilon|)]^{2/3} } \;.
\label{GWform}
\end{equation}

The density of zero-energy states $w$ depends sensitively on correlations between vacancies and decreases as $n_v$ is lowered.  The crossover energy $\epsilon_c$ is found to decrease rapidly with decreasing $w$, while $y$ (in fits to Eq.~(\ref{1Dform}) for $|\epsilon| > \epsilon_c$) decreases much more slowly. On comparing the corresponding crossover length scale $l_c$, defined as the mean spatial separation between nonzero energy modes with $|\epsilon| < \epsilon_c$, with $l_w \equiv w^{-1/2}$, the mean spatial separation between zero-energy states, we find that $l_c$ tracks $l_w$ up to a nonuniversal prefactor. Thus, our results imply that the $w \rightarrow 0$ limit of the DOS is singular and does not commute with the $\epsilon \rightarrow 0$ limit: For any $w>0$, the true asymptotic form $\rho_{\rm GW}(\epsilon)$ cannot be obtained from an extrapolation of results
obtained for $\epsilon_c < \epsilon \ll t$, which instead reflect the intermediate-energy
physics encoded in the form $\rho_{\rm 1D}(\epsilon)$.

Our work sheds light on an interesting question motivated by the results of Willans {\em et. al.}, who found a vacancy-induced DOS of the form $\rho_{\rm 1D}(\epsilon)$ at not-too-low energies 
in their study of Majorana excitations of Kitaev's honeycomb model~\cite{Willans_Chalker_Moessner_PRB}: Does a nonzero vacancy density lead to
low-energy properties qualitatively different from the asymptotic behaviour 
expected in the chiral orthogonal universality class of quasiparticle localization?
In recent work that addressed this question in the context of graphene~\cite{Hafner_etal,Ostrovsky_etal}, it was argued that vacancies lead to a new term in the low-energy field theory, which
causes the DOS to take on the form $\rho_{\rm 1D}(\epsilon)$, Eq.~(\ref{1Dform}), with $y=1/2$ at asymptotically low energies, rather than the asymptotic form $\rho_{\rm GW}(\epsilon)$, Eq.~(\ref{GWform}), expected on symmetry
grounds.  

Clearly, our conclusion is quite different, and raises
two perhaps more interesting questions: When $\epsilon_c \ll t$, are the crossover exponent $y$ and crossover energy $\epsilon_c$
 ``universally'' determined by the zero-mode density $w$, although the function $w(n_v)$ itself depends sensitively on microscopic details such as correlations between vacancies? Can this crossover be understood within a renormalization group description of the low-energy physics? Leaving these interesting questions for future work, we devote the remainder
of this Letter to an account of the calculations that lead us to our results, and thence,
to these questions.

We choose the lattice spacing of the honeycomb lattice as our unit of length
and measure all energies in terms of the hopping amplitude $t$, which is set by the bandwidth of the $\pi$-band of undoped graphene.
We focus on the compensated case, with
exactly $n_vL^2$ vacancies placed randomly on {\em each} sublattice of a finite $L\times L$ honeycomb lattice with $L^2$ unit cells ($2L^2$ sites).
The spectrum of single-particle states can be obtained by diagonalizing
the real symmetric matrix $H$
\begin{equation}
H = \left( \begin{array}{cc}
0 & T_{AB}\\
T^{\dagger}_{AB} & 0 \end{array} \right)
\end{equation}
where $T_{AB}$ is the $(1-n_v)L^2$-dimensional matrix of amplitudes for hopping from 
the undeleted sites of the $B$ sublattice to their undeleted $A$ sublattice neighbours,
and $T^{\dagger}_{AB}$ is the transpose of this matrix (the spin label of the electronic quasiparticles is dropped since we do not study magnetic properties or sources of spin-flip scattering in this Letter).
\begin{figure}
{\includegraphics[width=\hsize]{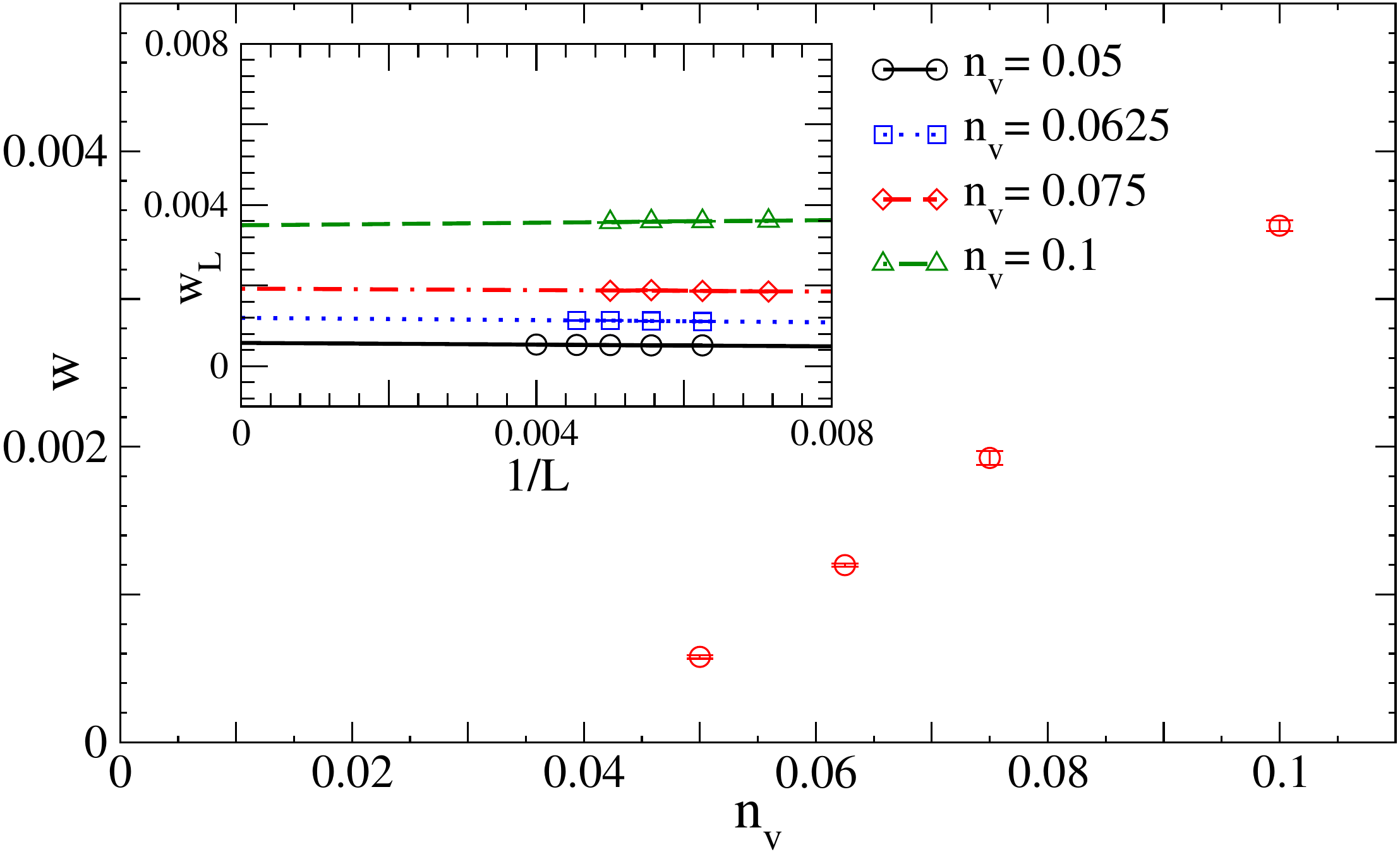}}
\caption{$w_L$, the density of zero modes in an $L \times L$ sample, tends to
a nonzero thermodynamic limit $w$ that depends on $n_v$, the concentration of vacancies.}
\label{Fig1}
\end{figure}

The purely off-block-diagonal form of $H$ reflects the ``chiral'' symmetry
of the problem, corresponding to the bipartite structure of the honeycomb lattice, which
guarantees that  every eigenstate with energy $\epsilon>0$ has a corresponding eigenstate at energy $-\epsilon$.
In order to eliminate zero modes of $H$ in the pure $L \times L$ lattice~\cite{Lieb_Loss,Ryu_Hatsugai,Brey_Fertig}, we choose even values of $L$ and impose antiperiodic boundary conditions along the $\hat{x}$
direction, while terminating the lattice in the $\hat{y}$ direction in a pair
of armchair edges. We also impose a nearest-neighbour and next-nearest-neighbour
exclusion constraint on the vacancies, and do not allow them to interrupt the armchair edges. These restrictions, along with the compensated nature of the vacancy disorder, eliminate all previously studied and well-understood sources of vacancy-induced~\cite{Pereira_dosSantos_CastroNeto,Brouwer_Racine_Furusaki_Hatsugai_Morita_Mudry} zero modes in the spectrum of $H$.

We find it convenient
to focus on the symmetric matrix $T^{\dagger}_{AB}T_{AB}$, which has a single eigenvalue $\epsilon^2$ for every pair of nonzero eigenvalues $(\epsilon,-\epsilon)$ of $H$.
Zero modes of $T^{\dagger}_{AB}T_{AB}$, with wavefunction living entirely on the $B$ sublattice, map on to exactly half of the zero modes in
the spectrum of $H$, while zero modes of the symmetric matrix  $T_{AB}T^{\dagger}_{AB}$, with
wavefunction living entirely on the $A$ sublattice, make up the other half of the null space
of $H$.
We use the ALGOL~\cite{ALGOL} routines of Martin and Wilkinson~\cite{Martin_Wilkinson} to compute the number ${\mathcal N}_{\Lambda}$ of eigenvalues of the banded matrix $T^{\dagger}_{AB}T_{AB}$ 
which are smaller in magnitude than some positive number $t^2 \times 10^{-\Lambda} $. Our implementation~\cite{Sanyal_Thesis} uses
calls to the GNU multiprecision library~\cite{GMP} for all arithmetic operations, including
comparison of the magnitudes of two numbers, and has been benchmarked against
routines from the LAPACK library~\cite{LAPACK} as well as C-translations (used in earlier work~\cite{Motrunich_Damle_Huse})
of the ALGOL routines of Martin and Wilkinson.
\begin{figure}
{\includegraphics[width=0.8\hsize]{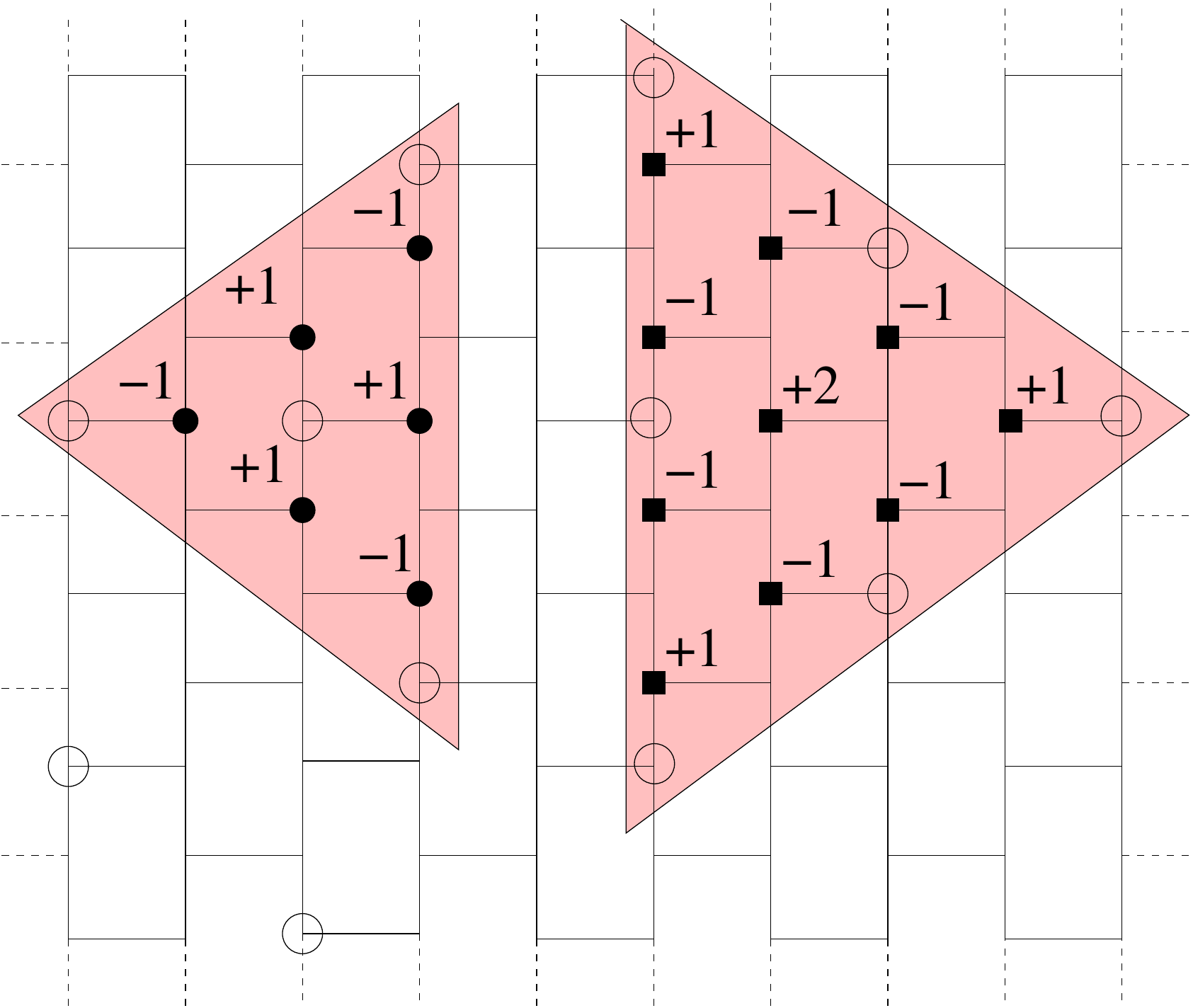}}
\caption{Four $B$-sublattice (six $A$-sublattice) vacancies arranged in a ``4-triangle'' pattern (``${\mathcal R}_6$'' motif) give rise to a zero mode of $H$ living on $A$-sublattice sites ($B$-sublattice sites) within the 4-triangle (${\mathcal R}_6$ motif). While hopping disorder eliminates the 4-triangle zero mode, it only changes the wavefunction of the ${\mathcal R}_6$ zero mode {\em without} changing its energy.}
\label{Fig2}
\end{figure}

Anticipating that the physics of interest to us spans many orders of magnitude in energy $\epsilon$,
we define the `log-energy' $\Gamma = \log_{10}(t/|\epsilon|)$,
and compute $N^{(i)}_{\rm tot}(\Gamma,L) \equiv {\mathcal N}^{(i)}_{\Lambda=2\Gamma}/L^2$ for the $i^{th}$ $L \times L$ random sample using values of log-energy drawn from
an equispaced grid ranging from $\Gamma \sim 1$ to $\Gamma \sim 100$. For large enough $\Gamma$, $N^{(i)}_{\rm tot} (\Gamma,L)$ plateaus out to a constant value 
which represents
the density of zero modes $w^{(i)}_L$ of that sample. For not-too-small $n_v$ ($n_v \geq 0.05$) for which we are able to access this plateau, we separately
keep track of $w_L^{(i)}$ and $N_L^{(i)}(\Gamma) \equiv N^{(i)}_{\rm tot} (\Gamma,L) - w^{(i)}_L$. From the position, $\Gamma_{\rm g}^{(i)}(L)$, of
the last downward step in $N^{(i)}_{\rm tot}(\Gamma,L)$, we also obtain the 
spectral gap $\epsilon^{(i)}_{\rm g}(L) \equiv t \times 10^{-\Gamma^{(i)}_{\rm g}(L)}$ corresponding to the
lowest pair of nonzero eigenvalues $\pm \epsilon^{(i)}_{\rm g}$ for that sample. Analyzing this data for
up to $3000$ samples for each value of $L$ and $n_v$, we obtain statistically reliable estimates of the corresponding disorder-averaged quantities $w_L$ and $N_L(\Gamma)$. The density of states $\rho_L(\epsilon)$ can then be obtained from $N_L$ using 
the relation $\rho_L(\epsilon) \equiv \frac{1}{2\epsilon}\frac{dN_L}{d\Gamma}$. Additionally, we estimate $f_L$, the probability that an $L \times L$ sample has at least
one pair of zero modes, and measure the histogram of $\Gamma_{\rm g}(L)$. The position
of the peak in the latter provides us an estimate of  $\Gamma_{\rm g}^{*}(L)$, the
most probable value of  $\Gamma_{\rm g}(L)$. For the smallest values of $n_v$, which require multiprecision computation at impracticably large $\Gamma$ in order to access the plateau in $N^{(i)}_{\rm tot} (\Gamma,L)$ (and thence, $w_L^{i}$), we instead compute
$\frac{dN_L}{d\Gamma}$ by numerical differentiation of $N^{(i)}_{\rm tot} (\Gamma,L)$.
\begin{figure}
{\includegraphics[width=\hsize]{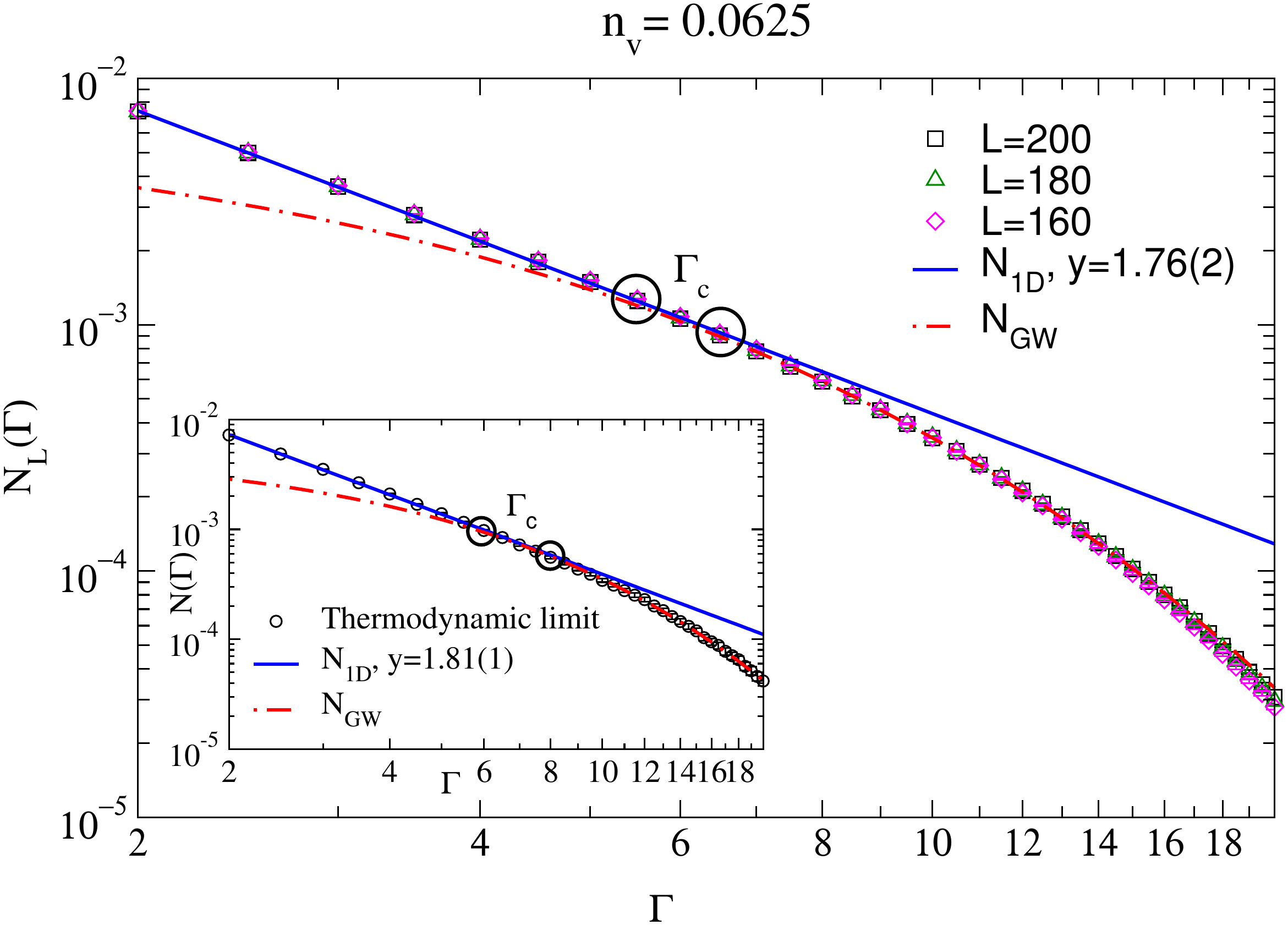}}
{\includegraphics[width=\hsize]{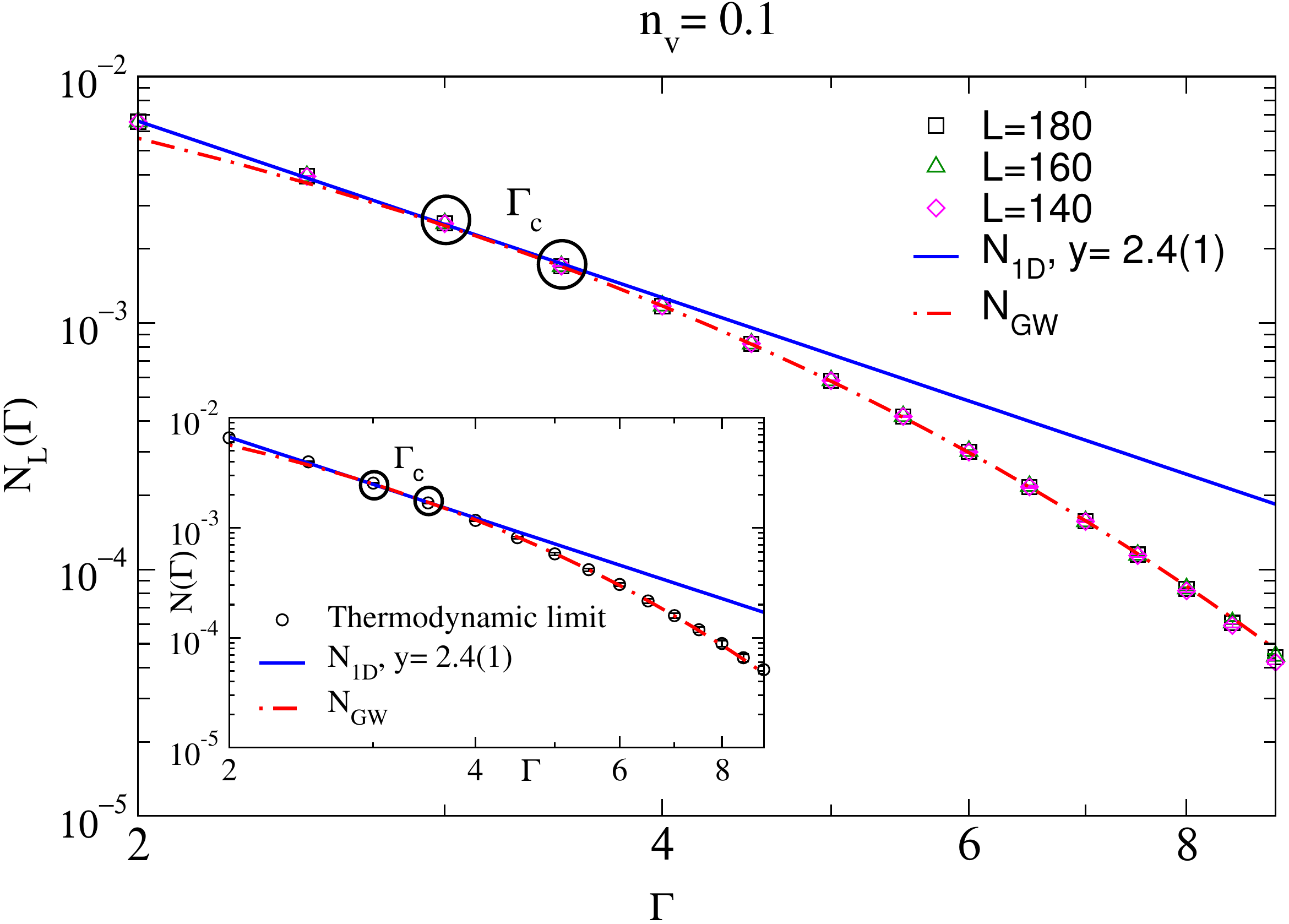}}
\caption{$N_L(\Gamma)$ at the three largest values of $L$ studied for $n_v=0.0625$ and $n_v=0.1$. 
Insets show $N(\Gamma)$ obtained by extrapolation to the thermodynamic limit.
Circles demarcate
the crossover region centered at the crossover scale $\Gamma_c$. Data for $\Gamma \lesssim \Gamma_c$ fits well to power-law form $N_{\rm 1D}(\Gamma)$ with the values
of $y$ indicated in each case, while the large-$\Gamma$
regime fits well to the modified Gade-Wegner form $N_{\rm GW}(\Gamma)$.}
\label{Fig3}
\end{figure}

Extrapolating our results for $f_L$ (Supplemental Material Section)
and $w_L$ (Fig.~\ref{Fig1}) to obtain $f \equiv \lim_{L \rightarrow \infty}f_L$ and $w \equiv \lim_{L \rightarrow \infty}w_L$, we find that $f=1$, and that $w$ depends sensitively
on $n_v$ (Fig.~\ref{Fig1}). 
To understand these results, we observe that  $T_{AB}T^{\dagger}_{AB}$ ($T^{\dagger}_{AB}T_{AB}$) {\em must}
have a zero mode, with wavefunction shown in Fig.~\ref{Fig2}, if four of the $B$-sublattice vacancies (six of the $A$-sublattice vacancies) are
arranged in the specific ``4-triangle'' pattern (``${\mathcal R}_6$ motif'') shown in Fig.~\ref{Fig2}, with no restrictions on the positions of the other vacancies. $H$ must therefore have a {\em pair of zero modes}  if a {\em single} 4-triangle or ${\mathcal R}_6$ motif occurs 
anywhere in the sample on either sublattice. Since there is a nonzero probability
of finding a 4-triangle at a given location, this already implies that a large enough sample will certainly have at least one zero mode, {\em i.e.}, $f=1$. Additionally, one has an elementary lower-bound
on $w_L^{(i)}$ in terms of the numbers $N_{\Delta_{4A}}^{(i)}$ and $N_{\Delta_{4B}}^{(i)}$ of 4-triangles on $A$ and $B$ lattices
in a given sample: $w_L^{(i)} \geq \left [ {\rm max}(N_{\Delta_{4A}}^{(i)},N_{\Delta_{4B}}^{(i)}) \right ]/L^2$, implying $w \geq n_{\Delta_4}$, where $n_{\Delta_4}$ is the ensemble averaged concentration of 4-triangles in the thermodynamic limit. When the vacancies obey the exclusion constraints described earlier, it is not possible to produce a similar zero mode with fewer than four vacancies (Supplemental Material Section). Thus, we expect $w \sim n_v^4$ in the $n_v \rightarrow 0$ limit.

While our lower bound can be strengthened somewhat by including larger versions of the 4-triangle motif (Supplemental Material Section), they do not change this limiting behaviour. However, our results (Fig.~\ref{Fig1}) suggest that this limiting behaviour sets in only for $n_v \ll 0.05$, for which a direct computation of $w$ would require access to impracticably large $\Gamma$. For $n_v \gtrsim 0.05$, 4-triangles are {\em not} the dominant contribution to $w$ (Supplemental Material Section), which we expect arises instead
from generalizations of the ${\mathcal R}_6$ motif: Such ``${\mathcal R}$-type'' regions have more undeleted sites belonging to one sublattice than the other, but are connected to the rest of the lattice only via sites belonging to the other sublattice.  Like the ${\mathcal R}_6$ zero mode, all such ${\mathcal R}$-type zero modes  are robust to disorder in the nearest-neighbour hopping amplitudes (Supplemental Material Section). Unlike zero modes associated with specific patterns like 4-triangles, these ${\mathcal R}$-type zero modes cannot be eliminated by any additional local constraints on the vacancy positions. They are therefore a {\em generic} feature of the diluted graphene lattice.
Thus we see that a nonzero concentration $n_v$ of vacancies leads to a density $w$ of zero modes of $H$, where $w$  depends sensitively on $n_v$, and on correlations in the positions of
vacancies, but remains generically nonzero even in the compensated case, including in the presence of hopping disorder.

Figure~\ref{Fig3} displays $N_L(\Gamma)$ for $n_v=0.0625$ and $n_v=0.1$ for the three largest sizes used in our extrapolations to the thermodynamic limit. Since we expect finite-size
effects to dominate for $\Gamma > \Gamma_{\rm g}^{*}(L)$, we estimate $ \Gamma_{\rm g}^{*}(L)$ from histograms of  $\Gamma_{\rm g}(L)$ (Supplemental Material Section) and restrict attention
to $\Gamma < \Gamma_{\rm g}^{*}(L_{\rm min})$, where $L_{\rm min}$,
the smallest of the sizes used in our extrapolations, is chosen large
enough that $f_{L_{\rm min}} \approx 1$ in order to ensure that the physics of zero modes
is correctly captured in all our analysis. In this range of $\Gamma$, we can
reliably extrapolate  (see Supplemental Material Section) from our data to obtain the thermodynamic limit $N(\Gamma)$
displayed in the inset of Fig.~\ref{Fig3}. Up to a fairly well-defined and readily-identified crossover scale $\Gamma_c(L) \equiv \log_{10}(t/|\epsilon_c(L)|)$, $N_{L}(\Gamma)$ is found to
fit well to a power-law form
$N_{\rm 1D}(\Gamma) \equiv c\Gamma^{-y}$. However, for larger $\Gamma$ beyond $\Gamma_c$, the asymptotic fall-off is clearly faster than a power law. $\Gamma_c(L)$ increases
slightly with $L$ over the range of $L$ studied, but saturates at large $L$ to a finite thermodynamic limit $\Gamma_c$ that
marks the presence of the same crossover in the limiting curve $N(\Gamma)$. Thus, $N(\Gamma)$ is again fit well by the power-law form $N_{\rm 1D}$ for $\Gamma \lesssim \Gamma_c$, but falls off much faster in the large-$\Gamma$ regime.
\begin{figure}
{\includegraphics[width=\hsize]{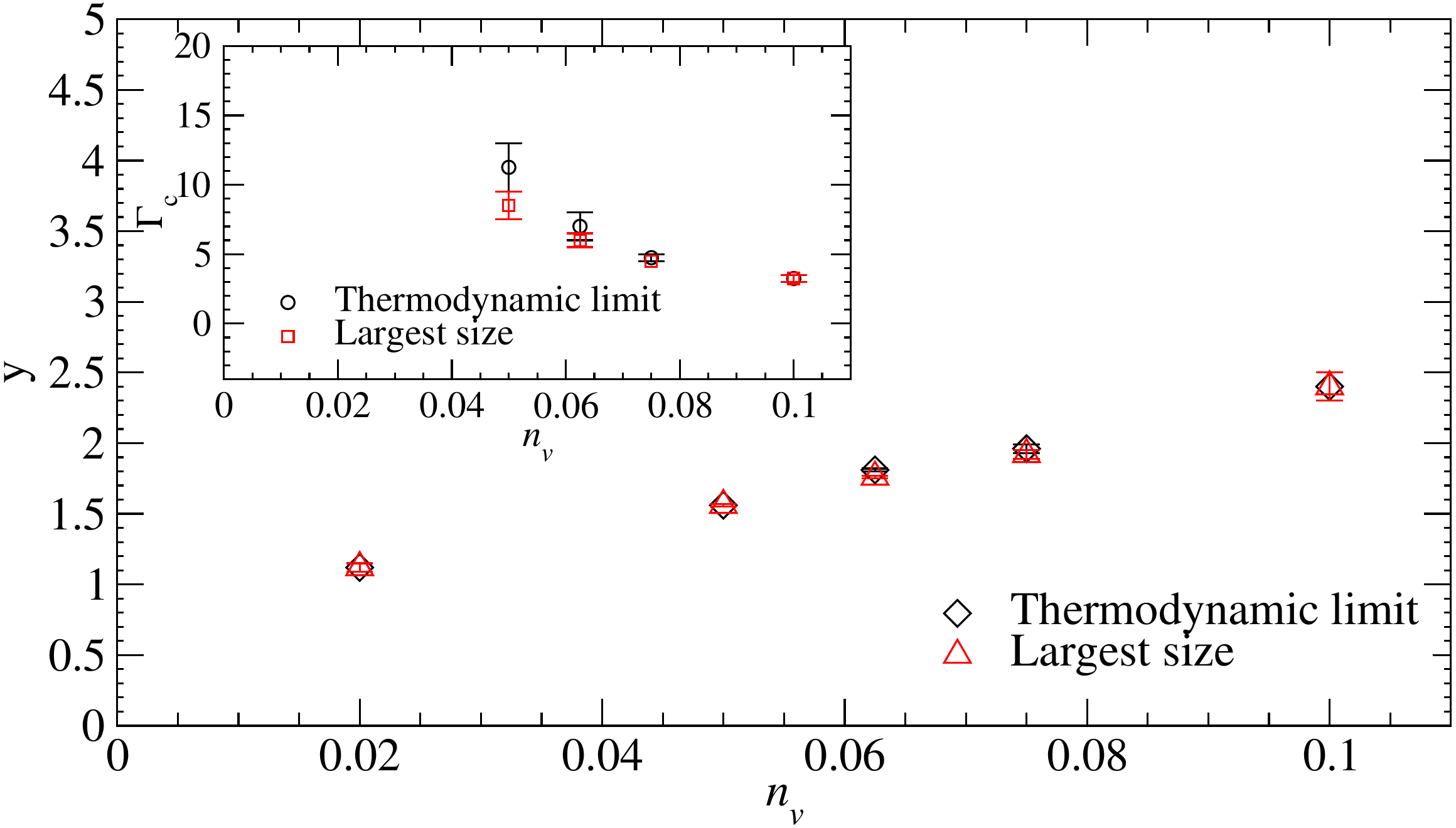}}
\caption{$n_v$ dependence of crossover scale $\Gamma_c$ and power-law exponent $y$ for samples with compensated  random dilution.}
\label{Fig4}
\end{figure}
\begin{figure}
{\includegraphics[width=\hsize]{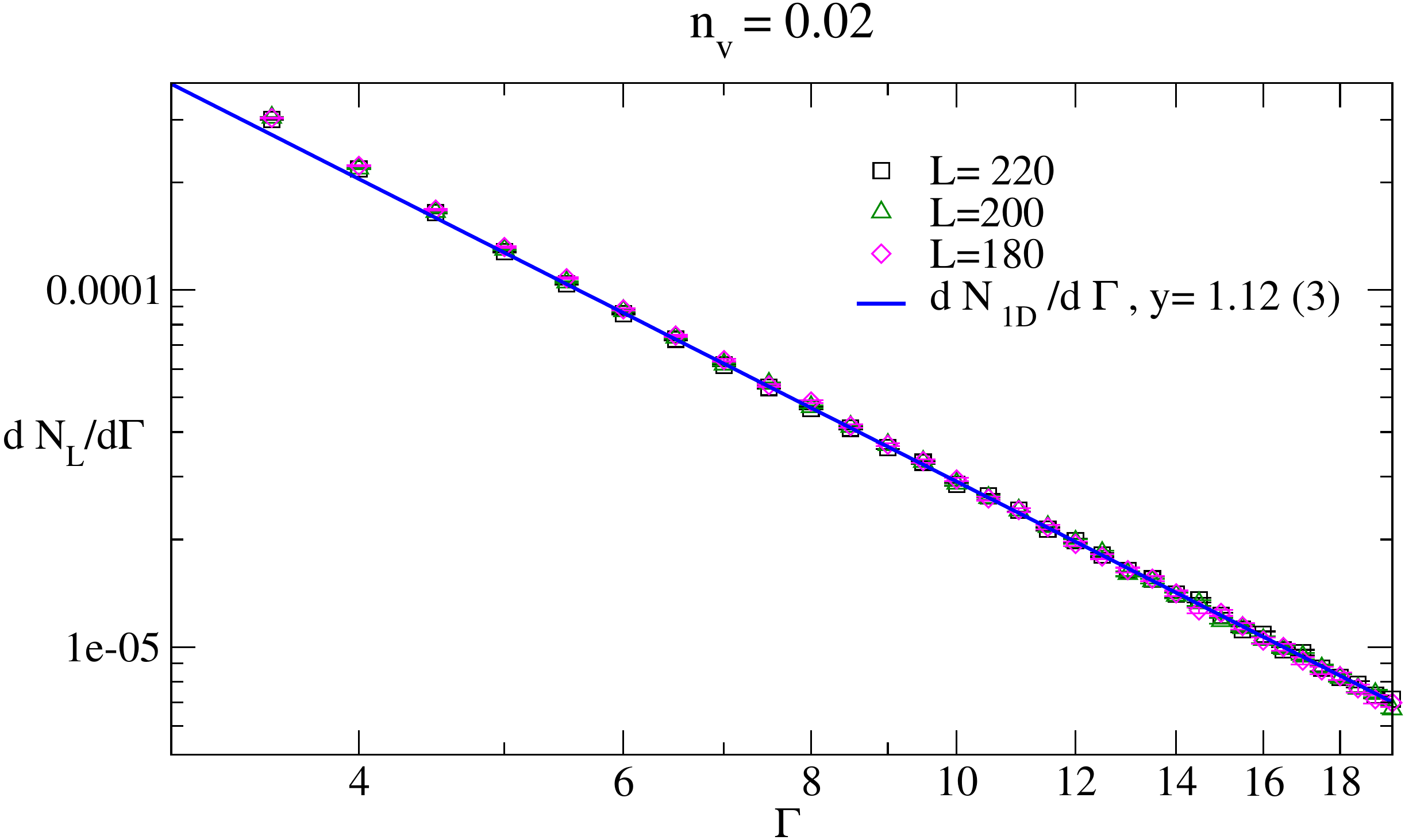}}
\caption{$\frac{dN_{L}(\Gamma)}{d\Gamma}$ at $n_v=0.02$ in the crossover regime 
converges to the thermodynamic limit for $L\sim 200$ and fits well to the form $\frac{dN_{\rm 1D}(\Gamma)}{d\Gamma}$, with a value of $y$ consistent with the trends established at larger $n_v$ for $\Gamma_c$ and $y$ (Fig.~{\protect{\ref{Fig4}}}). Based on these trends, we expect $N(\Gamma)$ to cross over to the asymptotic form $N_{\rm GW}$ at {\em much larger} values of $\Gamma$, for which we are unable to reliably compute $N(\Gamma)$ due to computational constraints.}
\label{Fig5}
\end{figure}

Given that $H$ belongs to the chiral orthogonal universality class, standard
universality arguments predict that $N(\Gamma)$ and $N_L(\Gamma)$ should, at large enough $\Gamma$, follow the  modified Gade-Wegner form~\cite{Gade,Gade_Wegner,Motrunich_Damle_Huse,Mudry_Ryu_Furusaki} $N_{\rm GW}(\Gamma) \equiv a\Gamma^{1/3}e^{-b\Gamma^{2/3}}$.
\begin{figure}
{\includegraphics[width=\hsize]{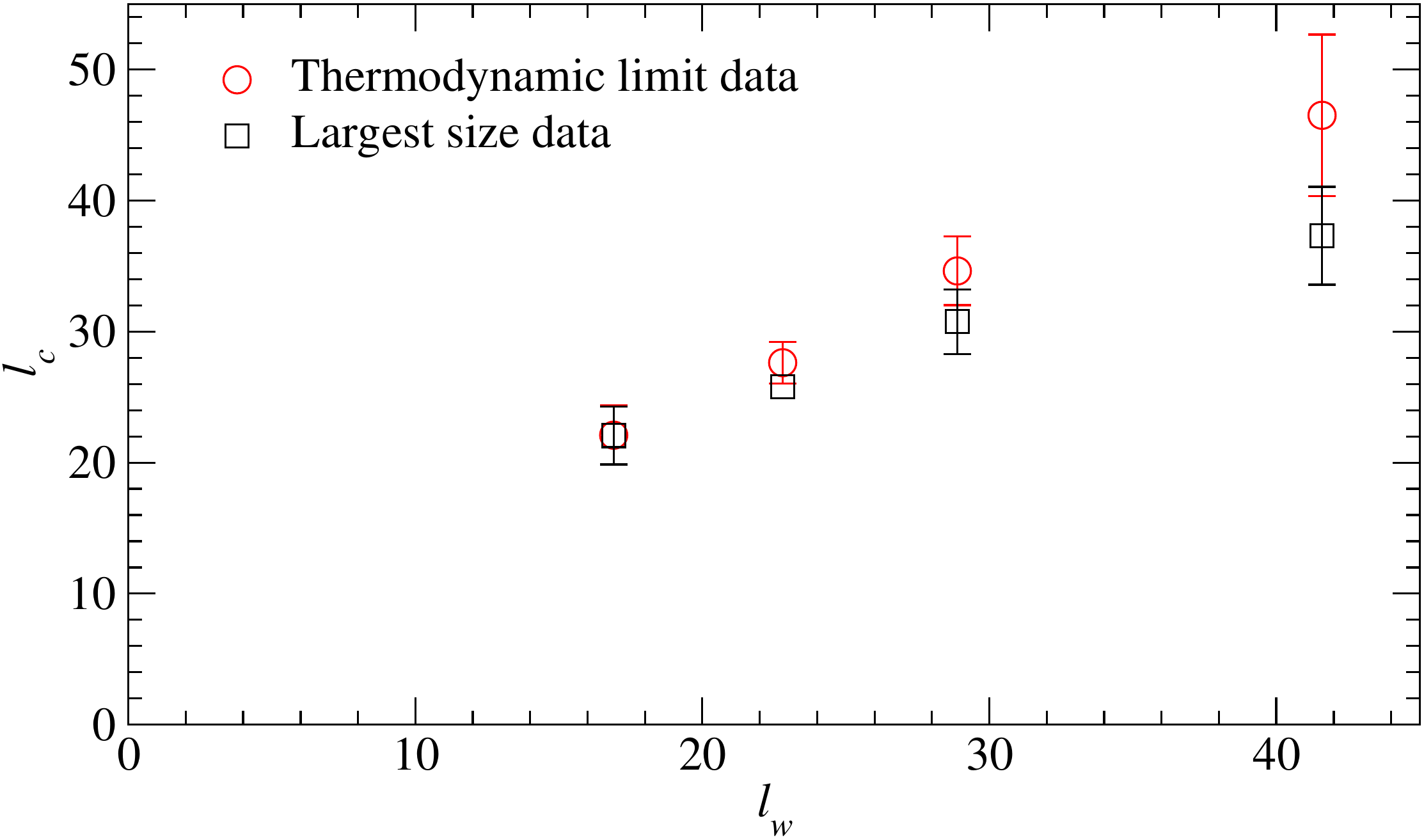}}
\caption{The crossover length-scale $l_c \equiv N(\Gamma_c)^{-1/2}$ tracks the mean spatial separation $l_w \equiv w^{-1/2}$ between zero modes reasonably well for
compensated random dilution. From left to right, the exhibited data points correspond to
vacancy densities $0.1$, $0.075$, $0.0625$, and $0.05$.
}
\label{Fig6}
\end{figure}
From Fig.~\ref{Fig3}, we see that this form indeed provides a very good
fit in the asymptotic large-$\Gamma$ regime. The same crossover is also visible at $n_v=0.05$ and $n_v=0.075$.
From Fig.~\ref{Fig4}, we see that $y$ decreases gradually with $n_v$, while $\Gamma_c$ increases extremely rapidly as we go to smaller values of $n_v$, thereby limiting our ability to directly study this crossover for $n_v \lesssim 0.05$. However, one can nevertheless reliably compute the exponent $y$ that characterizes the behaviour of $\rho(\epsilon)$ in
the intermediate regime $t \gg |\epsilon| \gg \epsilon_c$ (Fig.~\ref{Fig5}), and confirm that its value evolves
smoothly (Fig.~\ref{Fig4}) down to these small values of $n_v$. This strongly suggests that the crossover identified by us is an intrinsic and generic feature of the density of states for any nonzero $n_v$.

The corresponding crossover length scale $l_c \equiv N(\Gamma_c)^{-1/2}$, which represents the mean spatial separation between nonzero energy modes with $|\epsilon|/t < 10^{-\Gamma_c}$,
grows relatively slowly (Fig.~\ref{Fig6}) as $w$ is decreased, with $l_c \lesssim 50$ lattice units even at the smallest value of $w$ studied (corresponding to $n_v = 0.05$). This explains why our extrapolations to the thermodynamic limit using finite-size data with $L \sim 200$ remain reliable for all $n_v$ studied. From Fig.~\ref{Fig6}, which compares $l_c$ for the randomly diluted samples with $l_w \equiv w^{-1/2}$, the mean spatial separation between zero modes, we also see that $l_c$ tracks $l_w$ (up to a nonuniversal prefactor). This suggests that the crossover identified in this Letter is controlled primarily by the density of zero modes. Additional support for this idea comes from our study of samples diluted
with an equal number of randomly placed 4-triangles (instead of individual vacancies) on each sublattice (Supplemental Material Section), which show the same crossover, but with
very different values of $\epsilon_c$ and $y$ that are better predicted by the zero mode
density $w$ as opposed to the vacancy density. This then leads us to the questions identified earlier: Is the physics of this crossover ``universally controlled''
by the value of $w$ ({\em i.e.}, independent of correlations between vacancy-positions and
other microscopic details) in the limit of small $w$, and can it be understood via a renormalization group description of the low-energy physics?

\textit{Acknowledgements}
We thank M.~Barma and D.~Dhar for useful comments on a previous draft, and
gratefully acknowledge use of computational resources funded by DST (India) grant DST-SR/S2/RJN-25/2006, in addition to departmental computational resources of the Dept. of Theoretical Physics of the TIFR. KD and OM gratefully acknowledge hospitality of ICTS-TIFR (Bengaluru)  and IISc (Bengaluru) during completion of part
of this work. SS gratefully acknowledges funding from DST (India) and DAE -SRC (India) and support from IISc (Bengaluru) during completion of part of this work. OM also acknowledges support by the NSF through grant DMR-1206096.

%\newpage
%\widetext{
%\appendix*
\newpage
\onecolumngrid
\appendix
\section{Supplemental Material for ``Vacancy-induced low-energy states in undoped graphene"}

In this Supplemental Material, we present additional numerical evidence and analytical arguments which support the key findings described
in the main text.

\section{Additional numerical evidence}
%%%%%%%%%%%%%%%%%%%%%%%%%%%%%%%%%%%%%%%%%%%%%%%%%%%%%%%%%%%%%%%%%%%%%%%%%%%%%%%%%%%%%%%%%%%%%%%%%%%%%%%%%%%%%%%%%%%%%%%%%%%%%%
\subsection{Other concentrations}
\begin{figure}
{\includegraphics[width=10cm]{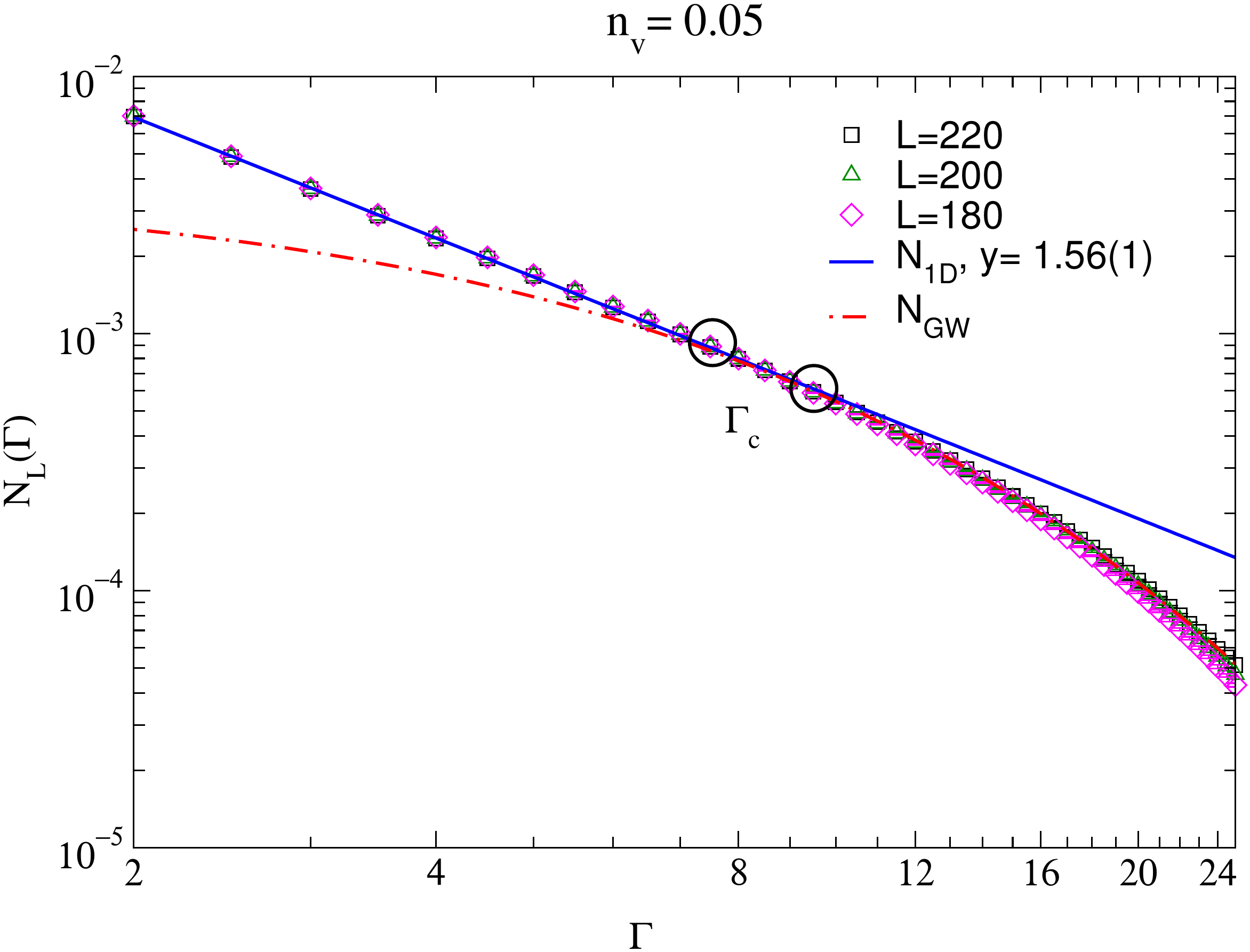}}
\caption{$N_L(\Gamma)$ at the three largest values of $L$ studied for $n_v=0.05$. Circles demarcate
the crossover region centered at the crossover scale $\Gamma_c$. Data for $\Gamma \lesssim \Gamma_c$ fits well to power-law form $N_{\rm 1D}(\Gamma)$ \textcolor{blue}(see text for details) with the value
of $y$ indicated in the figure, while the large-$\Gamma$
regime fits well to the modified Gade-Wegner form $N_{\rm GW}(\Gamma)$.}
\label{SuppFig1}
\end{figure}

\begin{figure}
{\includegraphics[width=10cm]{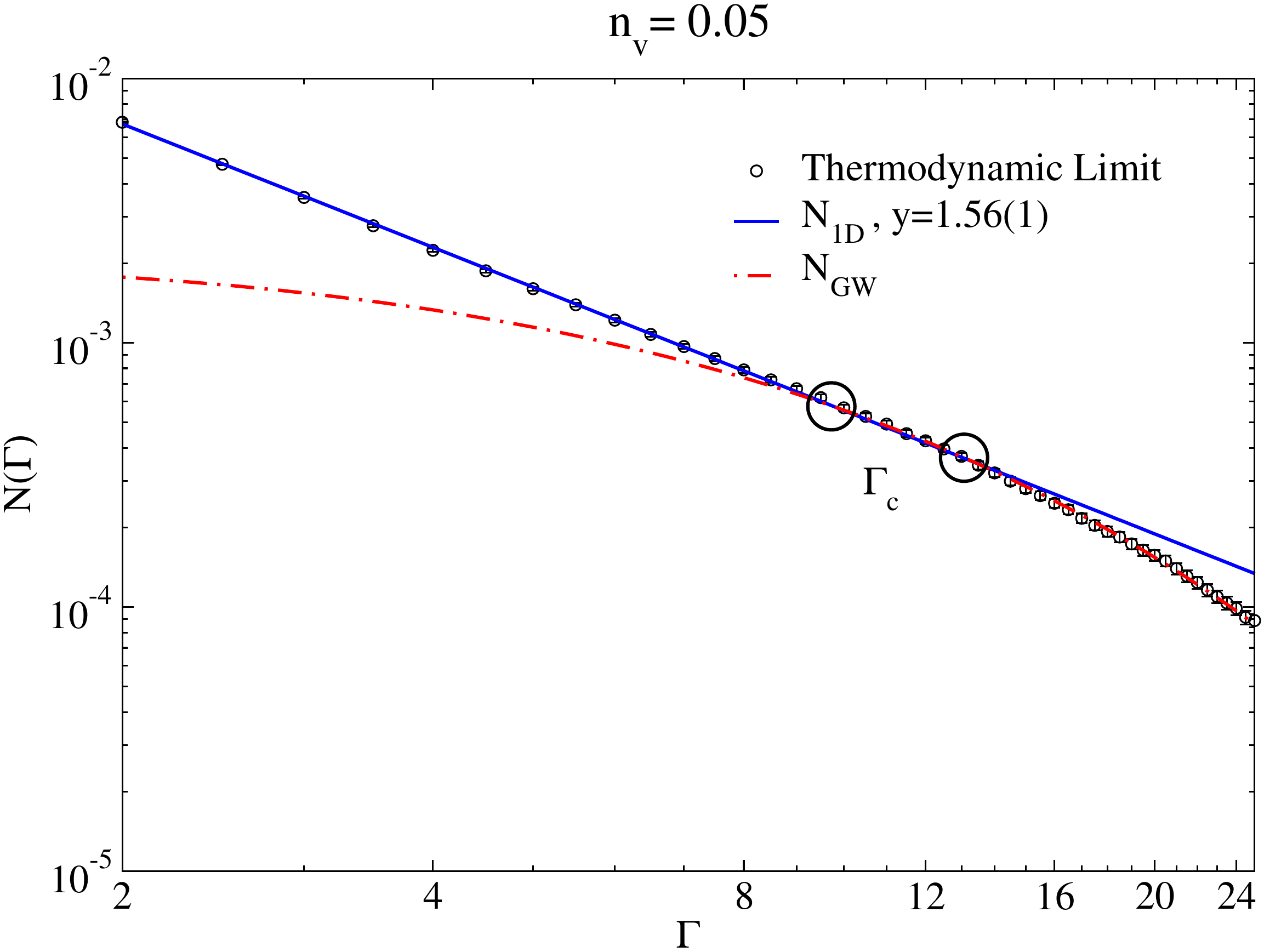}}
\caption{ $N(\Gamma)$,  the extrapolation to the thermodynamic limit of the finite-size data
from the previous figure. Again, circles demarcate
the crossover region centered at the crossover scale $\Gamma_c$. Data for $\Gamma \lesssim \Gamma_c$ fits well to power-law form $N_{\rm 1D}(\Gamma)$ with the value
of $y$ indicated in the figure, while the large-$\Gamma$
regime fits well to the modified Gade-Wegner form $N_{\rm GW}(\Gamma)$.}
\label{SuppFig2}
\end{figure}

\begin{figure}
{\includegraphics[width=10cm]{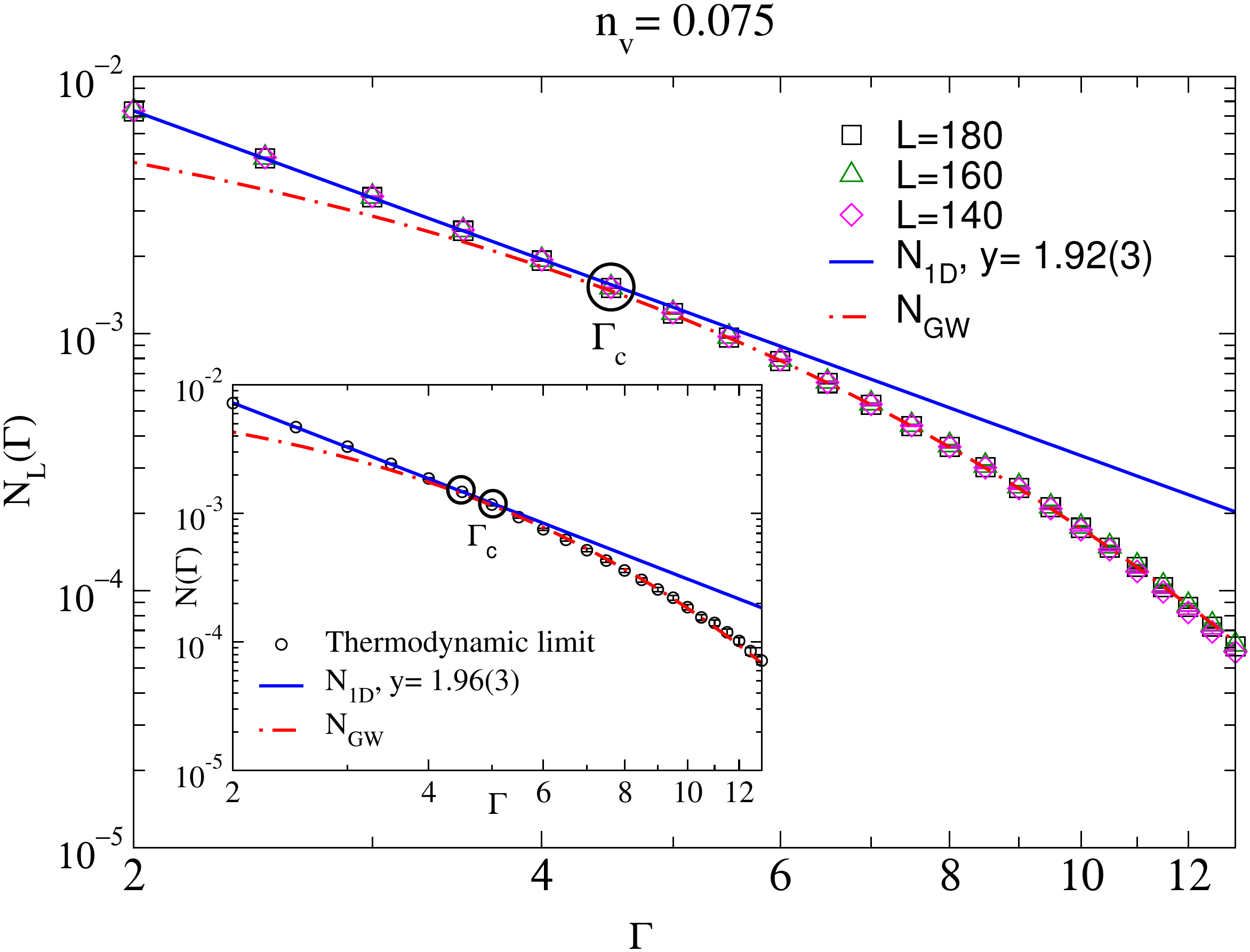}}
\caption{$N_L(\Gamma)$ at the three largest values of $L$ studied for $n_v=0.075$. Circles demarcate
the crossover region centered at the crossover scale $\Gamma_c$. Data for $\Gamma \lesssim \Gamma_c$ fits well to power-law form $N_{\rm 1D}(\Gamma)$ with the value
of $y$ indicated in the figure, while the large-$\Gamma$
regime fits well to the modified Gade-Wegner form $N_{\rm GW}(\Gamma)$. Inset shows
the extrapolation to the thermodynamic limit, in which the same crossover is clearly visible.}
\label{SuppFig3}
\end{figure}

\begin{figure}
{\includegraphics[width=10cm]{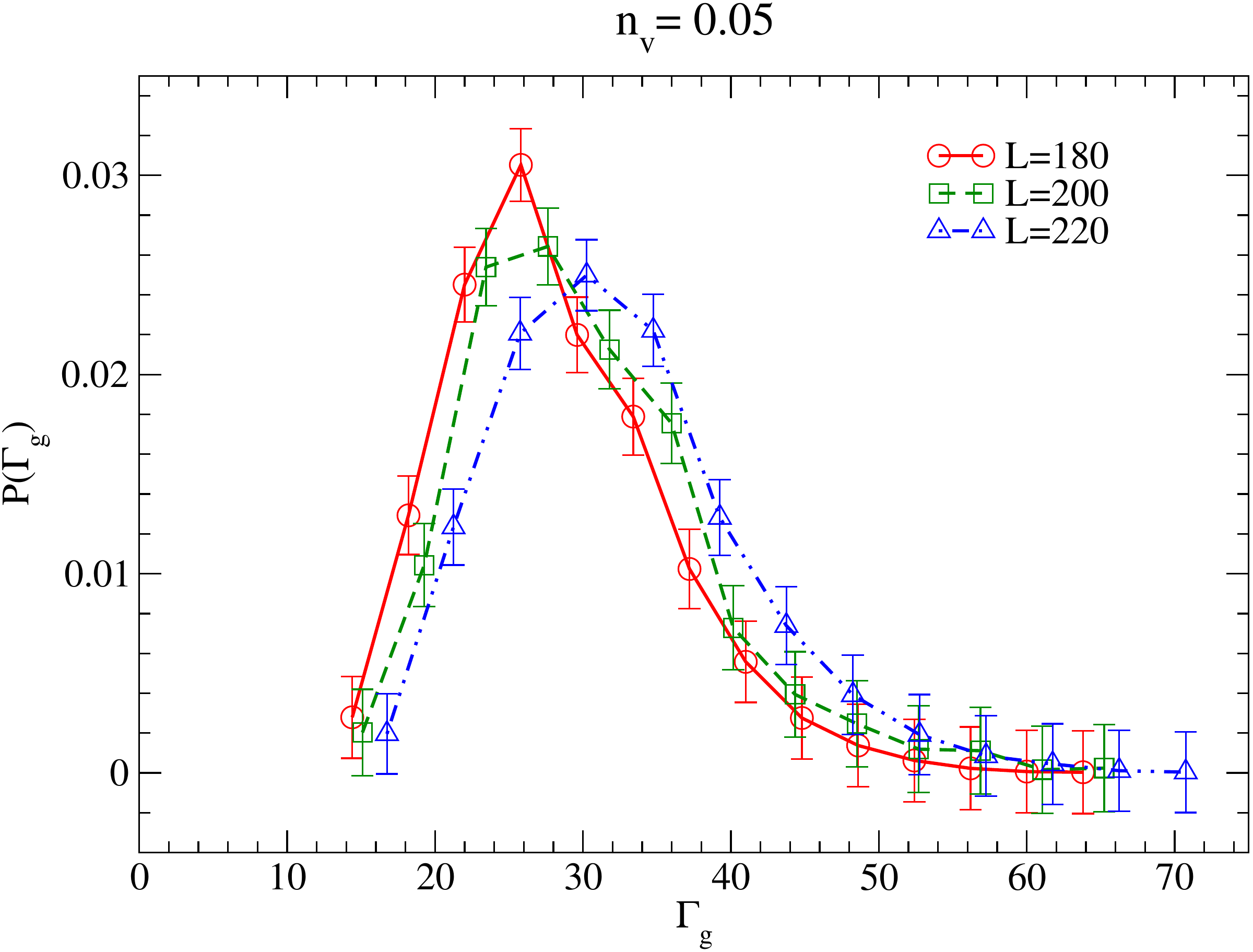}}
\caption{Histogram of $\Gamma_g$, corresponding to the lowest nonzero gap
for the three largest sizes studied at $n_v=0.05$.}
\label{SuppFig4}
\end{figure}

\begin{figure}
{\includegraphics[width=10cm]{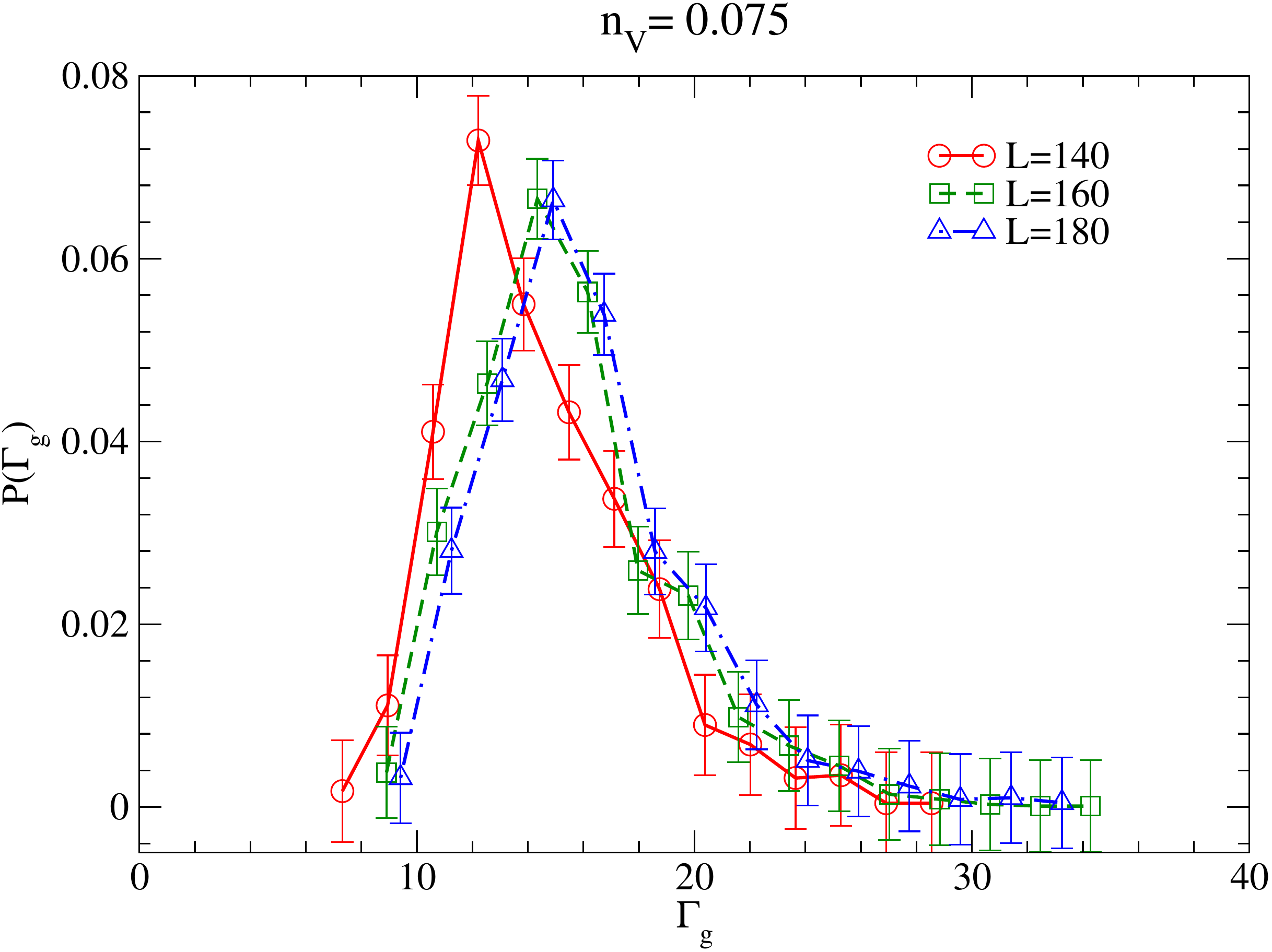}}
\caption{Histogram of $\Gamma_g$, corresponding to the lowest nonzero gap
for the three largest sizes studied at $n_v=0.075$.}
\label{SuppFig5}
\end{figure}

\begin{figure}
{\includegraphics[width=10cm]{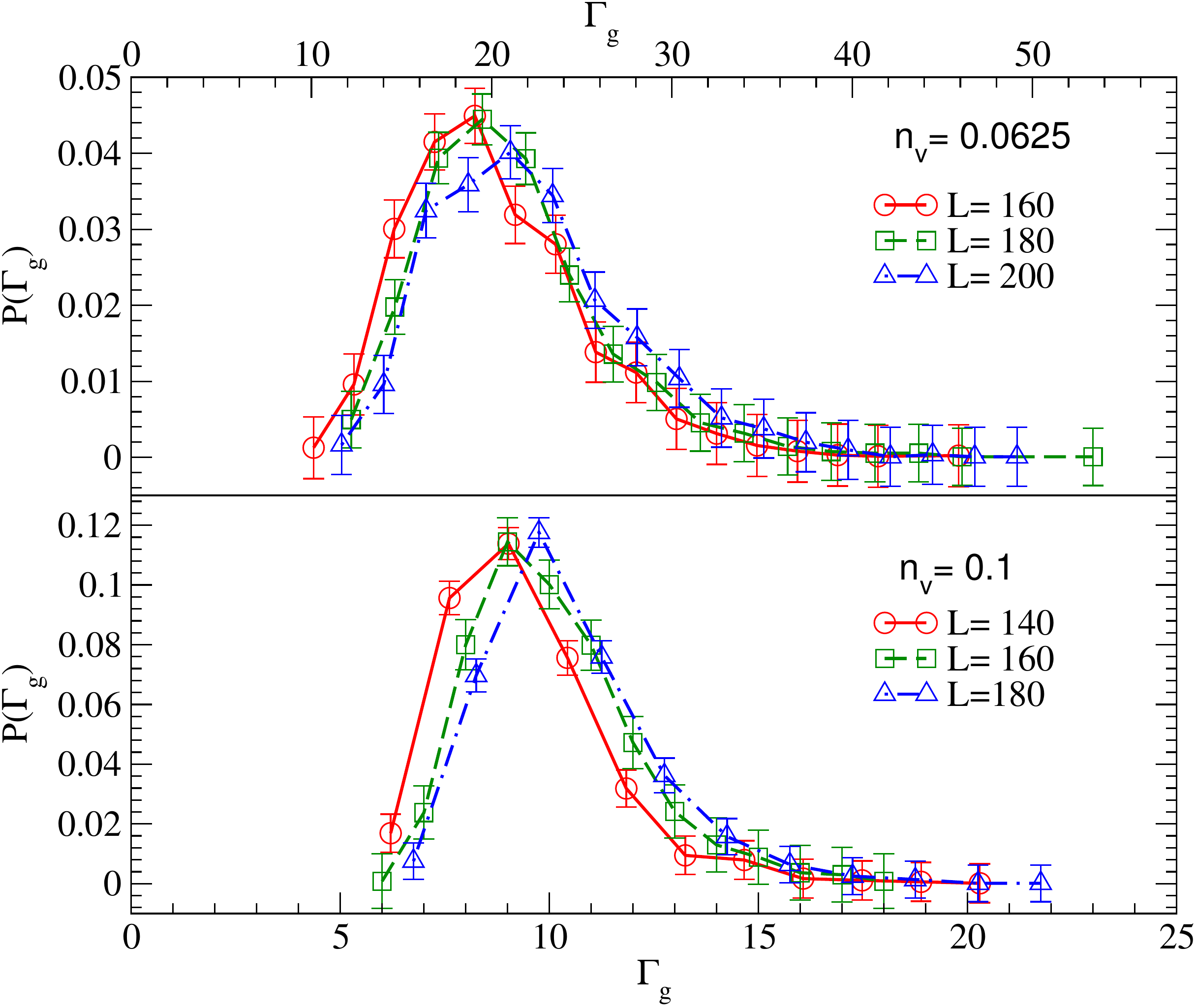}}
\caption{Histograms of $\Gamma_{\rm g}$ at the three largest values of $L$ studied for $n_v=0.0625$ and $n_v=0.1$.}
\label{SuppFig6}
\end{figure}

Figure~\ref{SuppFig1} displays $N_L(\Gamma)$ for $n_v=0.05$, for the three largest sizes studied. The corresponding extrapolation to the thermodynamic limit is shown in Fig.~\ref{SuppFig2}. The corresponding results for $n_v=0.075$ are displayed in Fig.~\ref{SuppFig3}.
In all these figures, we focus on $\Gamma < \Gamma_g^{*} (L_{\rm min})$, where $L_{\rm min}$
is the smallest size for which data is displayed, and $\Gamma_g^{*}$ is read off
from the peak in the histograms of $\Gamma_g$ shown in Fig.~\ref{SuppFig4} and Fig.~\ref{SuppFig5}. The corresponding histograms for $n_v=0.0625$ and $n_v=0.1$ are
displayed in Fig.~\ref{SuppFig6}

As is clear from these results for $n_v=0.05$ and $n_v=0.075$,  $N_{L}(\Gamma)$ is found to fit well to a power-law form
$N_{\rm 1D} \equiv c\Gamma^{-y}$ up to a fairly well-defined and readily-identified crossover scale $\Gamma_c(L) \equiv \log_{10}(t/|\epsilon_c(L)|)$.  However, beyond $\Gamma_c$, the asymptotic fall-off is clearly faster than a power law. While the increase of $\Gamma_c(L)$ with $L$ is more significant at the smallest concentration studied ($n_v=0.05$), it
is nevertheless clear that $\Gamma_c(L)$ does saturate to a finite value even in this
case. This is clear from the fact that the extrapolated thermodynamic density of states $N(\Gamma)$ (Fig.~\ref{SuppFig2}) also displays the same crossover seen in the finite-size data. In the large-$\Gamma$ regime beyond this crossover, the modified Gade-Wegner form $N_{GW}(\Gamma) \equiv  a\Gamma^{1/3}e^{-b\Gamma^{2/3}}$ is
seen to provide a very good fit of the data for both these concentrations.
The corresponding values of $\Gamma_c(L)$ and $\Gamma_c$, and of the best fit values of $y$, provide us additional points that fill in the curves shown in Fig.~4 and Fig.~6 of the
main text, which display the $n_v$ dependence of $y$ and $\Gamma_c$, and the
close relationship between $l_c \equiv N(\Gamma_c)^{-1/2}$ and  $l_w \equiv w^{-1/2}$.
Finally, we re-emphasize a point made already in the main text: Our computational constraints prevent us from accessing  the thermodynamic limit for the much larger values of  $\Gamma$ at which we expect to see the same crossover for the lowest concentration $n_v=0.02$.

\subsection{Extrapolations}
Since states at any finite $\Gamma$ ({\em i.e.}, away from the band center $\epsilon=0$) in such particle-hole symmetric hopping
problems are not critical, one expects the leading corrections
to the thermodynamic limit $N(\Gamma)$ at any finite $\Gamma$ to be regular rather than singular, similar to the finite-size
corrections expected in noncritical phases of matter ({\em i.e.}, away from critical points or critical lines).
Guided by this rationale, the thermodynamic limit $N(\Gamma)$ is obtained from $N_L(\Gamma)$ by performing a polynomial extrapolation in $1/L$ (note that we expect that the leading finite-size corrections
are ${\mathcal O}(1/L)$ rather than $\sim \exp(-L/\xi)$ because of ``surface'' contributions
associated with the semi-open boundary conditions we employ). 
Since we are careful to only use large enough sizes for which almost every sample has at least one zero mode ($f_L \approx 1$), our finite-size data is already rather close to the thermodynamic limit, leading to a rather small secular drift with increasing $L$.   In most cases, given the size of our error bars relative to the magnitude of this secular drift with $L$, the inclusion of the next-order term $c/L^2$ only results in an over-interpretation of statistical fluctuations. Therefore, a simple linear (in $1/L$) extrapolation $a + b/L$ has been used in
most cases.

\begin{figure}
{\includegraphics[width=10cm]{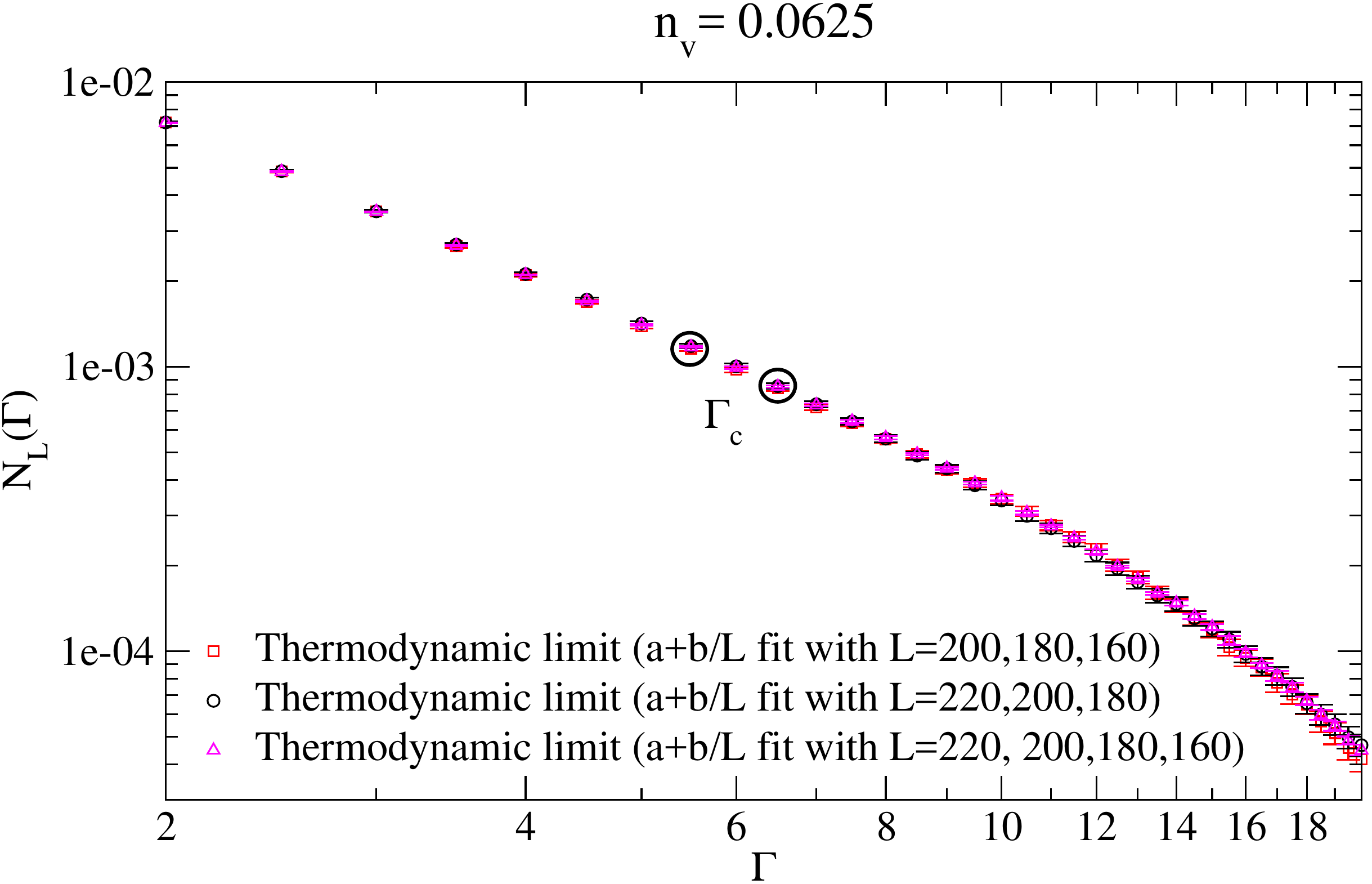}}
\caption{Three different extrapolations yield results for the thermodynamic limit
$N(\Gamma)$ that fall within the error bars of each other, confirming the reliability
and stability of our procedure to obtain the thermodynamic limit for the representative case of $n_v=0.0625$.}
\label{SuppFig7}
\end{figure}

\begin{figure}
{\includegraphics[width=8cm]{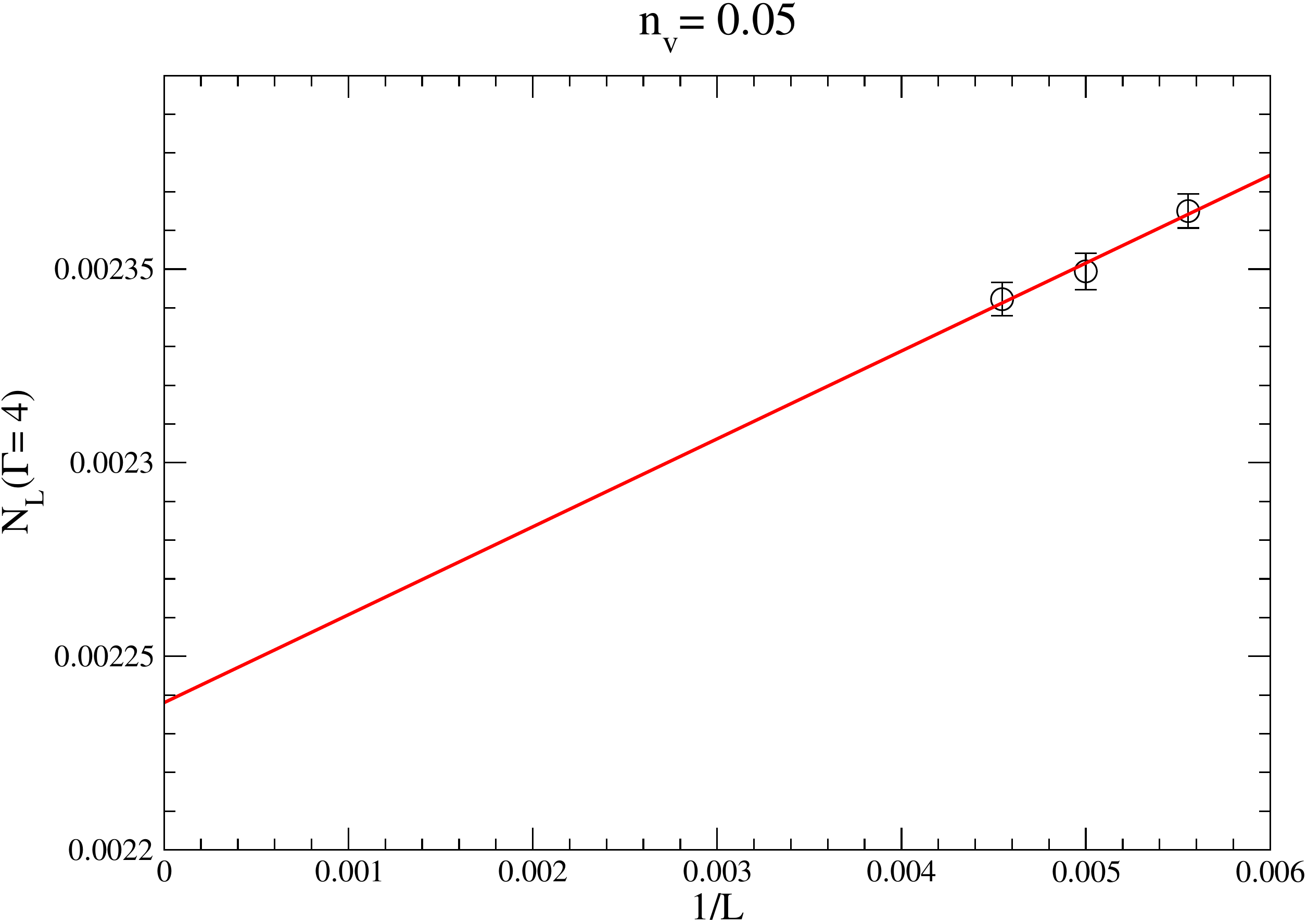}}
{\includegraphics[width=8cm]{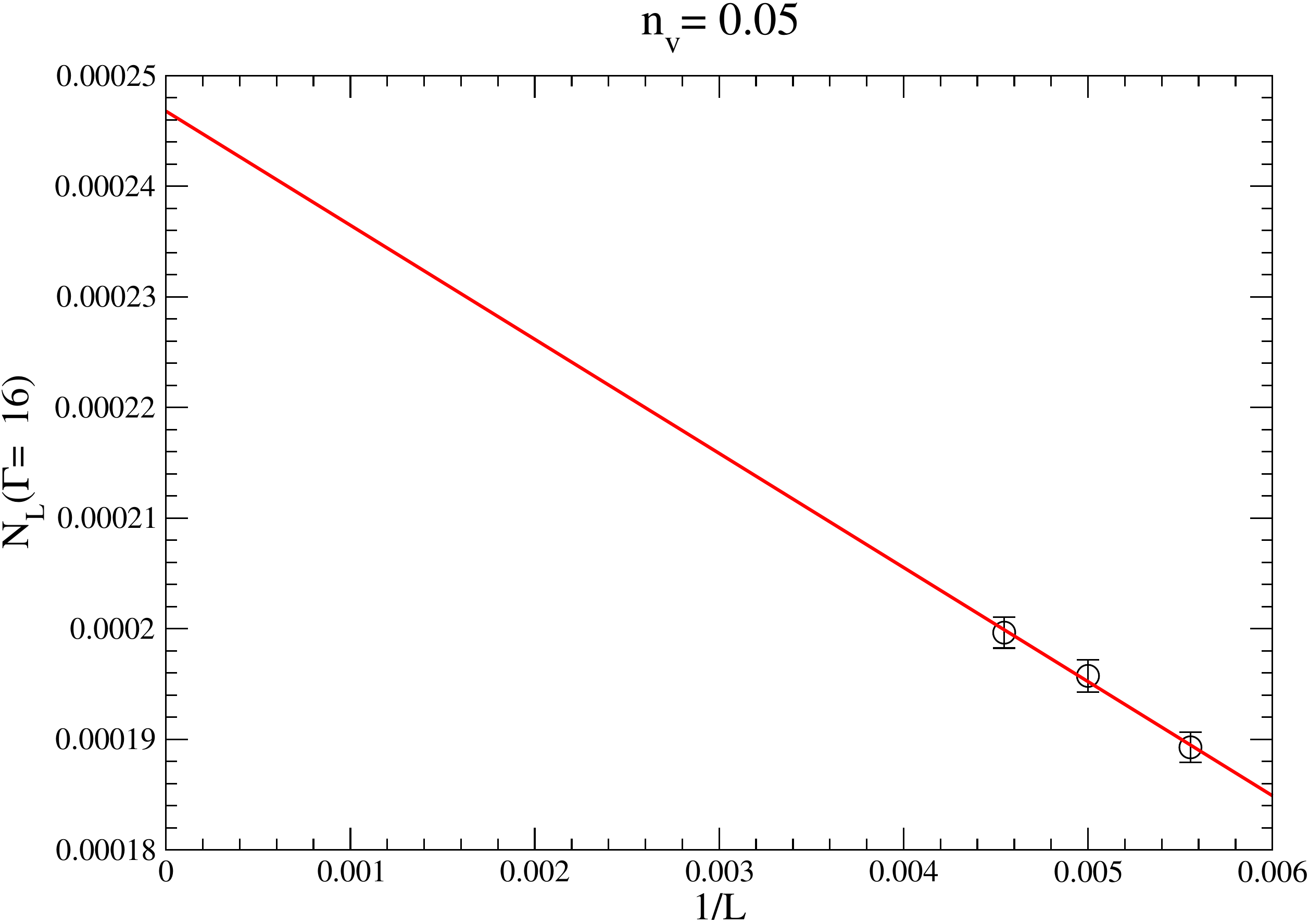}}
\caption{Examples of extrapolation of $N_L(\Gamma)$ to the thermodynamic
limit at $n_v=0.05$. For this concentration, $\Gamma_c \approx 11$ (see Fig.~4 in the main text), and the left panel illustrates the extrapolation for $\Gamma < \Gamma_c$, while the right panel is for $\Gamma > \Gamma_c$.
Note in particular that our extrapolation for $\Gamma > \Gamma_c$ is very likely an overstimate, so one can be fairly confident that $N(\Gamma)$ in the thermodynamic limit drops below 
$N_{1D}(\Gamma)$, ruling out a fit to this form for $\Gamma > \Gamma_c$.
}
\label{SuppFig8}
\end{figure}

\begin{figure}
{\includegraphics[width=8cm]{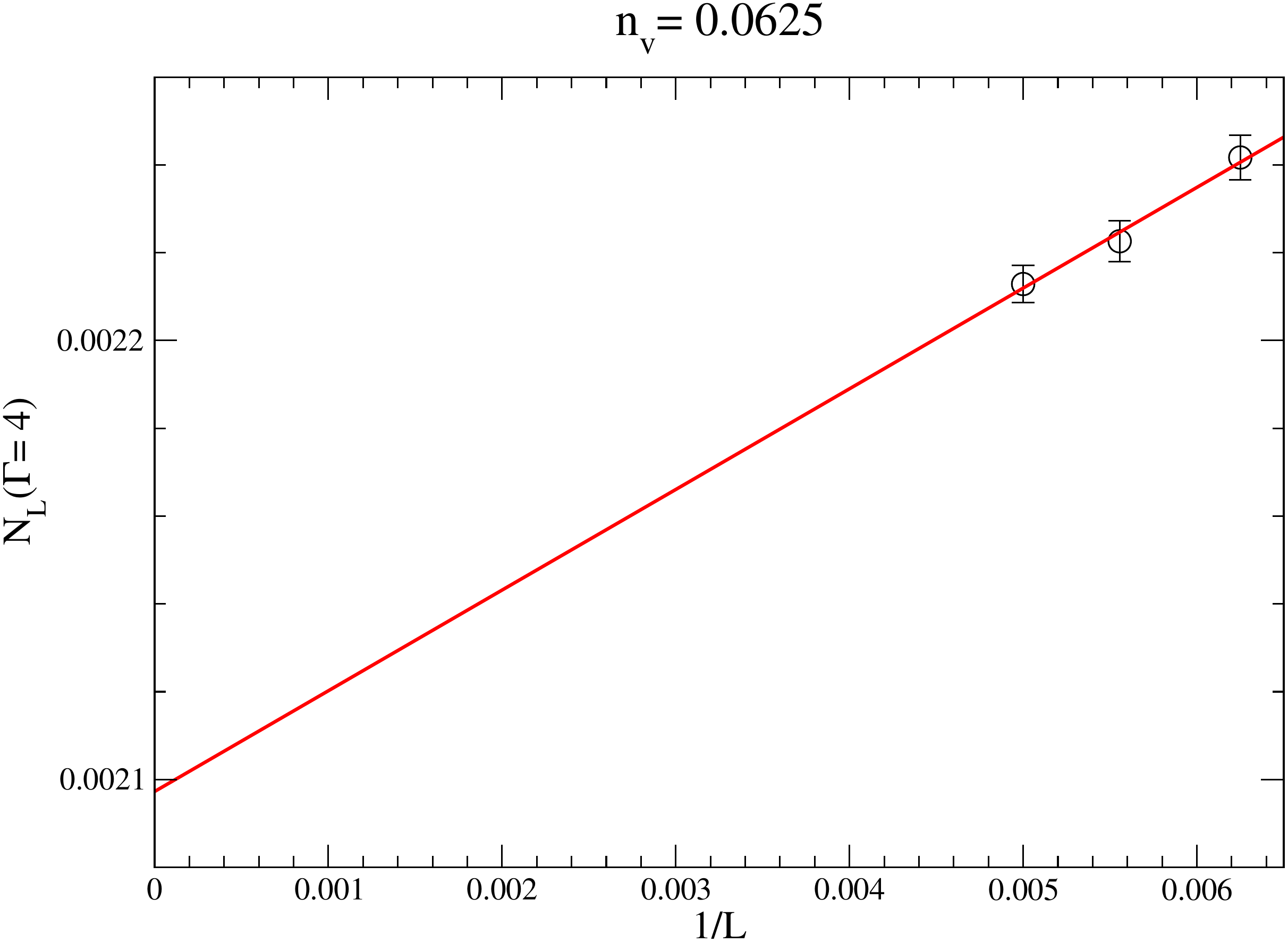}}
{\includegraphics[width=8cm]{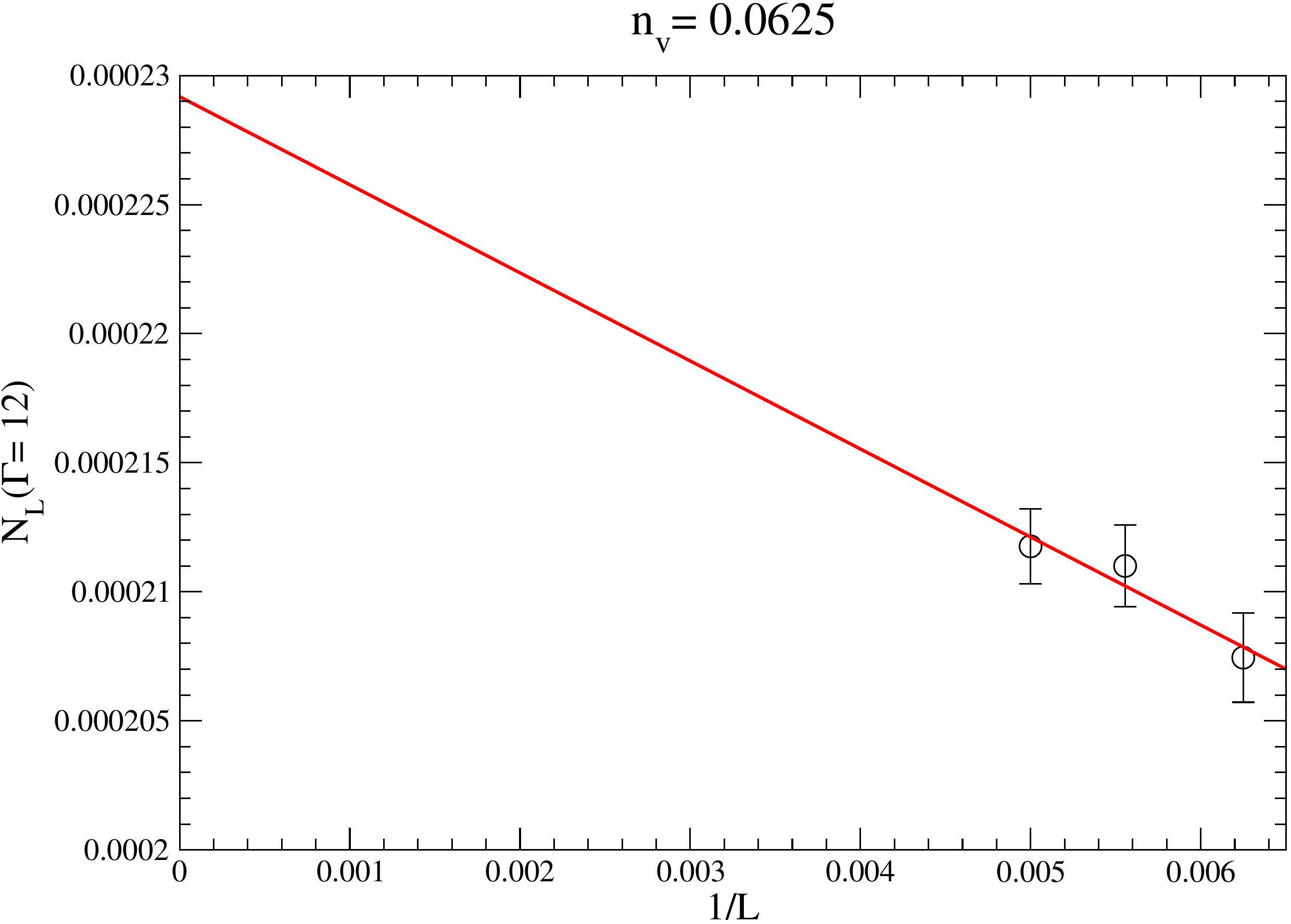}}
\caption{Examples of extrapolation of $N_L(\Gamma)$ to the thermodynamic
limit at $n_v=0.0625$. For this concentration, $\Gamma_c \approx 7$ (see Fig.~4 in the main text), and the left panel illustrates the extrapolation for $\Gamma < \Gamma_c$, while the right panel is for $\Gamma > \Gamma_c$. Note in particular that our extrapolation for $\Gamma > \Gamma_c$ is very likely an overstimate, so one can be fairly confident that $N(\Gamma)$ in the thermodynamic limit drops below 
$N_{1D}(\Gamma)$, ruling out a fit to this form for $\Gamma > \Gamma_c$. Extrapolations at other values of $\Gamma$, as well as extrapolations including a larger size ($L=220$) are shown in Fig.~{\protect{\ref{SuppFig20}}}.
}
\label{SuppFig9}
\end{figure}

\begin{figure}
{\includegraphics[width=8cm]{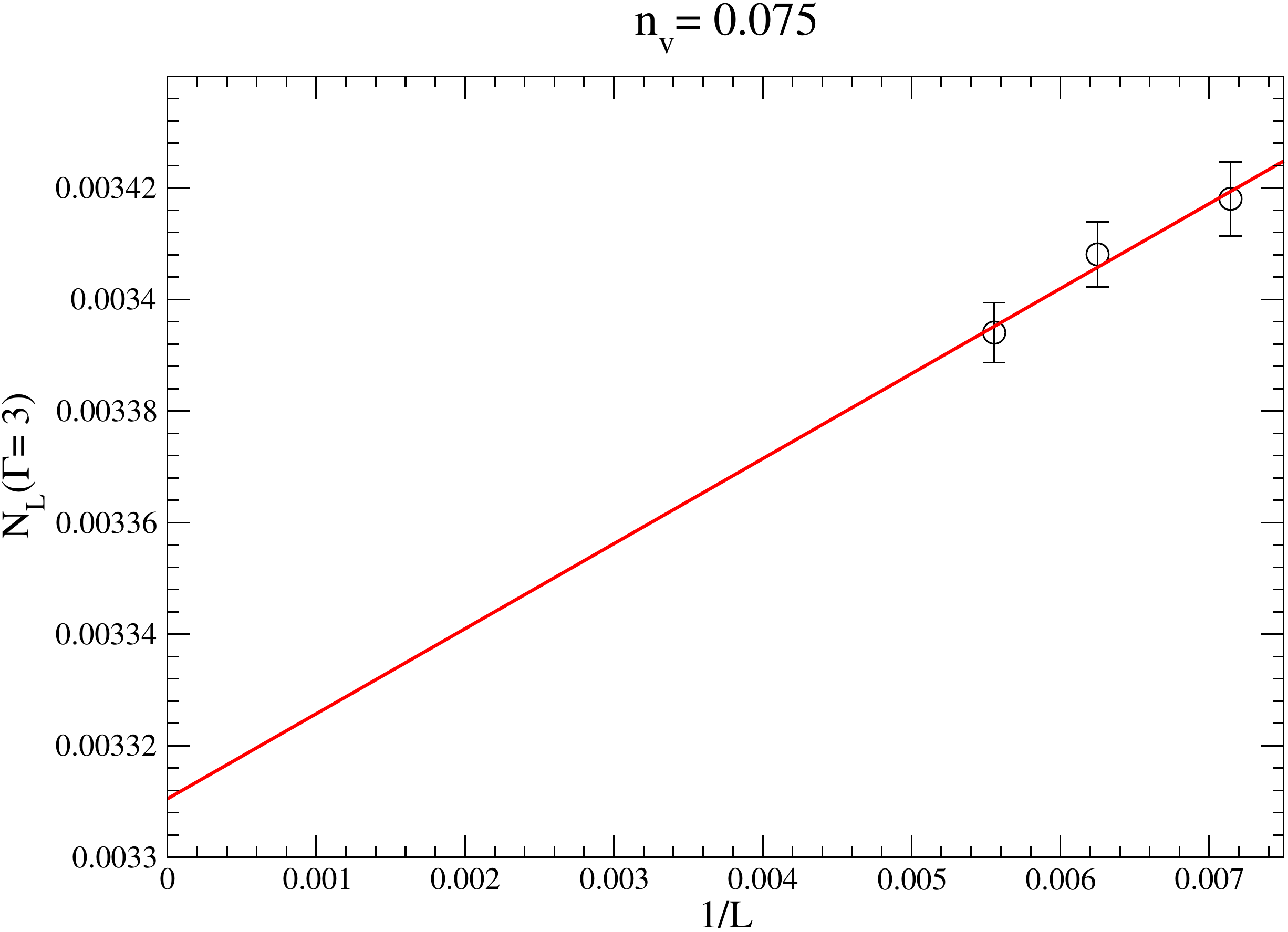}}
{\includegraphics[width=8cm]{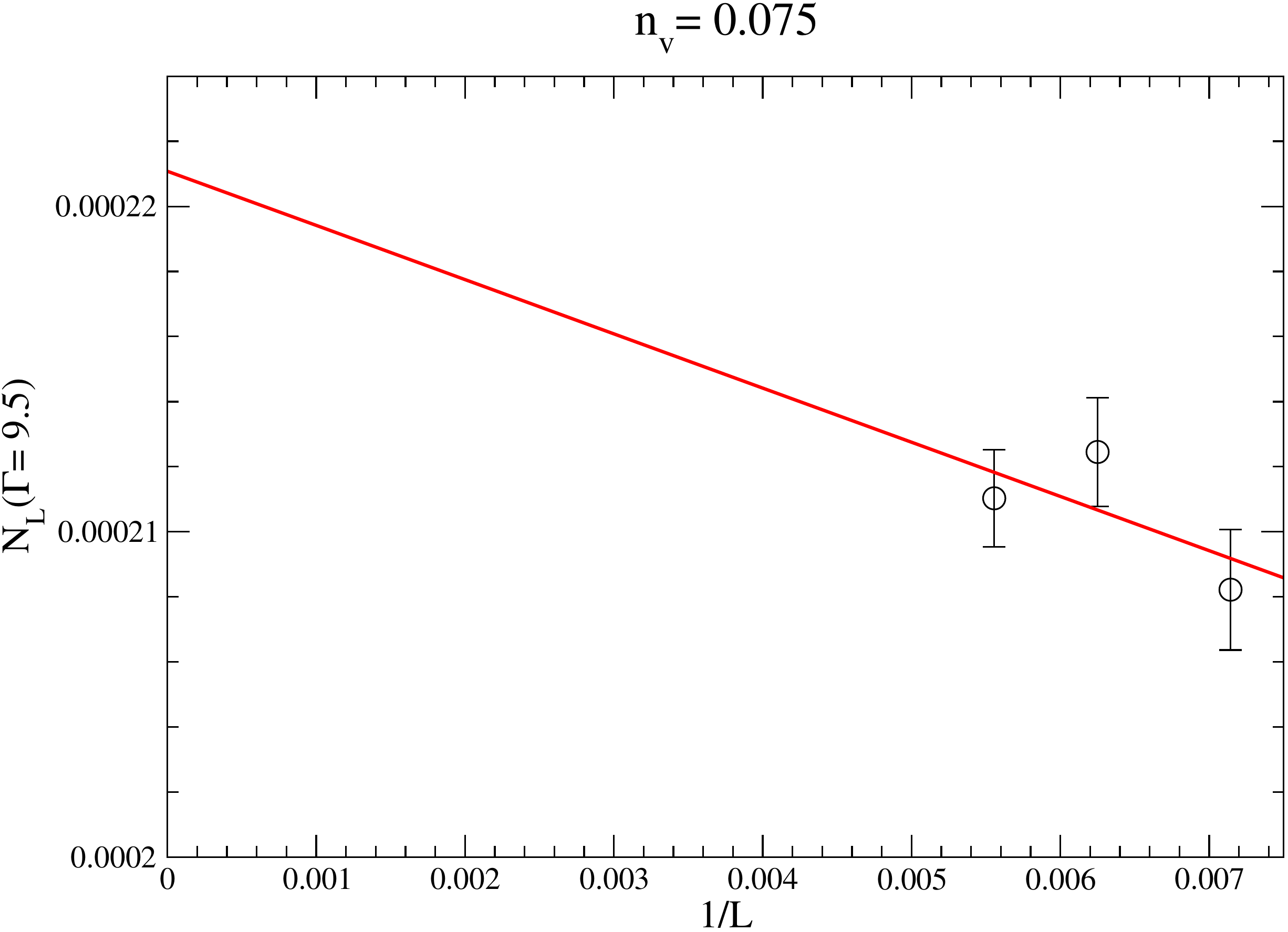}}
\caption{Examples of extrapolation of $N_L(\Gamma)$ to the thermodynamic
limit at $n_v=0.075$.  For this concentration, $\Gamma_c \approx 5$ (see Fig.~4 in the main text), and the left panel illustrates the extrapolation for $\Gamma < \Gamma_c$, while the right panel is for $\Gamma > \Gamma_c$. Note in particular that our extrapolation for $\Gamma > \Gamma_c$ is very likely an overstimate, so one can be fairly confident that $N(\Gamma)$ in the thermodynamic limit drops below 
$N_{1D}(\Gamma)$, ruling out a fit to this form for $\Gamma > \Gamma_c$.
}
\label{SuppFig10}
\end{figure}

\begin{figure}
{\includegraphics[width=8cm]{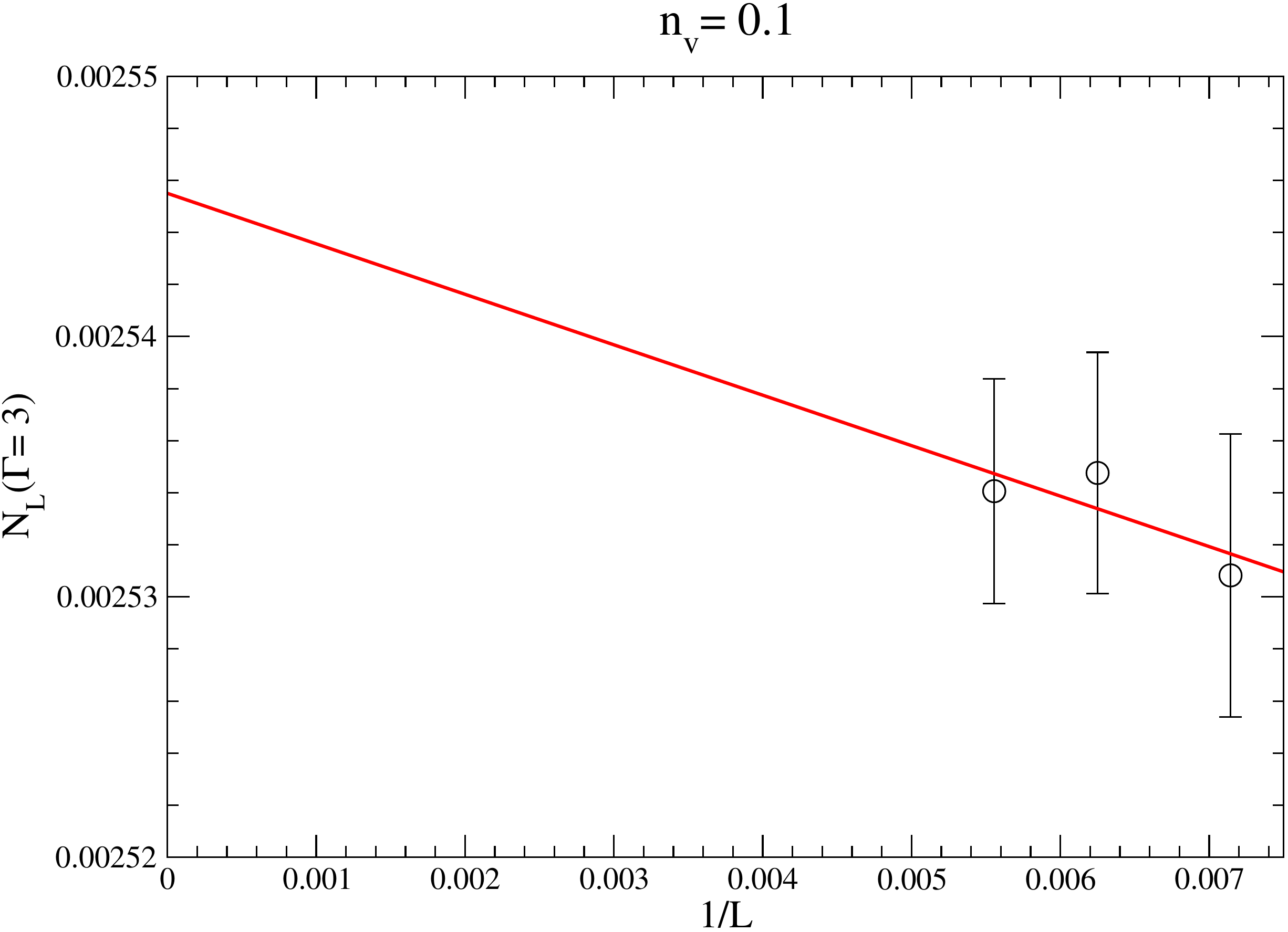}}
{\includegraphics[width=8cm]{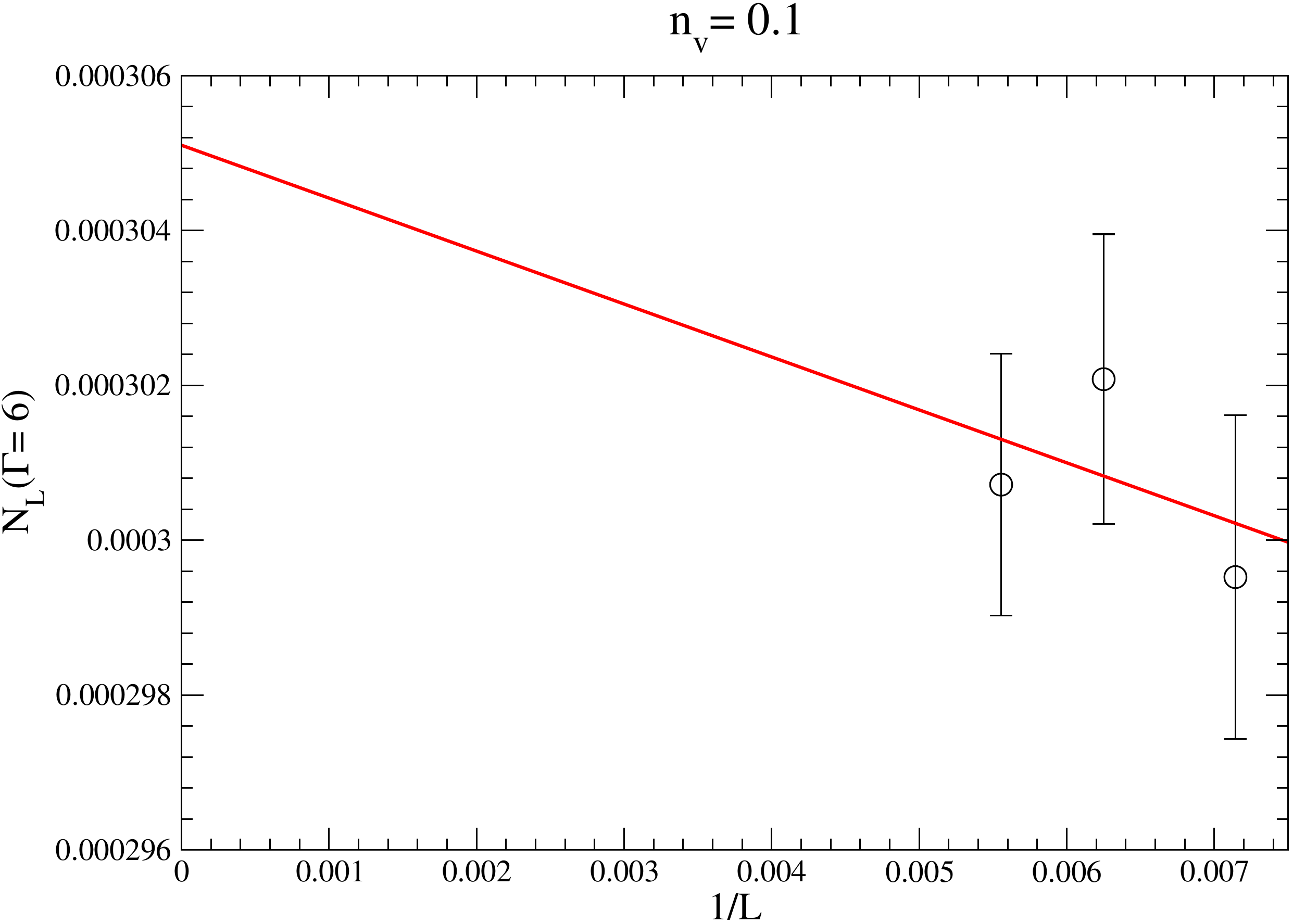}}
\caption{Examples of extrapolation of $N_L(\Gamma)$ to the thermodynamic
limit at $n_v=0.1$. For this concentration, $\Gamma_c \approx 4$ (see Fig.~4 in the main text), and the left panel illustrates the extrapolation for $\Gamma < \Gamma_c$, while the right panel is for $\Gamma > \Gamma_c$. Note in particular that our extrapolation for $\Gamma > \Gamma_c$ is very likely an overstimate, so one can be fairly confident that $N(\Gamma)$ in the thermodynamic limit drops below 
$N_{1D}(\Gamma)$, ruling out a fit to this form for $\Gamma > \Gamma_c$.
}
\label{SuppFig11}
\end{figure}

We have also tested the stability of this extrapolation procedure to the inclusion of
data at larger sizes. For the representative case of $n_v=0.0625$, this
is shown Fig.~\ref{SuppFig7}, which is devoted to a comparison of  the thermodynamic limit obtained in the main text using sizes $L=200,180,160$ with two other alternatives: A linear extrapolation from three sizes $L=220,200,180$, and a linear extrapolation from four sizes $L=220,200,180,160$.  As is clear from this figure, all three extrapolations ({\em i.e.}, the one
used in the main text as well as the other alternatives which use data at a larger size) yield extrapolated values that lie within the error-bars of each other. Further, there is no systematic trend that suggests that any one of these extrapolations yields a consistently higher or lower value of $N(\Gamma)$ at all $\Gamma$. Details of
all three extrapolations, for each value of $\Gamma$, are also shown as a separate multi-page figure (Fig.~\ref{SuppFig20}) placed at the end of this supplemental section for ease of inspection. Some examples of extrapolations used to arrive at $N(\Gamma)$ from data for $N_{L}(\Gamma)$ at other concentrations
are also shown in Figs.~\ref{SuppFig8},~\ref{SuppFig9},~\ref{SuppFig10},~and~\ref{SuppFig11}. From this careful and detailed study, we conclude that our approach indeed
allows us to reliably obtain the thermodynamic limit curve $N(\Gamma)$. 
\begin{figure}
{\includegraphics[width=10cm]{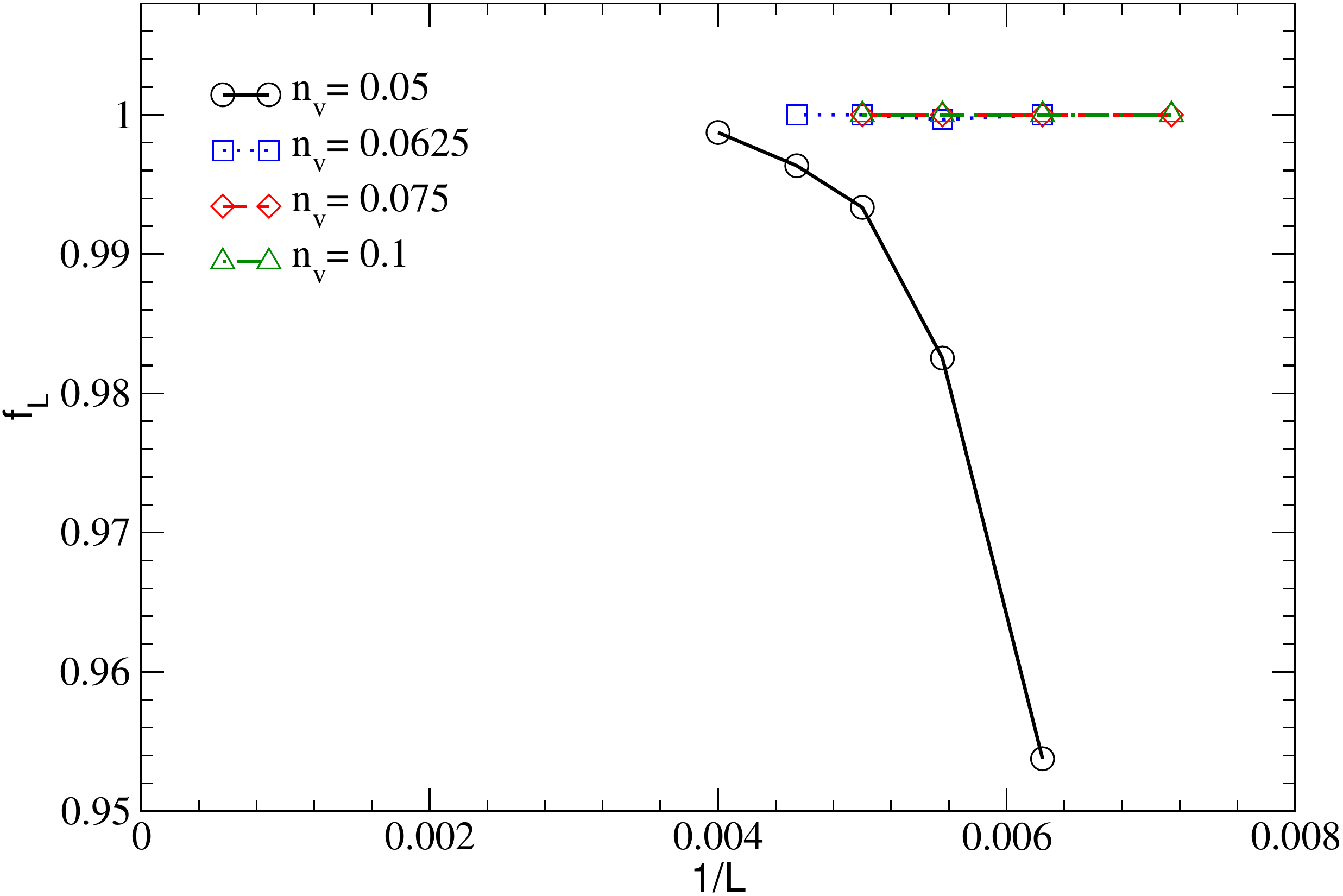}}
\caption{Probability $f_L$ that an $L\times L$ sample has at least one pair of zero modes tends to $1$ in the thermodynamic limit for each concentration studied. Due to our
computational constraints, we have been unable to obtain such data at $n_v=0.02$,
where we expect the density of zero modes to be much lower, but $f_L$ to still
tend to $1$ in the thermodynamic limit (based on the analytical arguments given
in the main text).}
\label{SuppFig12}
\end{figure}

\section{Further analysis of zero modes}
\begin{figure}
{\includegraphics[width=10cm]{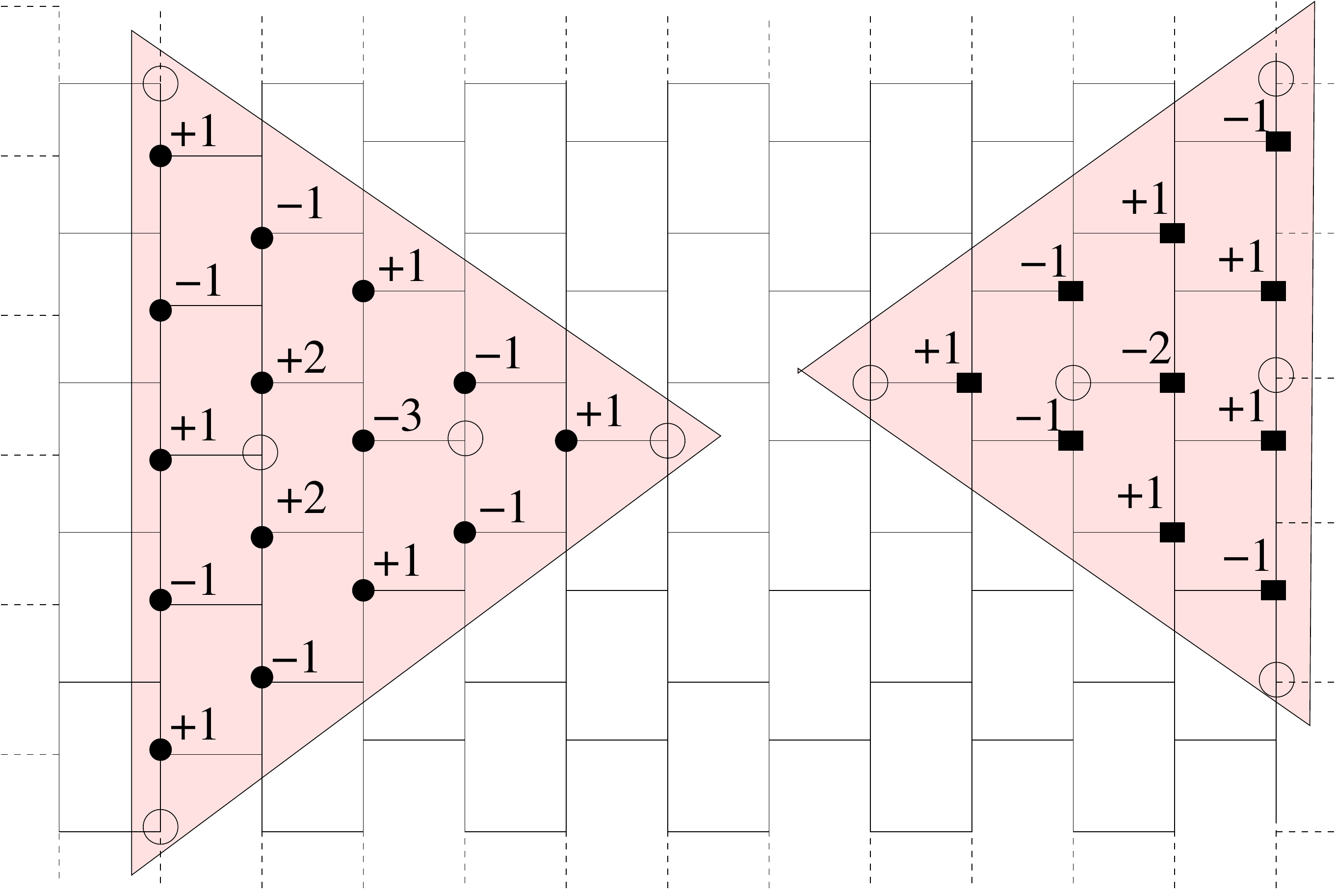}}
\caption{Two kinds of 5-vacancy clusters (``5-triangles'') that host an exact zero
mode, with the corresponding wavefunction marked. Open circles correspond to vacancies.}
\label{SuppFig13}
\end{figure}

\begin{figure}
{\includegraphics[width=10cm]{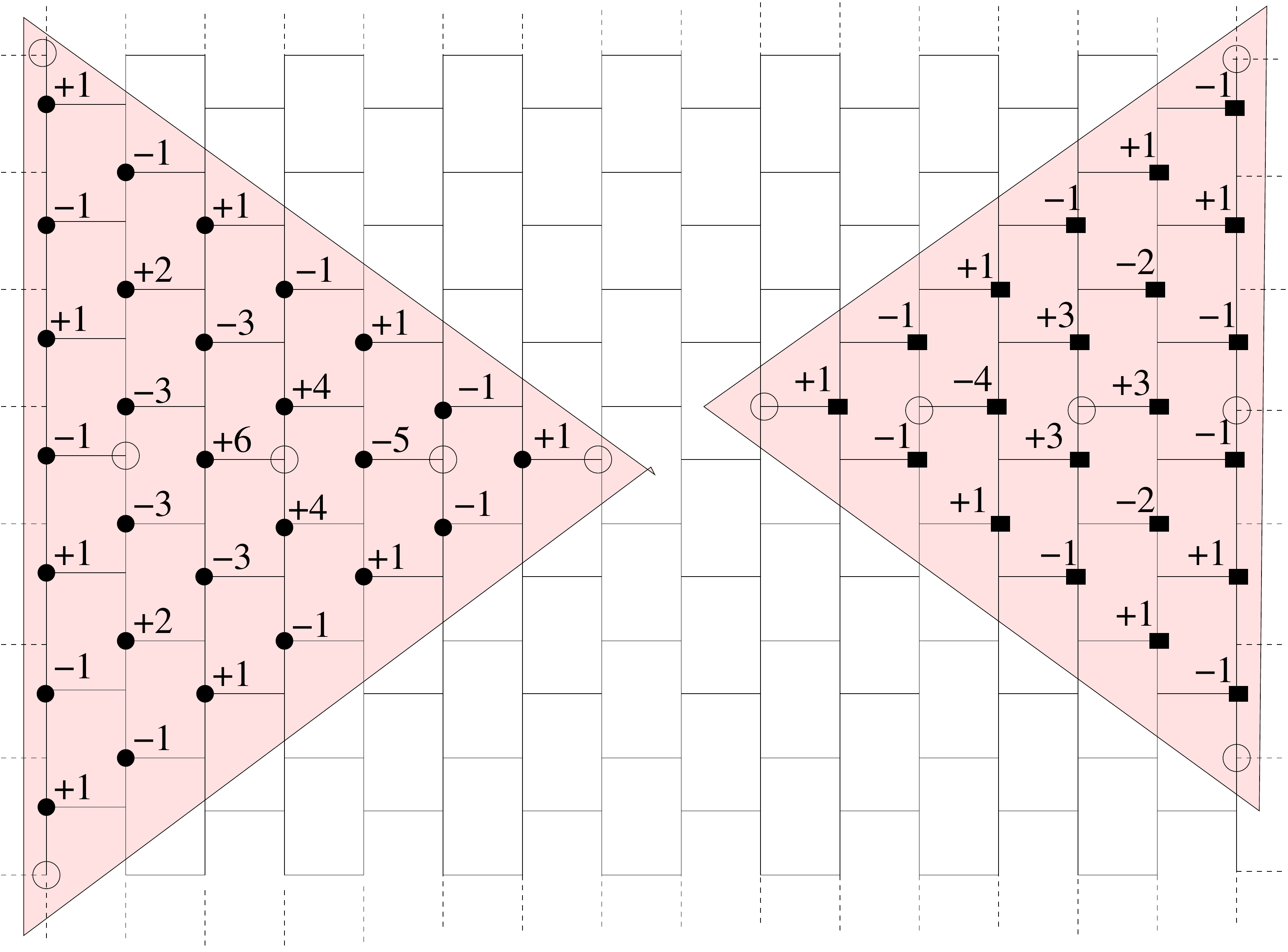}}
\caption{Two kinds of 6-vacancy clusters (``6-triangles'') that host an exact zero
mode, with the corresponding wavefunction marked. Open circles correspond to vacancies.}
\label{SuppFig14}
\end{figure}

Our data for $f_L$, the probability that an $L \times L$ sample has at least one
zero mode, is shown in Fig.~\ref{SuppFig12}. Clearly, $f_L$ tends to $1$ as
$L \rightarrow \infty$, as already mentioned in the main text. This is consistent
with the analytical argument in the main text, which also provides a simple
rigorous lower bound for the density of zero modes. 
The 4-triangle zero mode used in this argument is the first term in an
infinite series in $n_v$, with higher powers of $n_v$  arising from
bigger patterns consisting of a larger number of impurities in specific locations relative
to each other.
In Fig.~\ref{SuppFig13} and Fig.~\ref{SuppFig14},
we show
a few examples of zero mode constructions that contribute to this series.
However, as already noted in the main text, terms in this series do not give the dominant contribution to $w$ at the not-too-small values of $n_v$ studied by us in this work. Indeed,
we have explicitly measured the density of 4-triangles and checked that it is significantly smaller than the density of zero modes for all $n_v$ at which we have computed $w$  (including
$n_v = 0.05$). Additionally, we have enumerated all possible clusters of fewer than
four impurities and verified that it is not possible to produce a similar zero mode
with fewer than four vacancies in a cluster so long as the exclusion constraints
outlined in the main text are in place.
\begin{figure}
{\includegraphics[width=10cm]{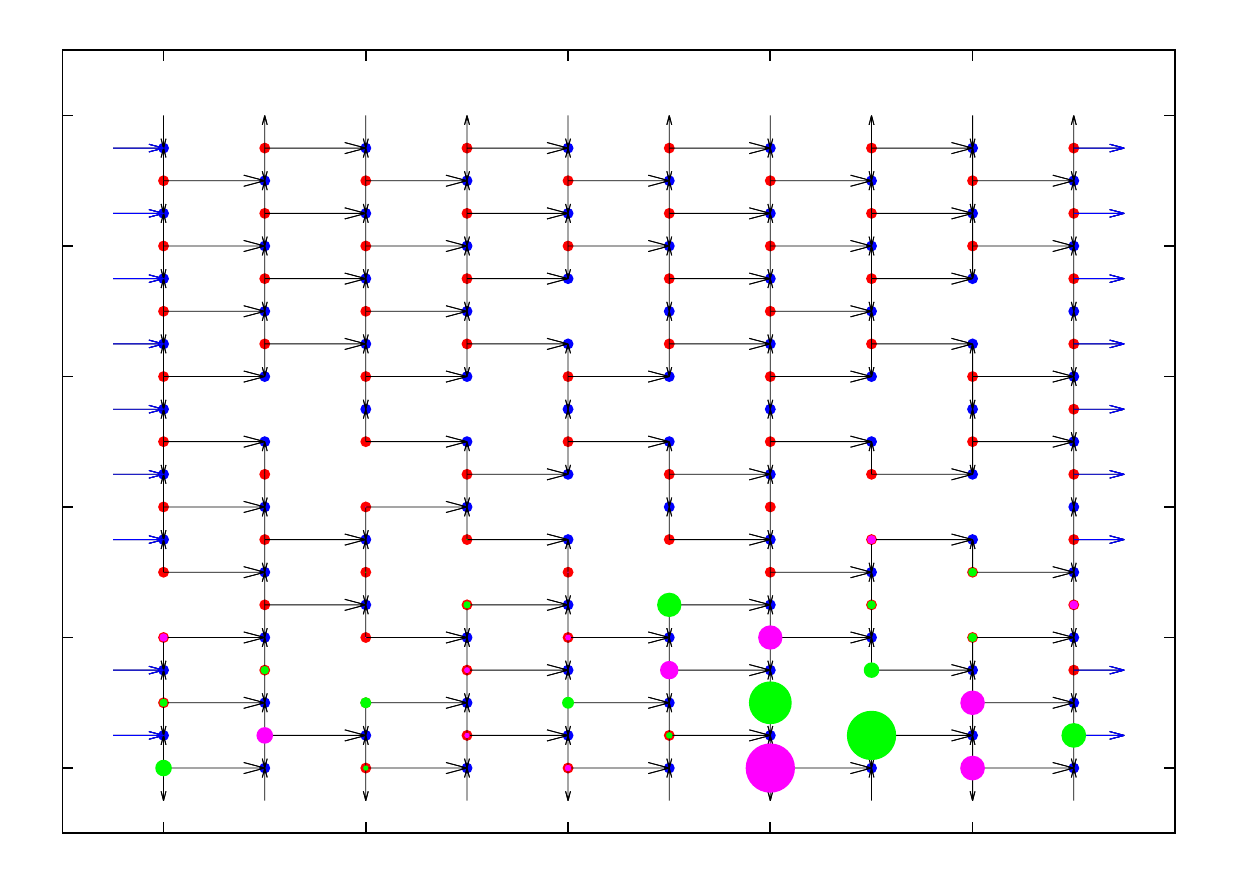}}
\caption{An $L \times L$ semi-open sample (of the type consistently used in all our numerical work) with $L=10$, with vacancies represented by missing lattice sites. This sample
provides a simple example of a zero mode that does not seem to arise from any of
the regular arrangements of vacancies used in our zero mode constructions. The actual
wavefunction of this zero mode is represented by color-coded circles. The size of
the circle corresponds to the magnitude of the wavefunction at the corresponding site,
while the two different colors represent opposite signs for the wavefunction at
the corresponding sites.}
\label{SuppFig15}
\end{figure}
\begin{figure}
{\includegraphics[width=10cm]{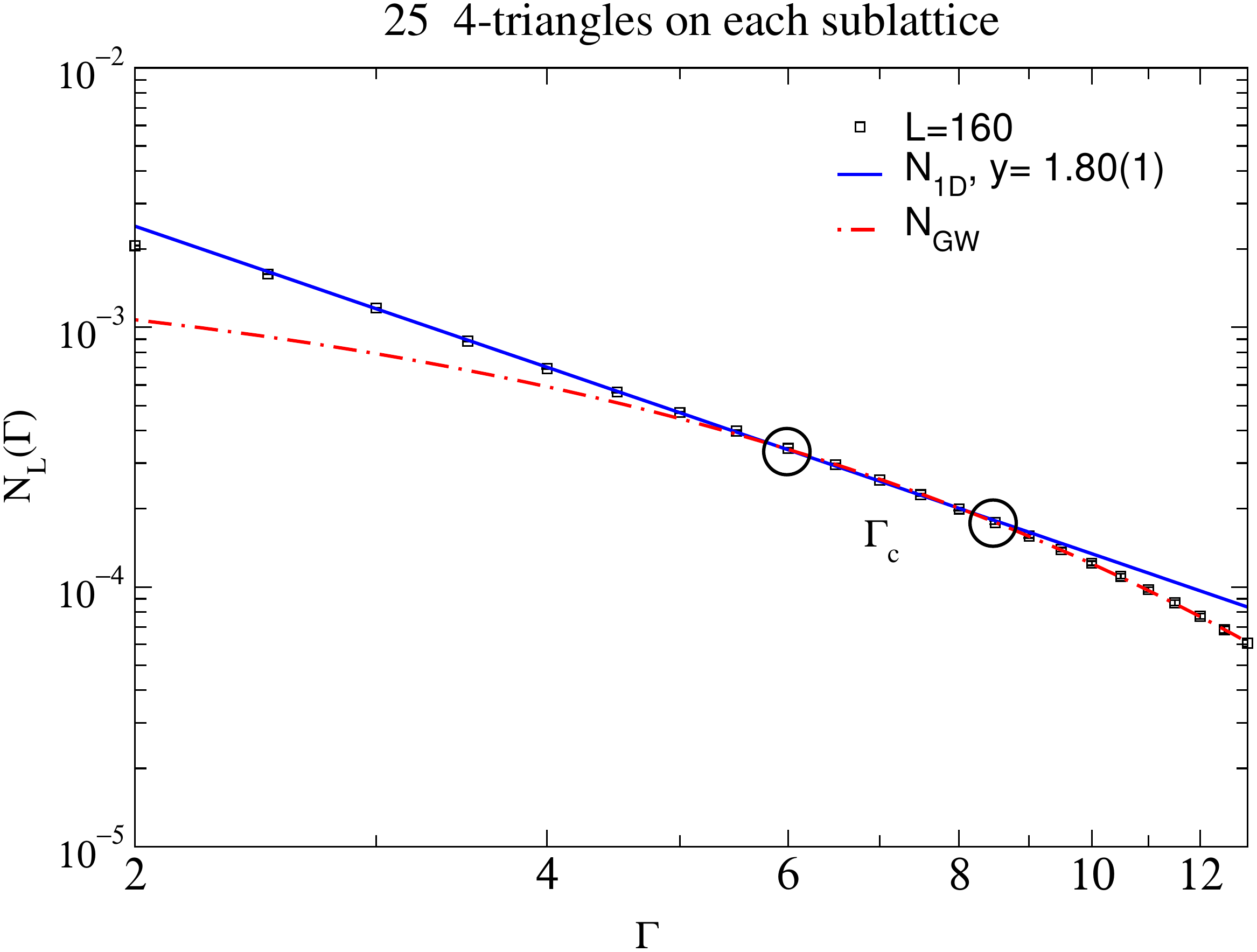}}
\caption{$N_L(\Gamma)$ in the toy model in which a $L=160$ sample
is diluted with $25$ randomly placed 4-triangles on each sublattice. Circles demarcate
the crossover region centered at the crossover scale $\Gamma_c$. Data for $\Gamma \lesssim \Gamma_c$ fits well to power-law form $N_{\rm 1D}(\Gamma)$ with the value
of $y$ indicated in the figure, while the large-$\Gamma$
regime fits well to the modified Gade-Wegner form $N_{\rm GW}(\Gamma)$.}
\label{SuppFig16}
\end{figure}
\begin{figure}
{\includegraphics[width=10cm]{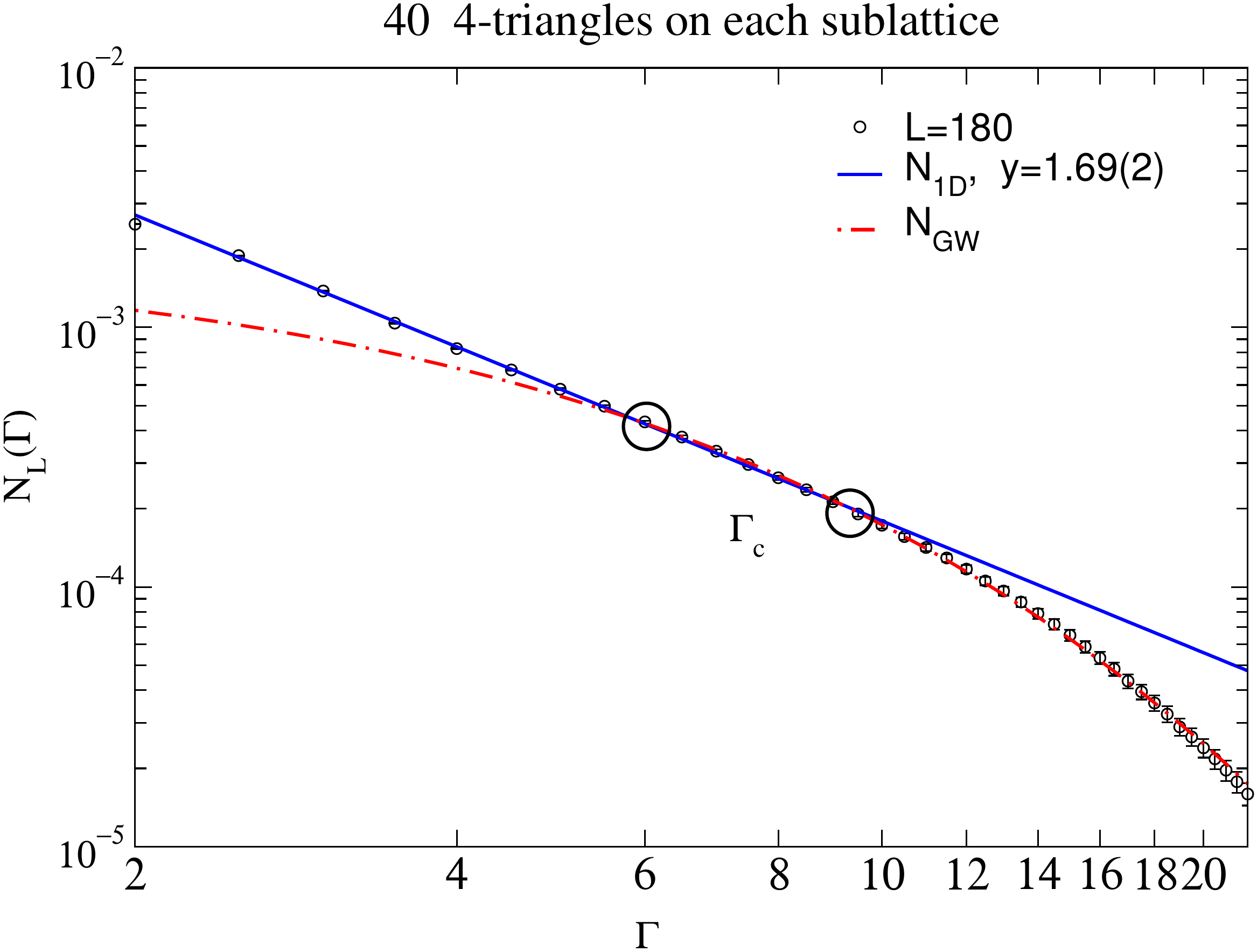}}
\caption{$N_L(\Gamma)$ in the toy model in which a $L=180$ sample
is diluted with $40$ randomly placed 4-triangles on each sublattice. Circles demarcate
the crossover region centered at the crossover scale $\Gamma_c$. Data for $\Gamma \lesssim \Gamma_c$ fits well to power-law form $N_{\rm 1D}(\Gamma)$ with the value
of $y$ indicated in the figure, while the large-$\Gamma$
regime fits well to the modified Gade-Wegner form $N_{\rm GW}(\Gamma)$.}
\label{SuppFig17}
\end{figure}

The zero mode associated with the ${\mathcal R}_6$ motif described in the main text also generalizes in an obvious way to yield
a series of zero modes that all survive the effects of bond disorder in a manner completely
analogous to the ${\mathcal R}_6$ zero mode. These ${\mathcal R}_n$ zero modes ($n>6$) live on larger and larger equilaterial triangles (with zig-zag edges) which are connected to the rest of the lattice only via $B$ ($A$) sublattice sites but have more undeleted $A$ ($B$) sublattice sites than $B$ ($A$) sublattice sites, allowing a zero mode to exist within the triangle for generic realizations of bond-disorder. As in the case of the ${\mathcal R}_6$ zero
mode described in the main text, this robustness to disorder follows from the fact that the number of free components of the wavefunction of any such mode is one more than
the number of zero-energy equations that they must satisfy.

\begin{figure}
{\includegraphics[width=10cm]{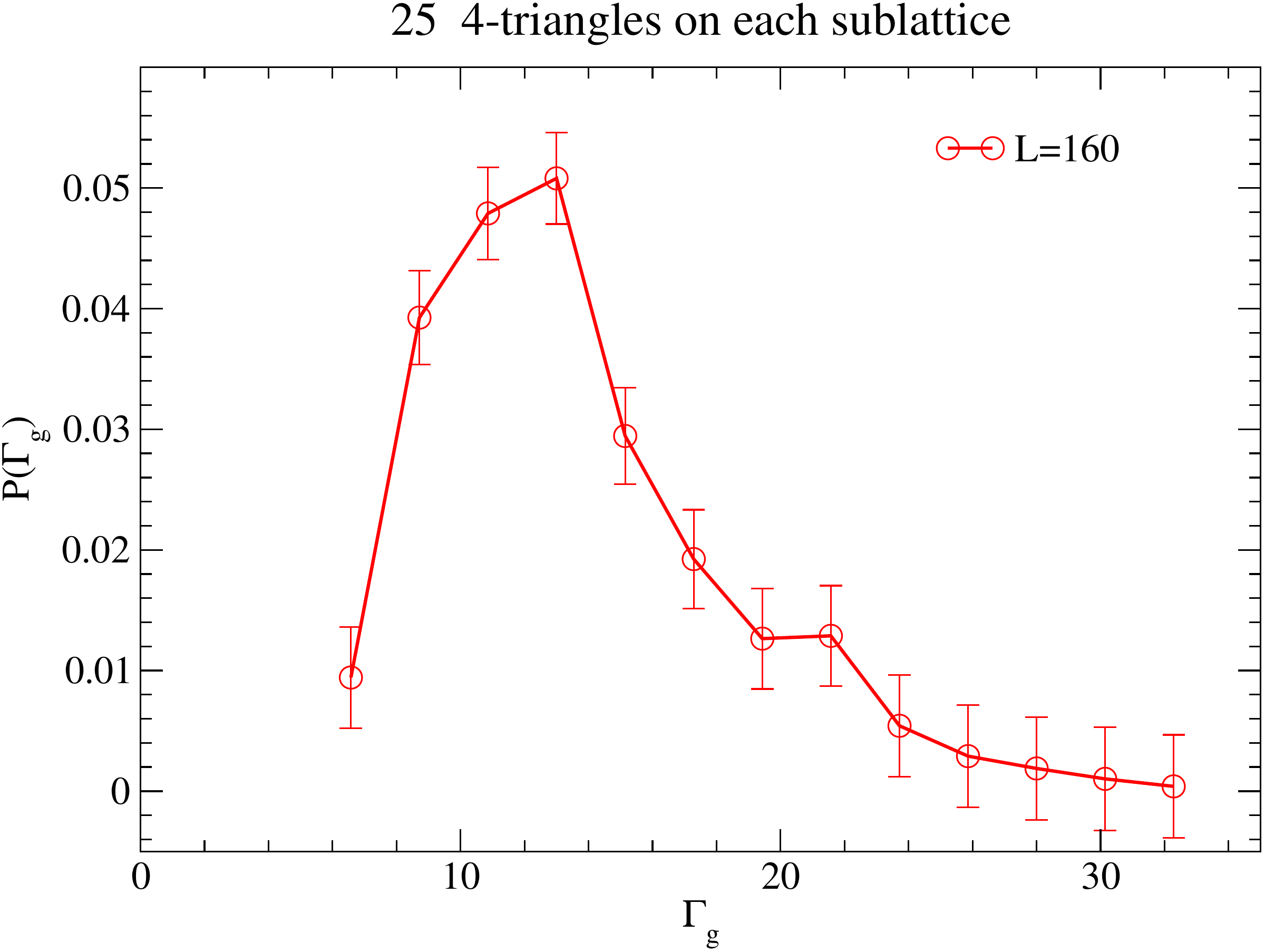}}
\caption{Histogram of $\Gamma_g$, corresponding to the lowest nonzero gap
for the $L=160$ sample diluted with $25$ randomly placed 4-triangles on each sublattice.}
\label{SuppFig18}
\end{figure}

We have also found other simple examples of such ``${\mathcal R}$-type'' zero modes that live near the armchair boundary
and are not associated with a specific regular arrangement of vacancies.
Instead, as already mentioned earlier, these modes appear to generically live in a region ${\mathcal R}$ which connects to the rest of the lattice only via $B$ ($A$) sublattice sites belonging to ${\mathcal R}$, although it has more undeleted
$A$ ($B$) sublattice sites than $B$ ($A$) sublattice sites. In such a region,  $T_{AB}T^{\dagger}_{AB}$ ($T^{\dagger}_{AB}T_{AB}$)
has a zero mode living on the $A$ ($B$) sublattice sites, simply because the number
of constraints that need to be satisfied by this zero mode wavefunction is smaller
than the number of $A$ ($B$) sublattice sites on which this zero mode lives.
As already noted, this feature also guarantees that such zero modes survive the effects
of disorder in the nearest-neighbour hopping amplitudes.
One example of such a mode is shown in Fig.~\ref{SuppFig15}.
We believe that bulk versions of such more general ${\mathcal R}$-type zero modes provide the dominant
contribution to $w$ for the values of $n_v$ studied by us, which is why our lower-bound on $w$ (obtained by thinking in terms of Fig.~2 in the main text)
 substantially underestimates $w$ at such not-too-small values of $n_v$. Clearly, no additional local correlations among impurities can entirely eliminate such more general ${\mathcal R}$-type zero modes . Therefore, a non-zero density of zero-energy modes is expected to be a generic
feature of such systems. However, we have been unable to convert this observation into an improved lower-bound.
\begin{figure}
{\includegraphics[width=10cm]{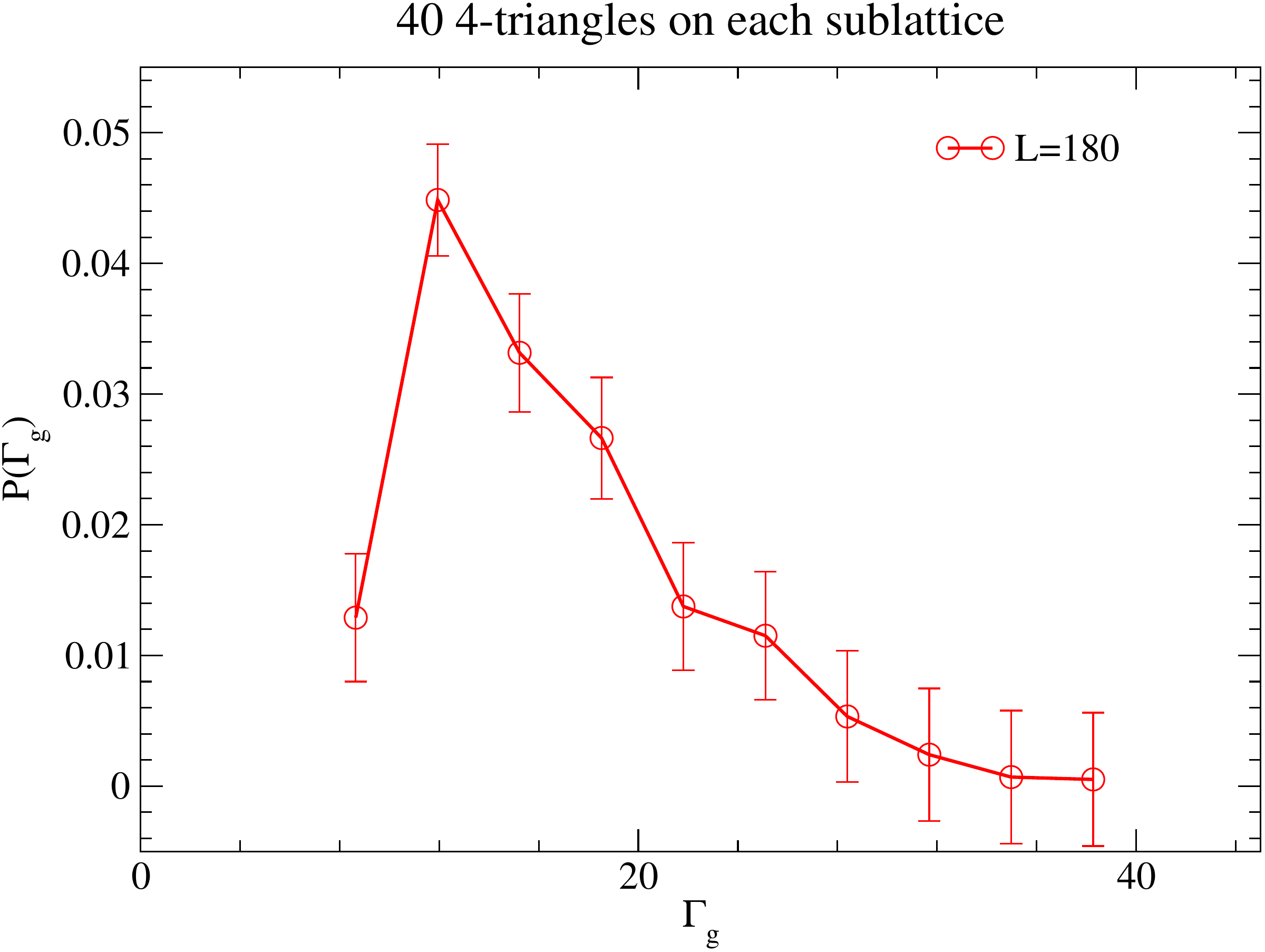}}
\caption{Histogram of $\Gamma_g$, corresponding to the lowest nonzero gap
for the $L=180$ sample diluted with $40$ randomly placed 4-triangles on each sublattice.}
\label{SuppFig19}
\end{figure}

\section{Dilution by 4-triangles}
Finally, we provide an illustration of the importance of spatial correlations
between vacancies via a simple toy model in which vacancies enter the sample
only in groups of four, arranged as a 4-triangle at random locations in the sample (as in Fig. 3 of the main text).
In Figs.~\ref{SuppFig16}~and~\ref{SuppFig17}, we respectively display the density of states
of $L \times L$ samples with $L=160$ and $L=180$. The $L=160$ sample is
diluted by $25$ 4-triangles placed at random on each sublattice, while the $L=180$ sample is diluted
with $40$ 4-triangles placed at random on each sublattice. The former sample
corresponds to a ``bare'' value of $n_v\approx 0.0039$, while the latter sample
corresponds to a bare value of $n_v\approx 0.0049$. These values of
$n_v$ are an order of magnitude different from the values of $n_v$ studied by us in
the main part of our work (in which the impurities are uncorrelated except for exclusion constraints designed to prevent the occurrence of ``trivial'' zero modes). However,
since all vacancies go in as part of a 4-triangle, $w \approx 9.76 \times 10^{-4}$ for
the $L=160$ sample and $w \approx 1.23 \times 10^{-3}$ for the $L=180$ sample. The values of $w$ are thus very similar to those obtained in our independently diluted
samples with $n_v$ in the range $0.05$---$0.06$. 

From Figs.~\ref{SuppFig16}~and~\ref{SuppFig17}, we see that the density of states
again undergoes a crossover that is qualitatively the same as the crossover
identified in our main study. However, the corresponding $\Gamma_c$ is much smaller
({\em i.e.}, the energy scale $|\epsilon_c|$ is much larger) than one would have expected based on the value of the overall vacancy concentration $n_v$ (had the vacancies been independent as in the main study). Similarly, the value of $y$ is also very different from
the (extrapolated) value of $y$ one would have expected at such small $n_v$. 
The corresponding histograms
of $\Gamma_g$ are shown in Figs.~\ref{SuppFig18}~and~\ref{SuppFig19}.
From these figures, we see that $\Gamma_g^{*}$, corresponding to the position of the peak in the histogram of $\Gamma_g$, is significantly smaller than one would have expected based on the overall
vacancy concentration $n_v$ (had the vacancies been independent, as in the main study).
This provides a simple illustration of the importance of spatial correlations between
vacancies in setting the lowest gap scale $\Gamma_g^{*}$, and the density of zero modes
$w$. It also emphasizes that the crossover identified by us is a robust and generic
aspect of the low-energy physics of vacancy-disorder.

Finally, we note that the values of $\Gamma_c$ and $y$ in the case of dilution by
4-triangles are apparently predicted much better by the value of $w$ (as opposed to the $n_v$). This raises the interesting questions already alluded to in the main text:
Are $\Gamma_c$ and $y$ determined in a ``universal'' way ({\em i.e.}, independent
of short-ranged correlations between vacancies and other such microscopic details) by the value of the zero-mode density $w$ in the limit of small but nonzero $w$? Can this
dependence be understood in terms of a low-energy effective theory or renormalization
group approach?

\begin{figure} 
\centerline{\includegraphics[width=0.3\columnwidth]{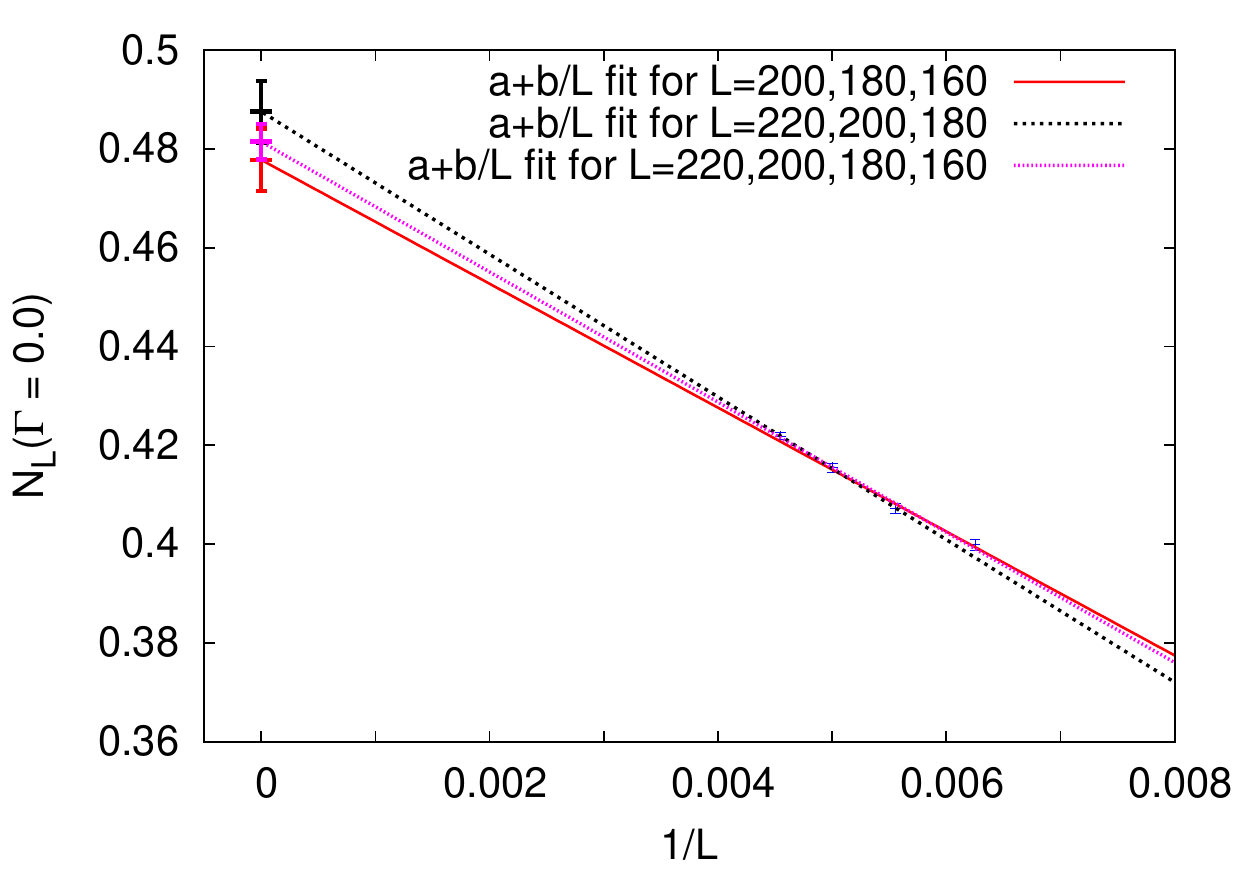}~
\includegraphics[width=0.3\columnwidth]{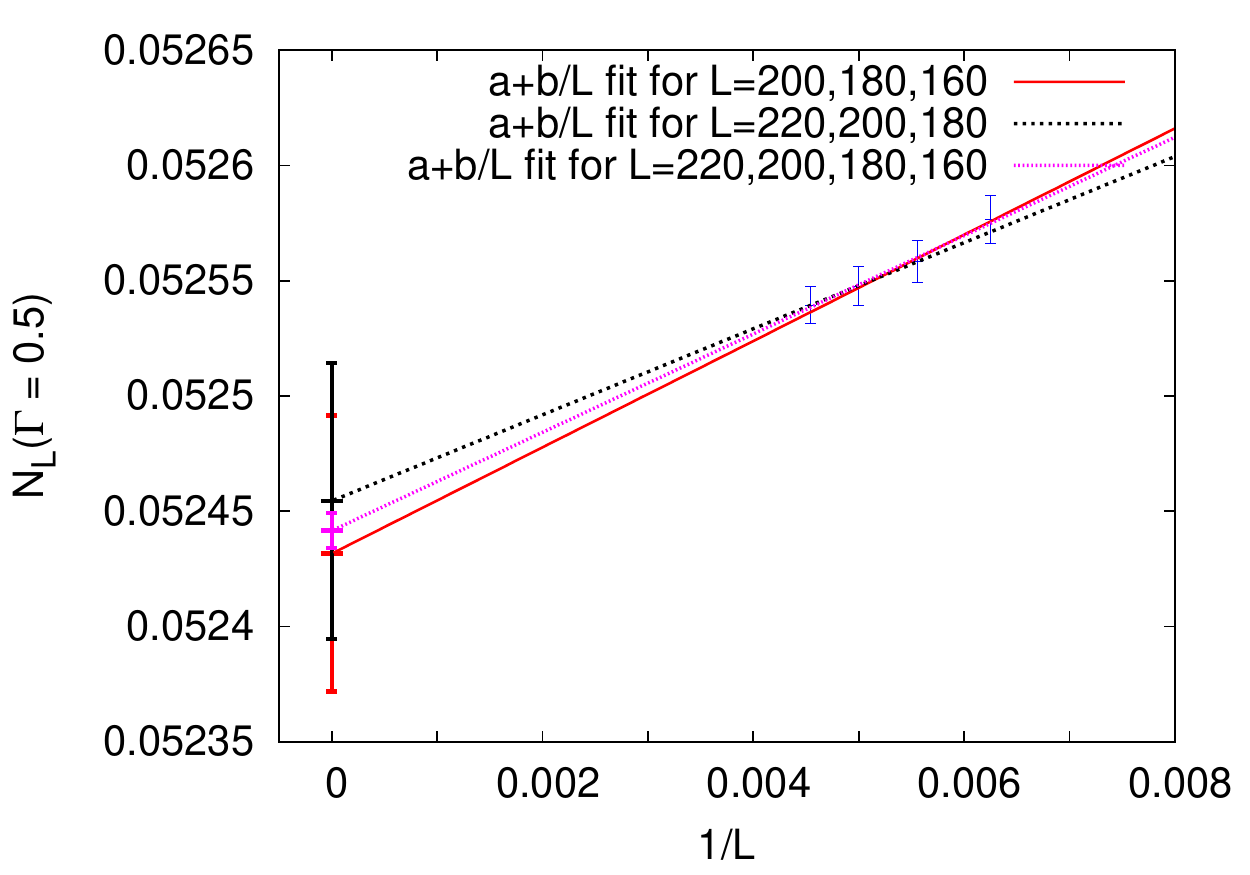}~
\includegraphics[width=0.3\columnwidth]{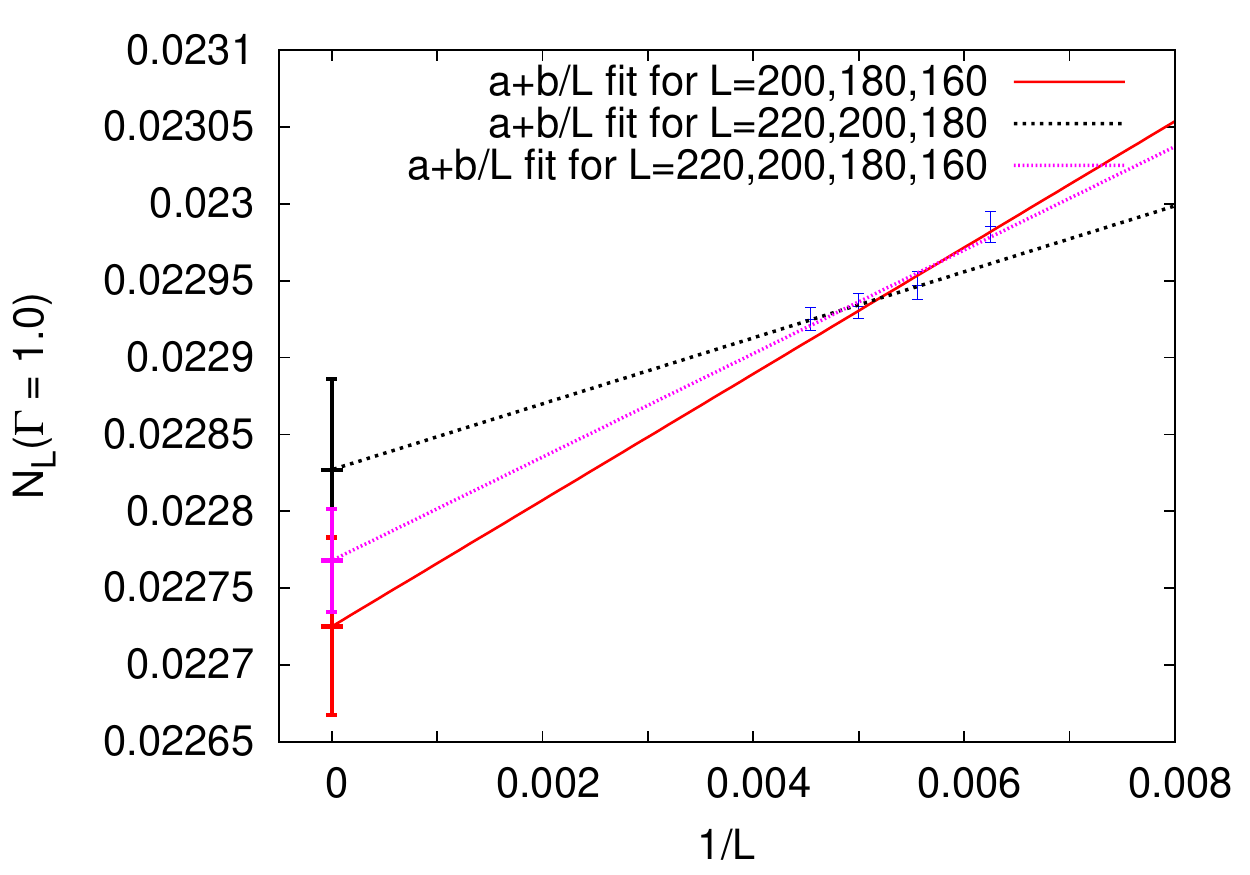}}
\centerline{\includegraphics[width=0.3\columnwidth]{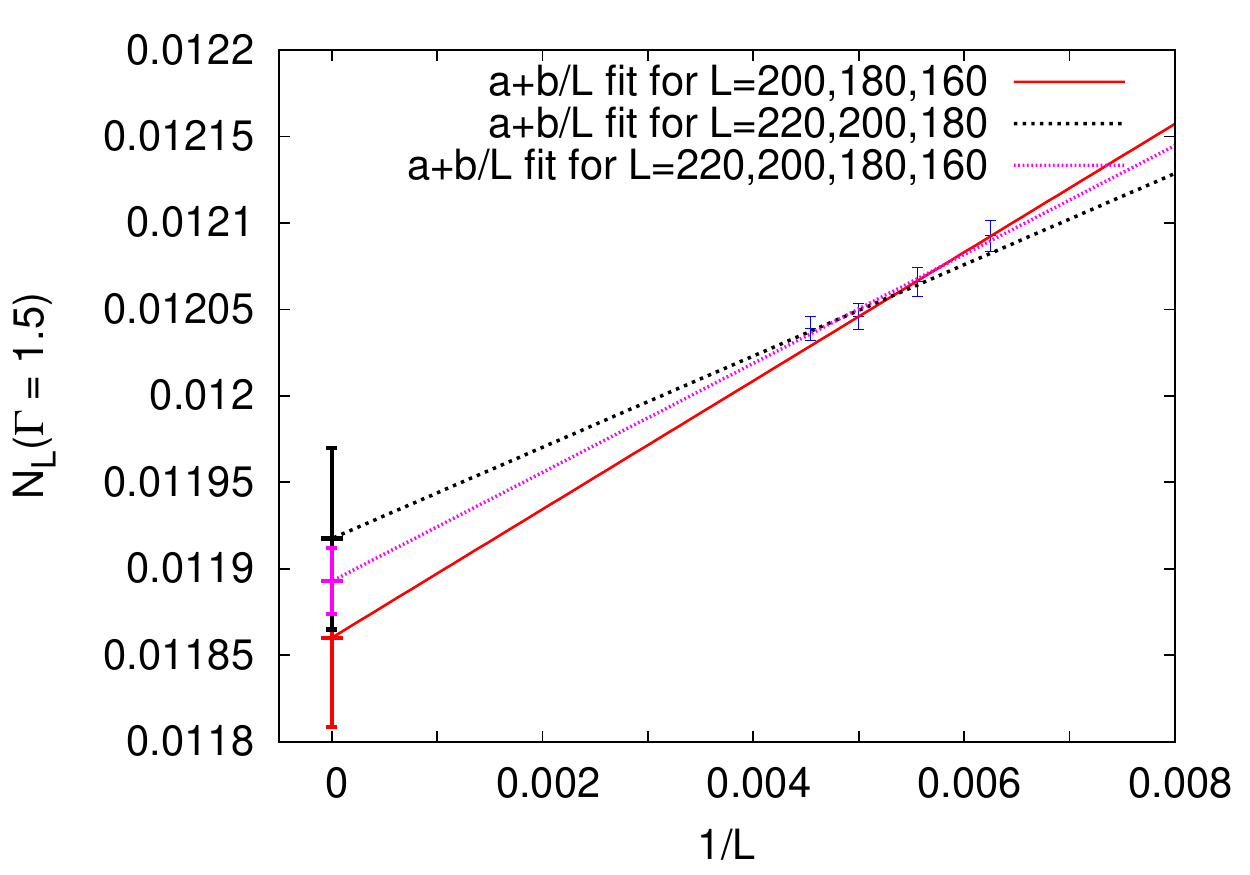}~
\includegraphics[width=0.3\columnwidth]{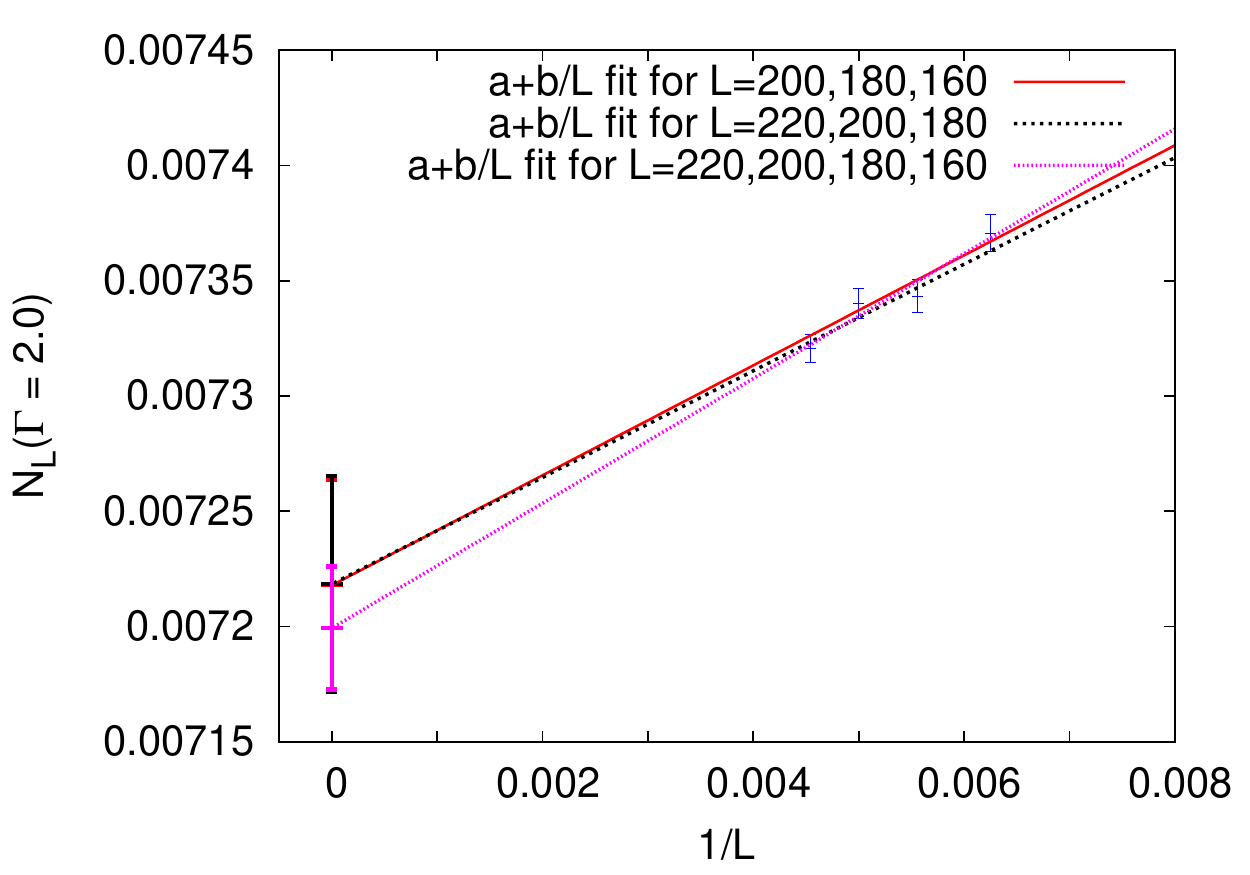}~
\includegraphics[width=0.3\columnwidth]{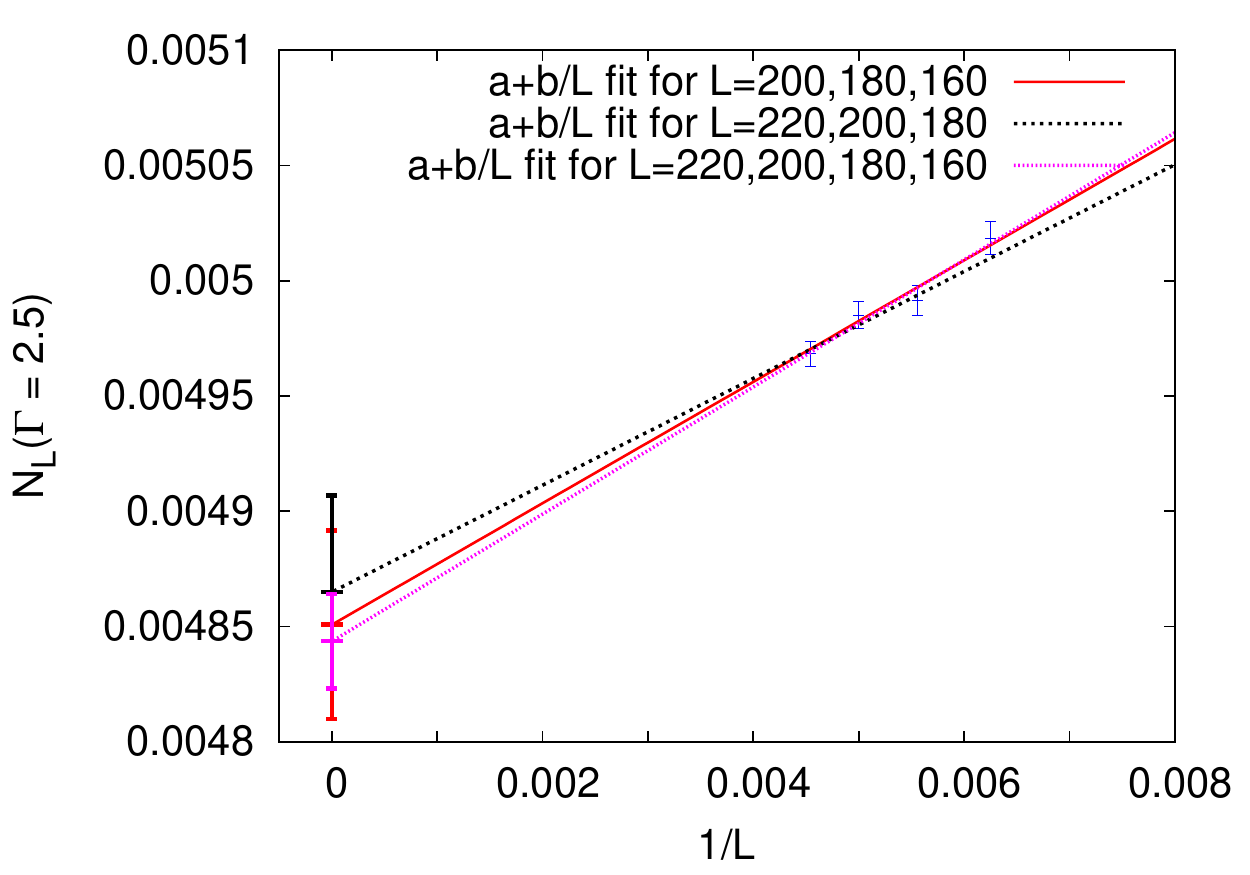}}
\centerline{\includegraphics[width=0.3\columnwidth]{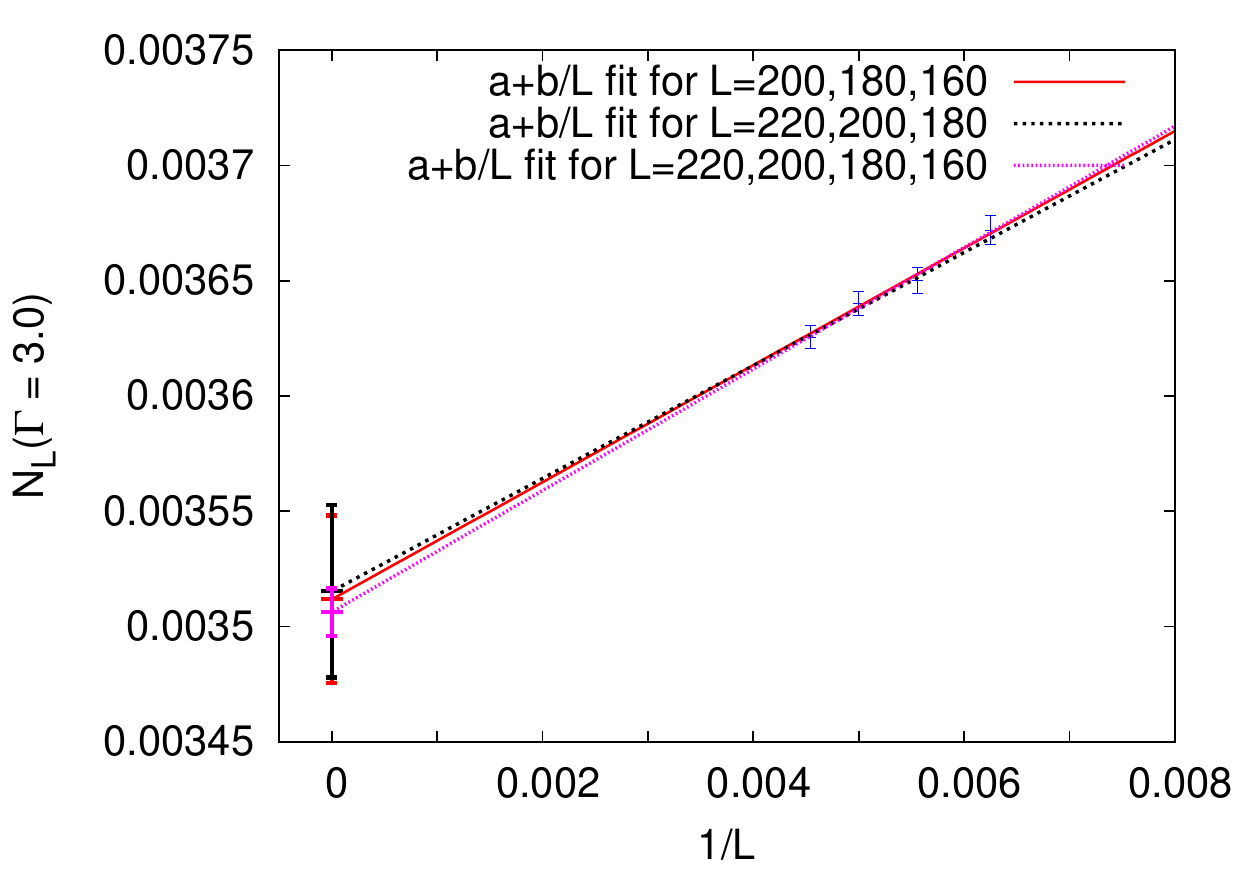}~
\includegraphics[width=0.3\columnwidth]{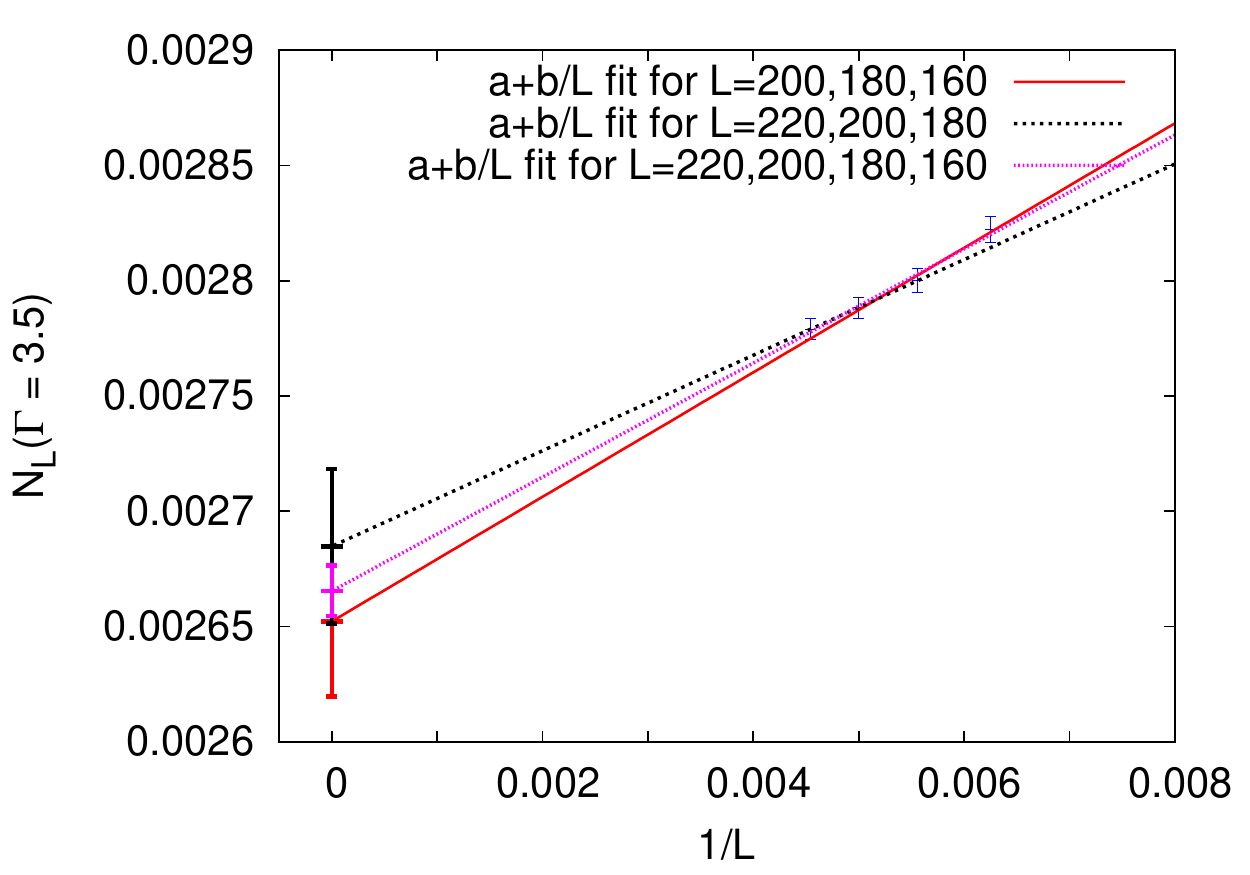}~
\includegraphics[width=0.3\columnwidth]{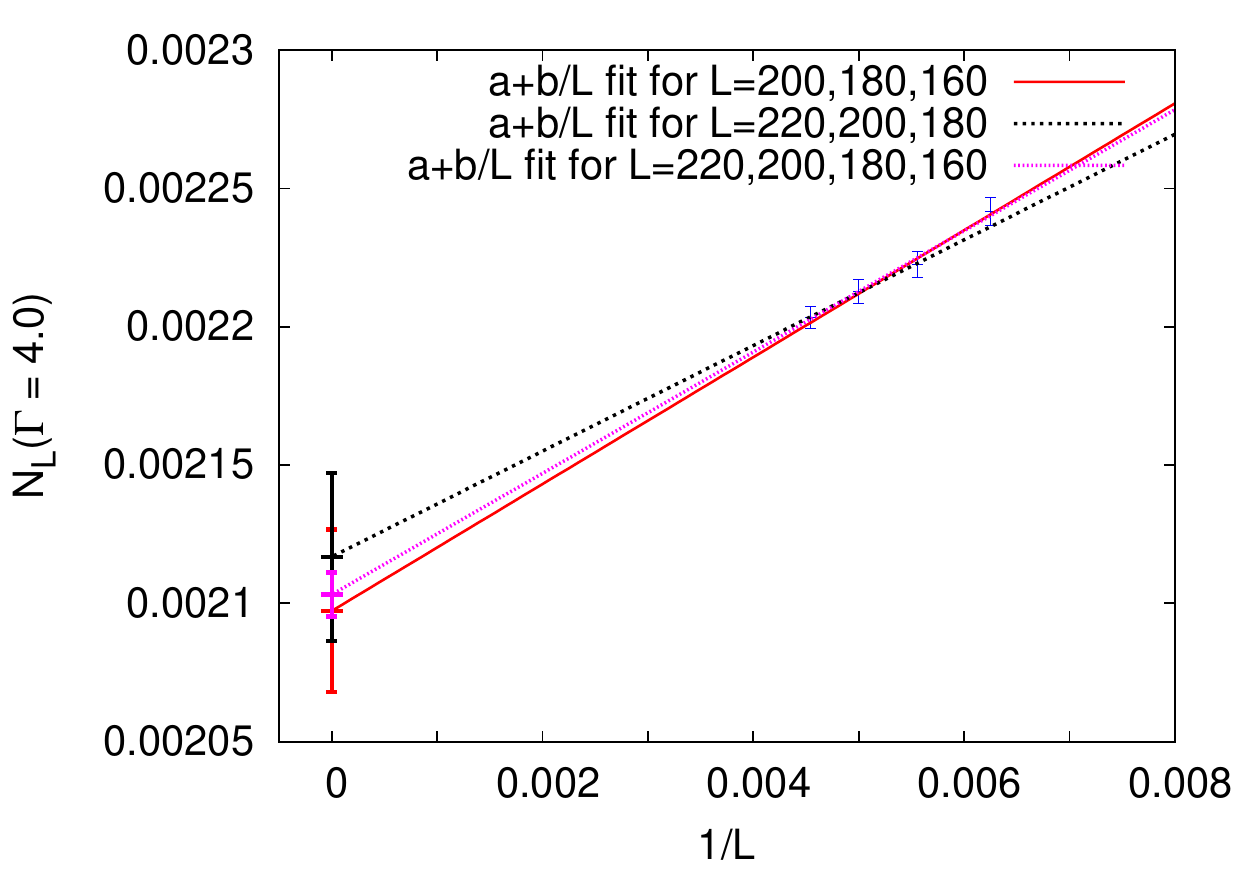}}
\centerline{\includegraphics[width=0.3\columnwidth]{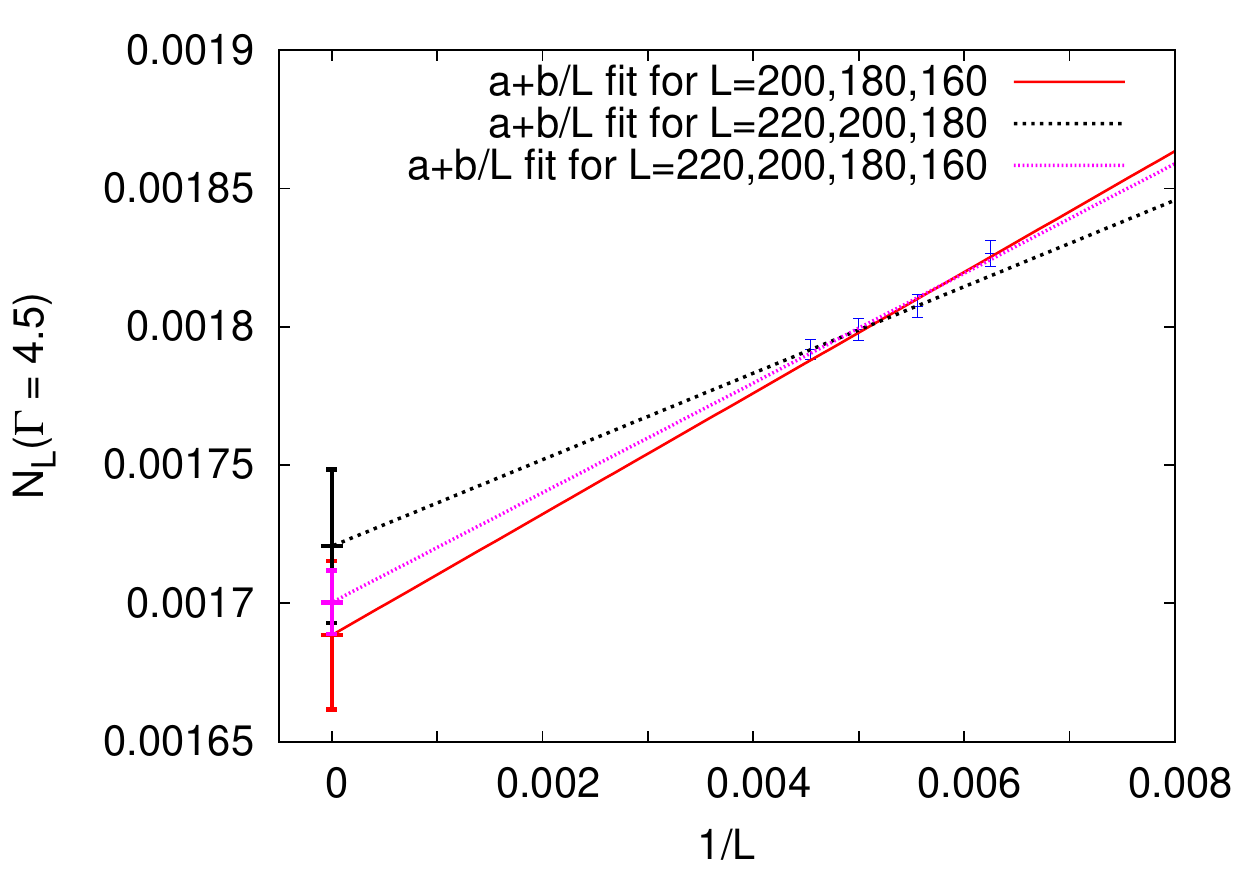}~
\includegraphics[width=0.3\columnwidth]{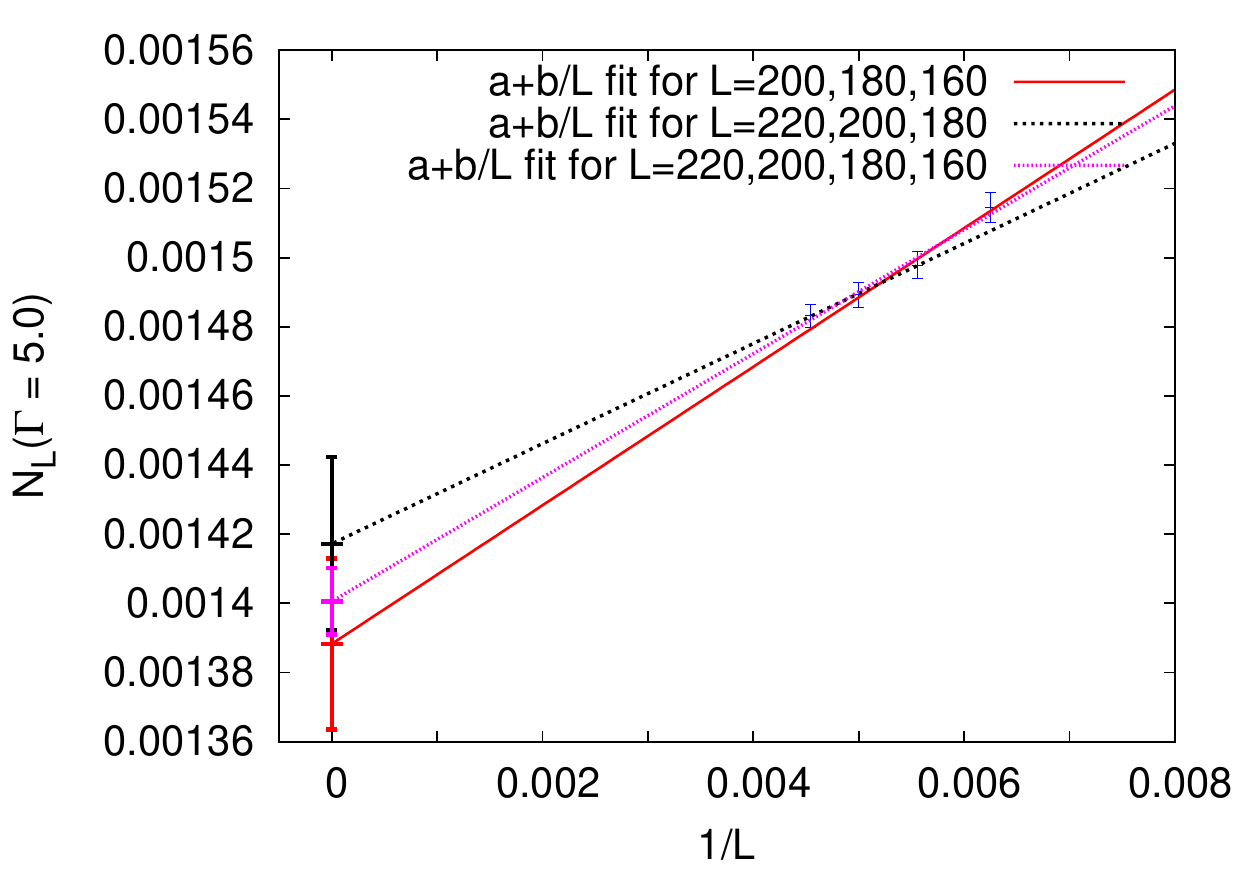}~
\includegraphics[width=0.3\columnwidth]{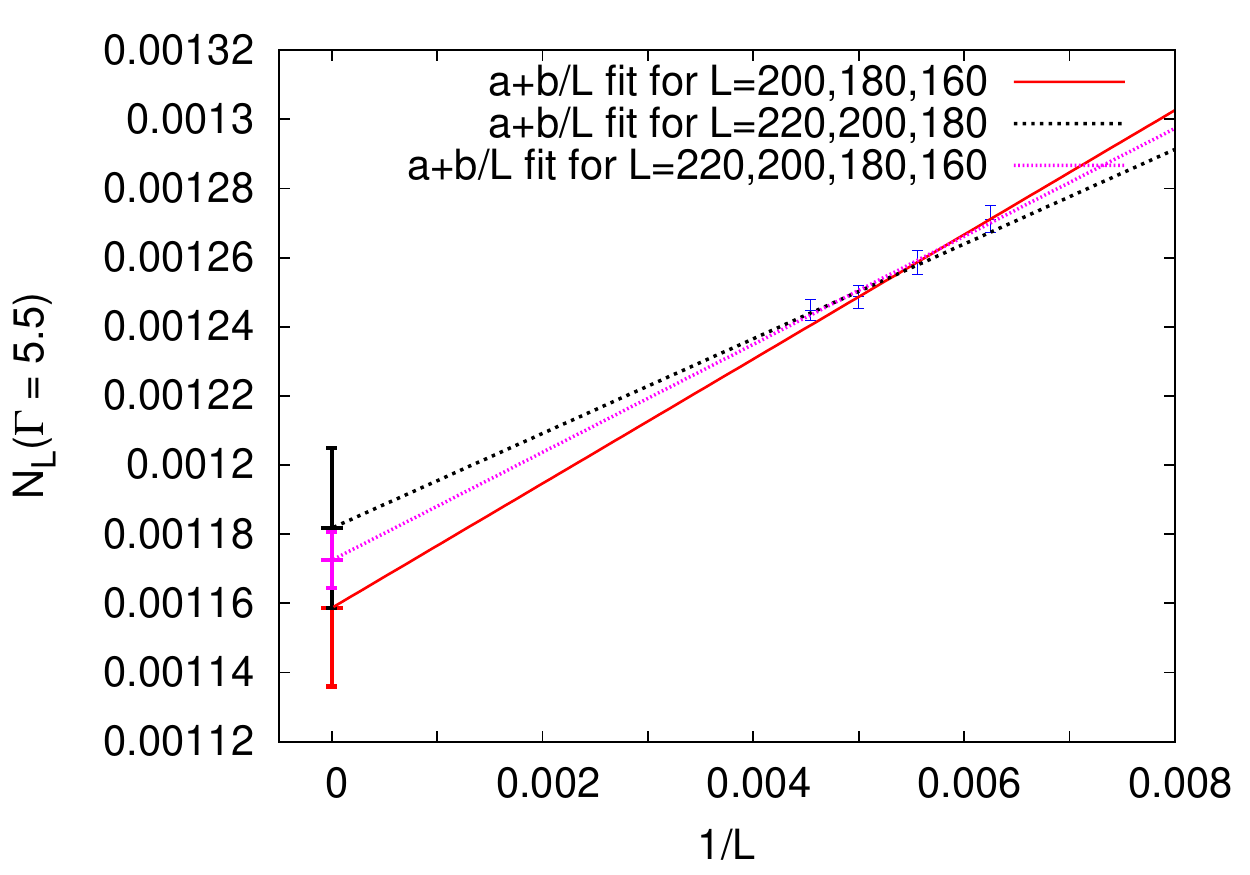}}
\centerline{\includegraphics[width=0.3\columnwidth]{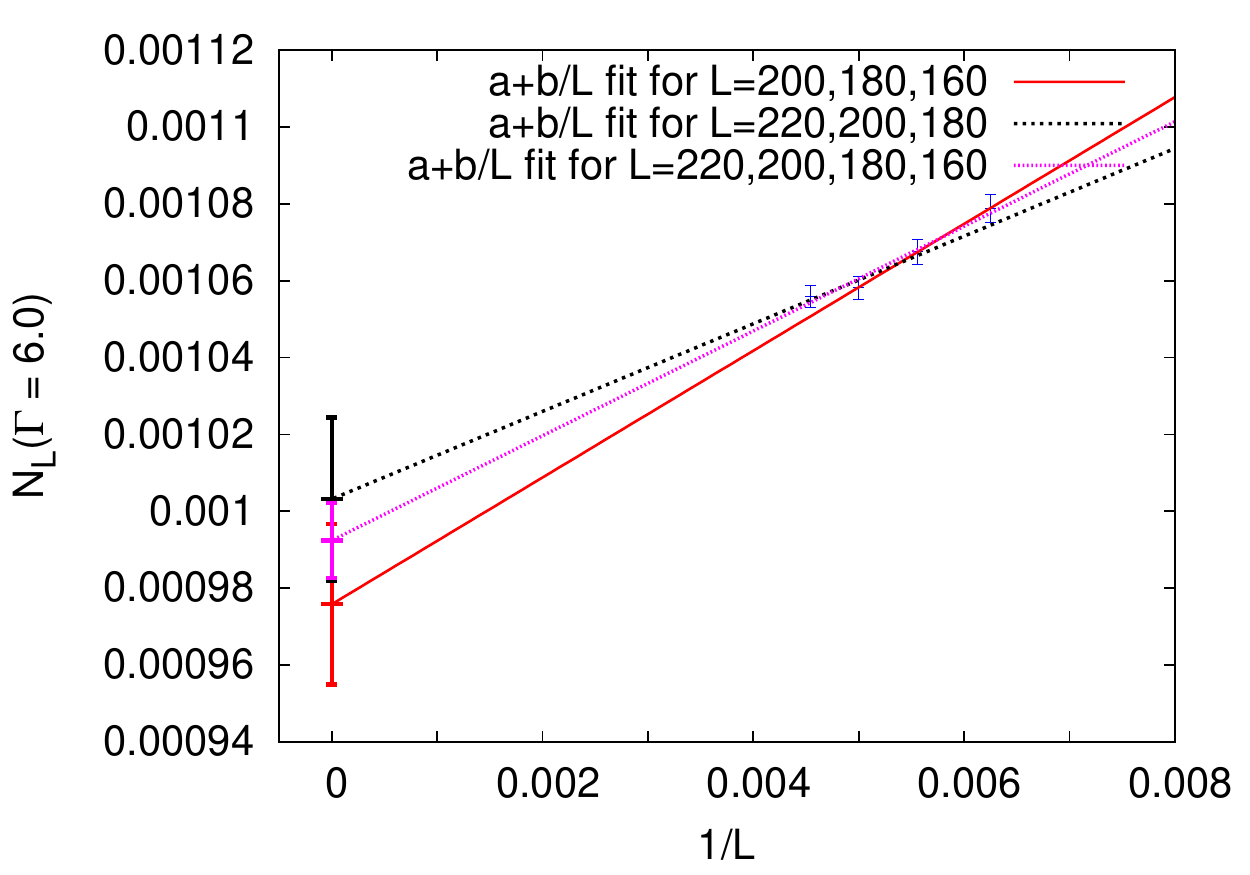}~
\includegraphics[width=0.3\columnwidth]{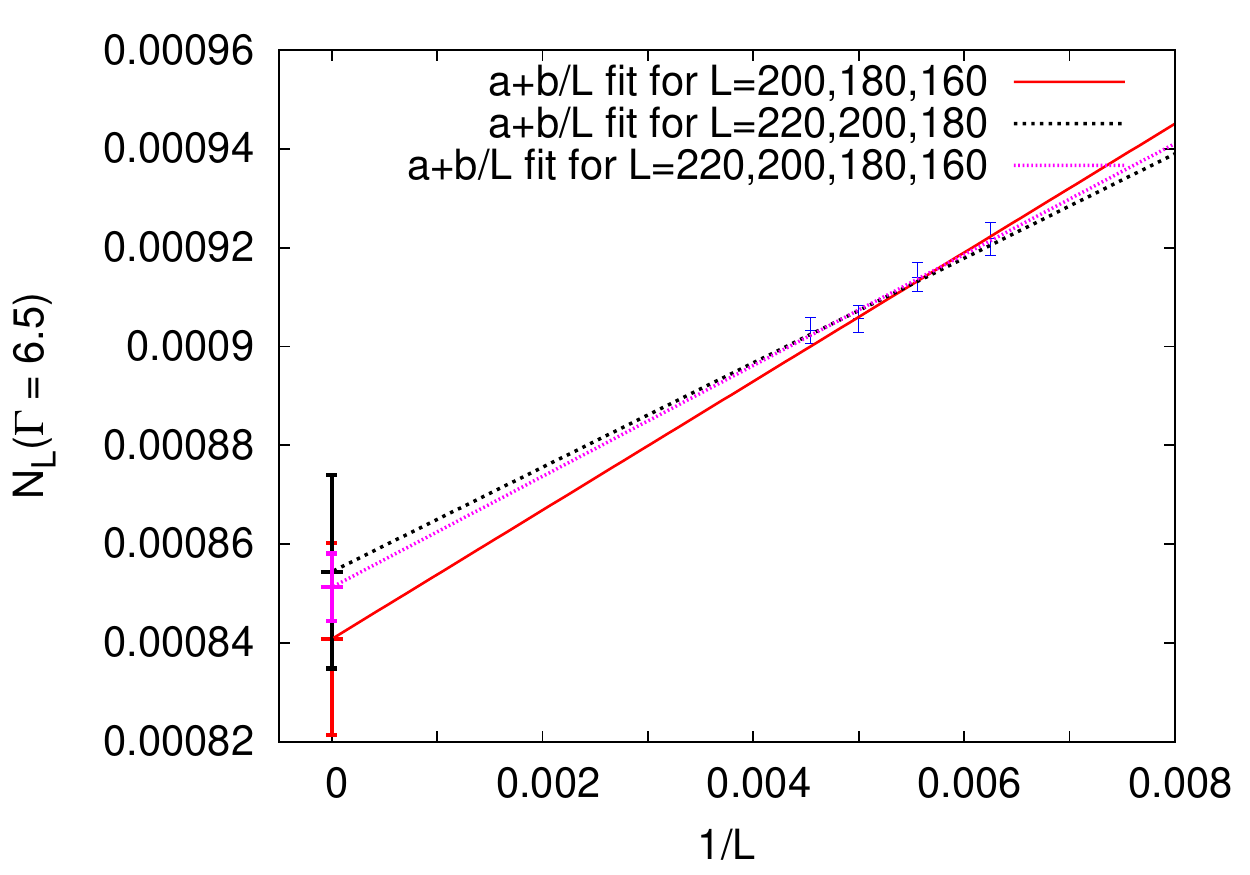}~
\includegraphics[width=0.3\columnwidth]{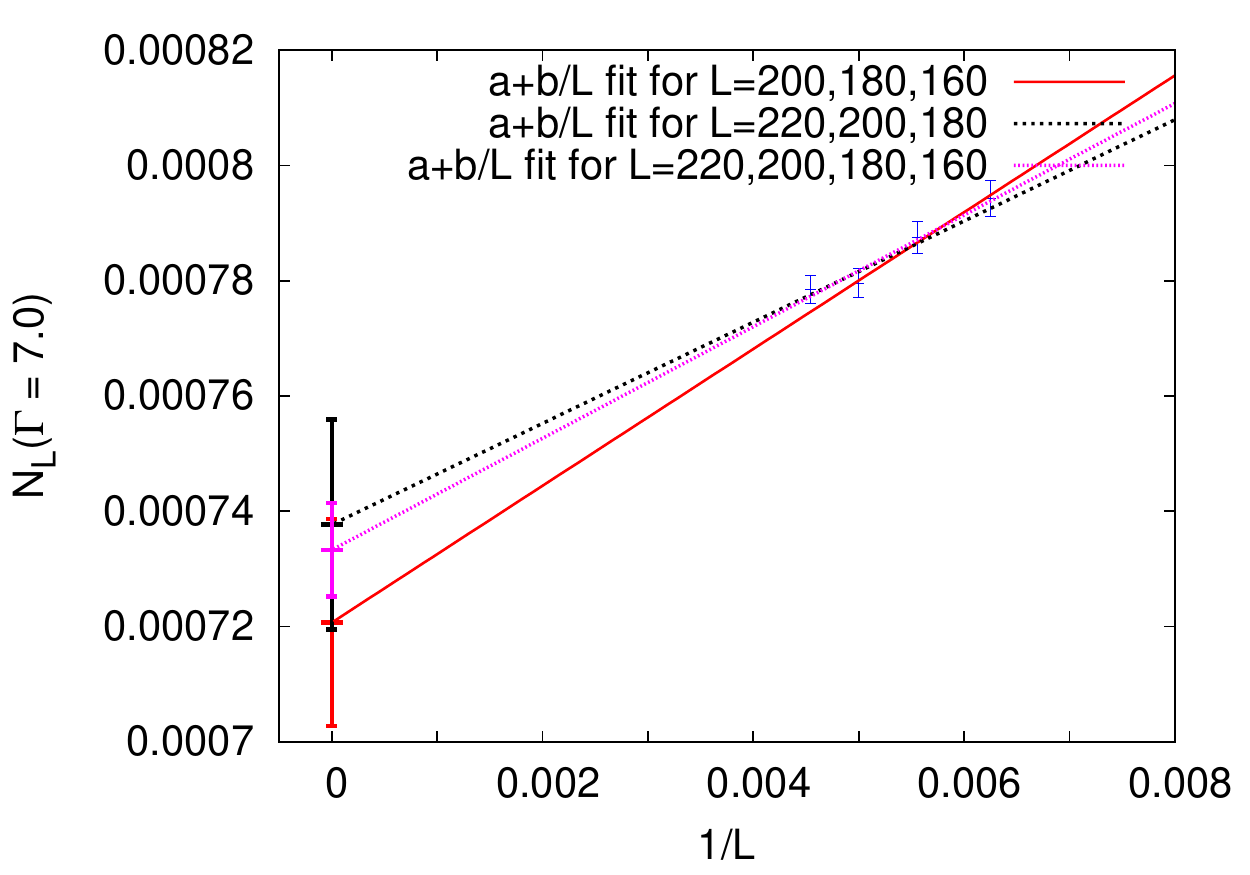}}
\centerline{\includegraphics[width=0.3\columnwidth]{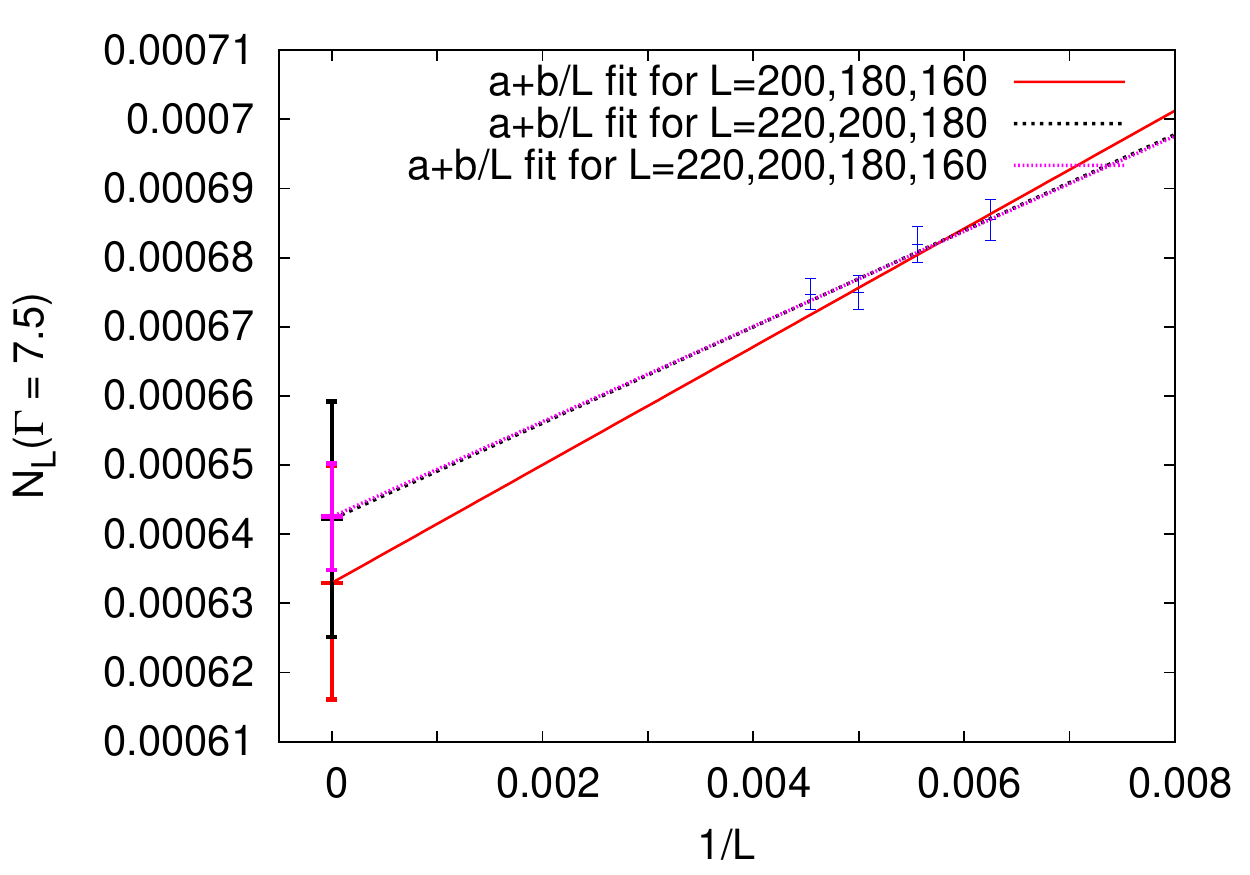}~
\includegraphics[width=0.3\columnwidth]{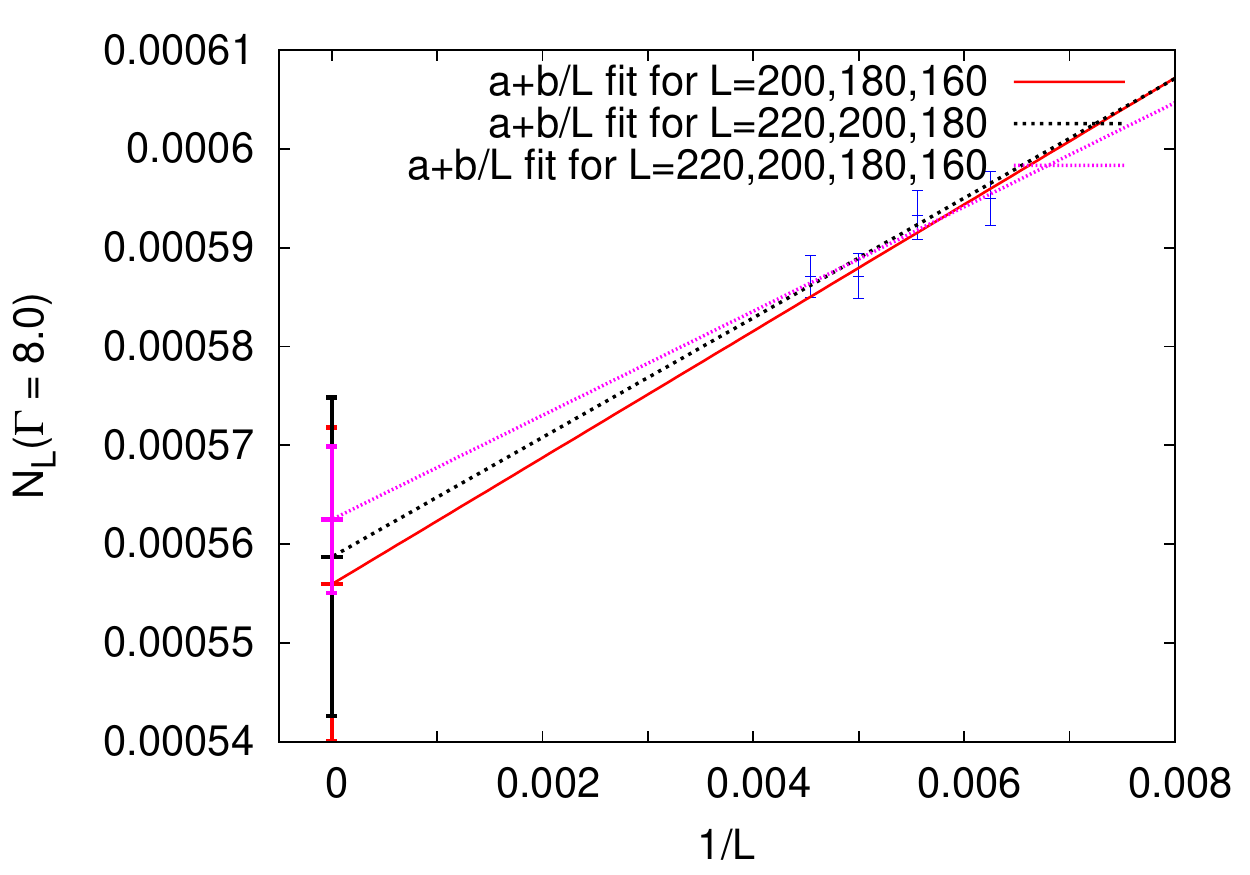}~
\includegraphics[width=0.3\columnwidth]{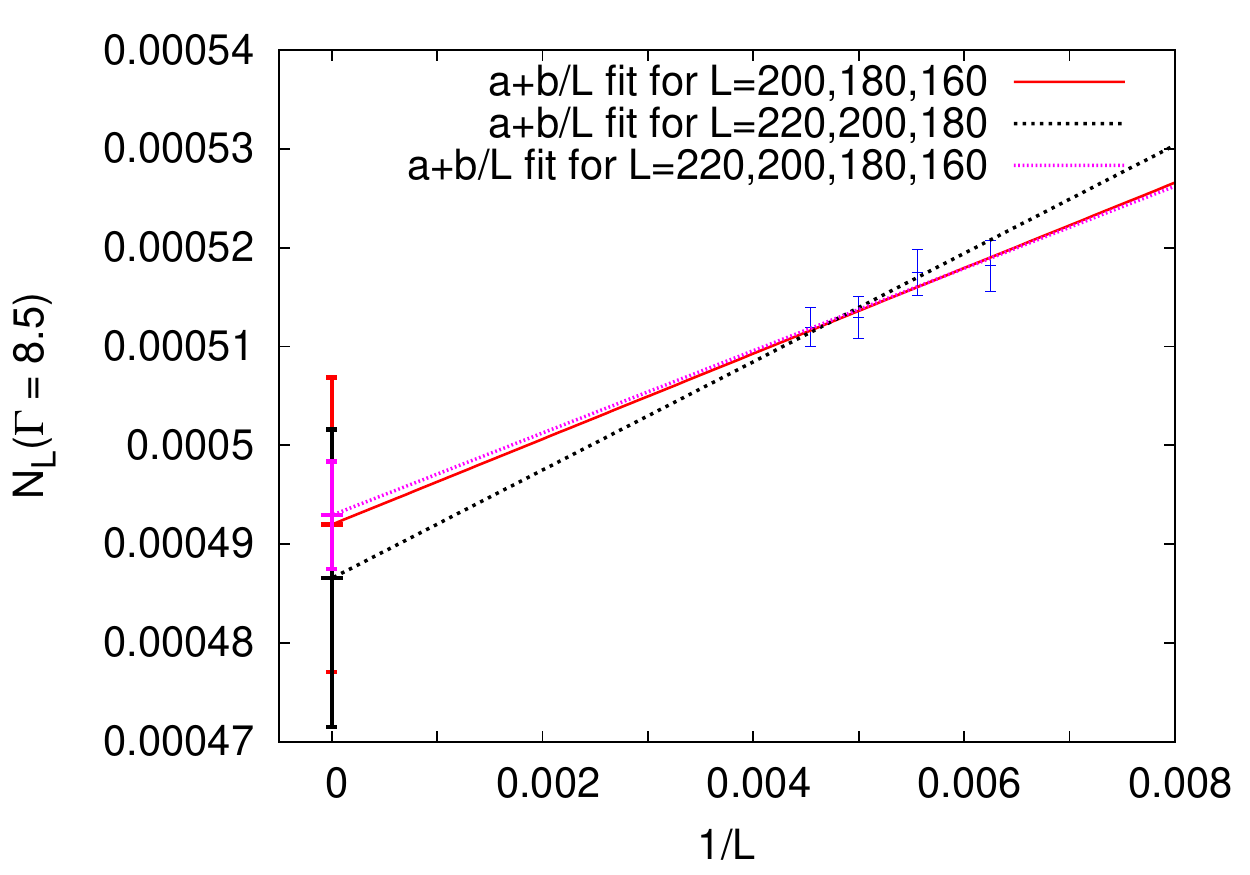}}
\centerline{\includegraphics[width=0.3\columnwidth]{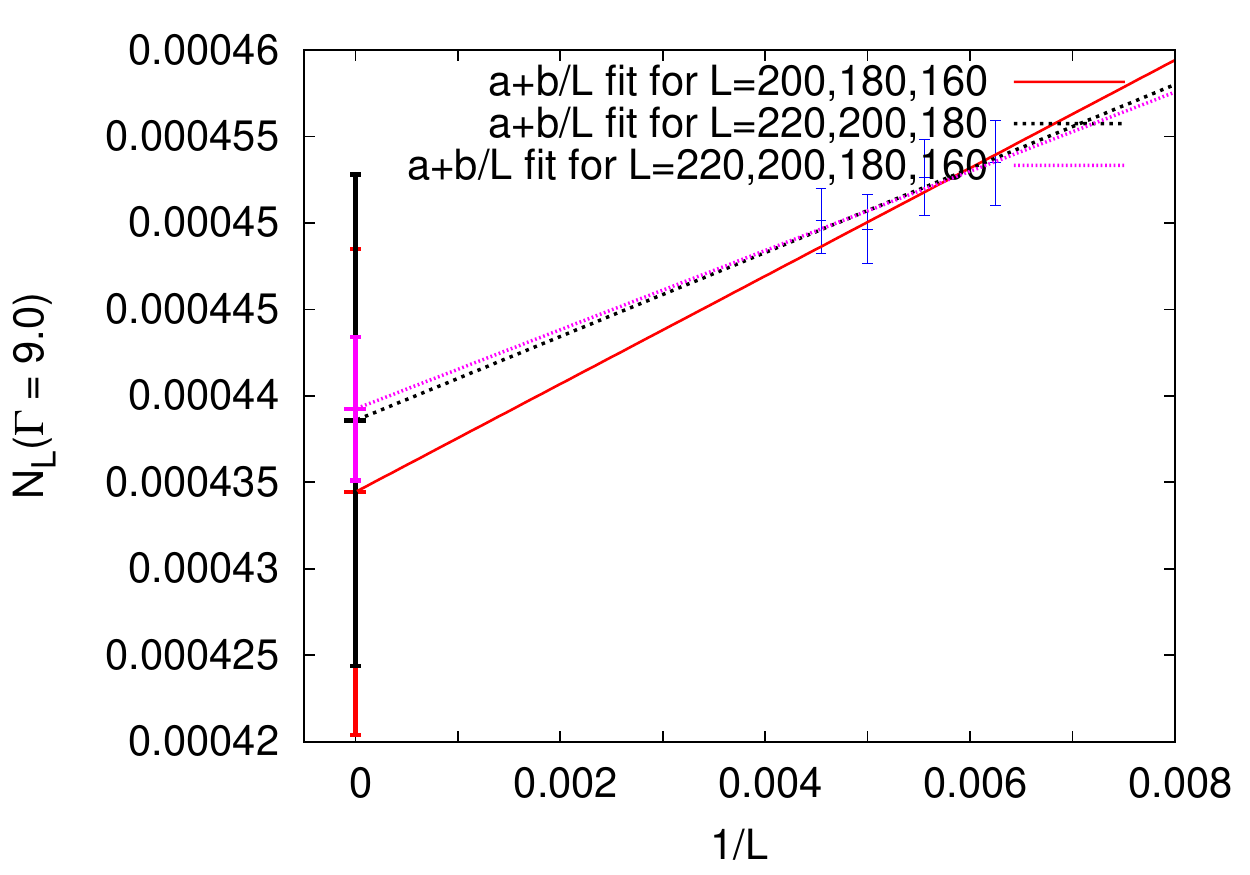}~
\includegraphics[width=0.3\columnwidth]{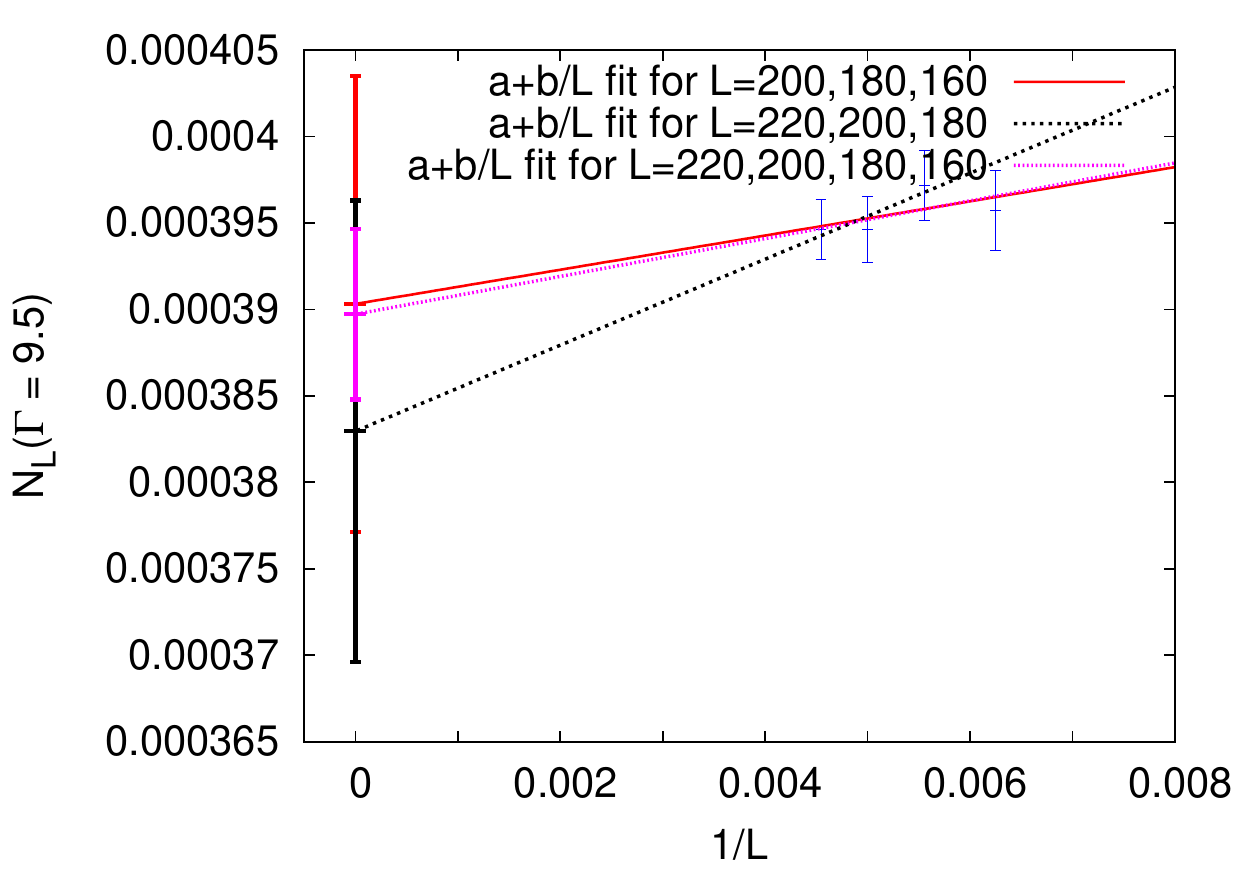}~
\includegraphics[width=0.3\columnwidth]{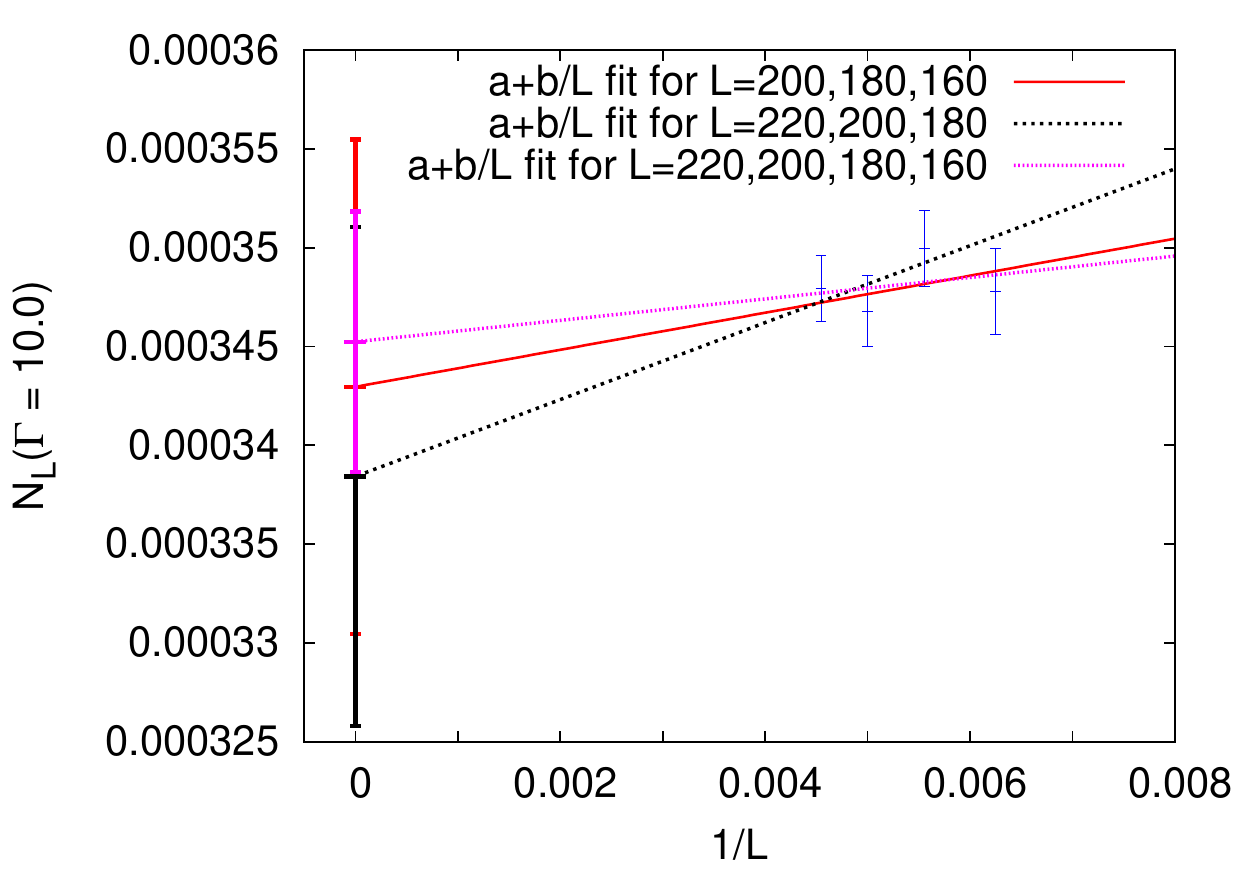}}
\end{figure}

\begin{figure} 
\centerline{\includegraphics[width=0.3\columnwidth]{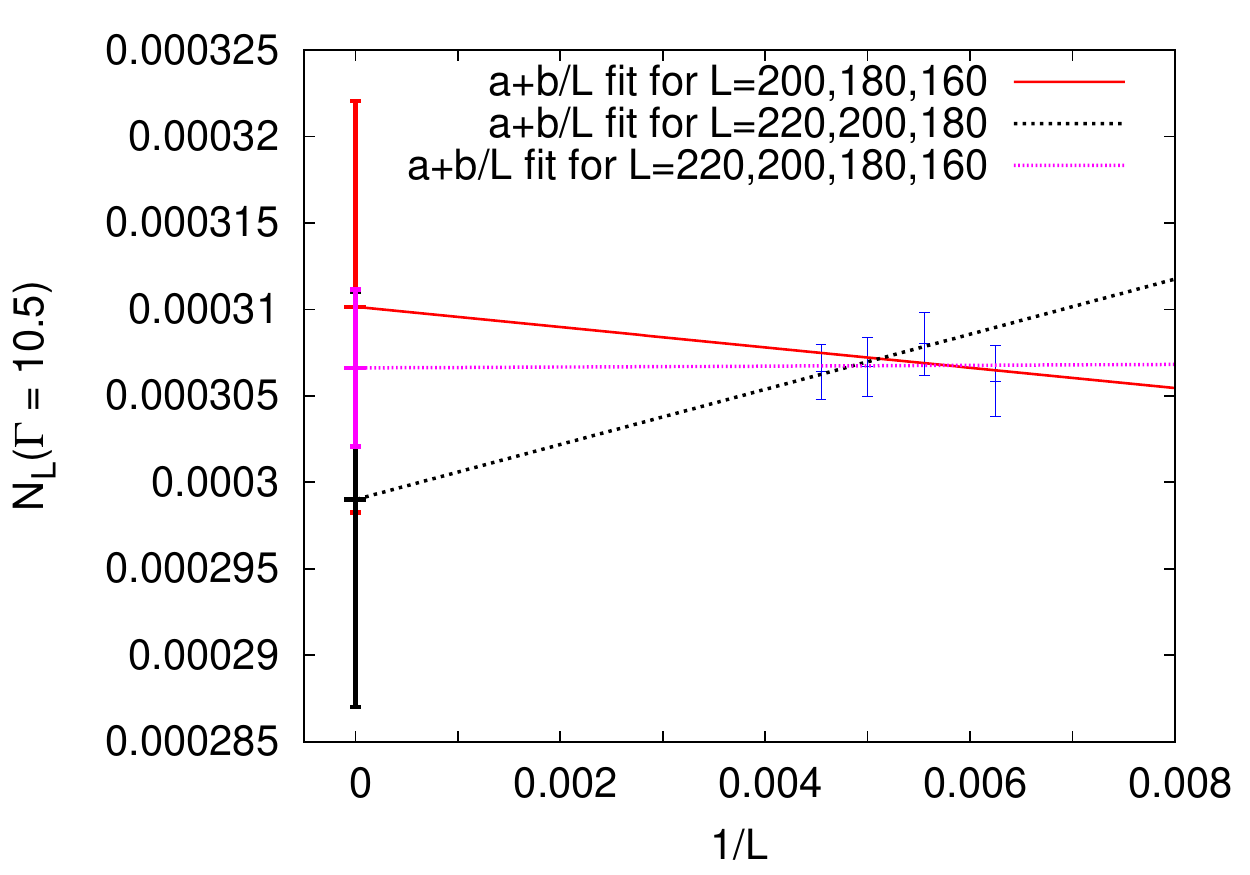}~
\includegraphics[width=0.3\columnwidth]{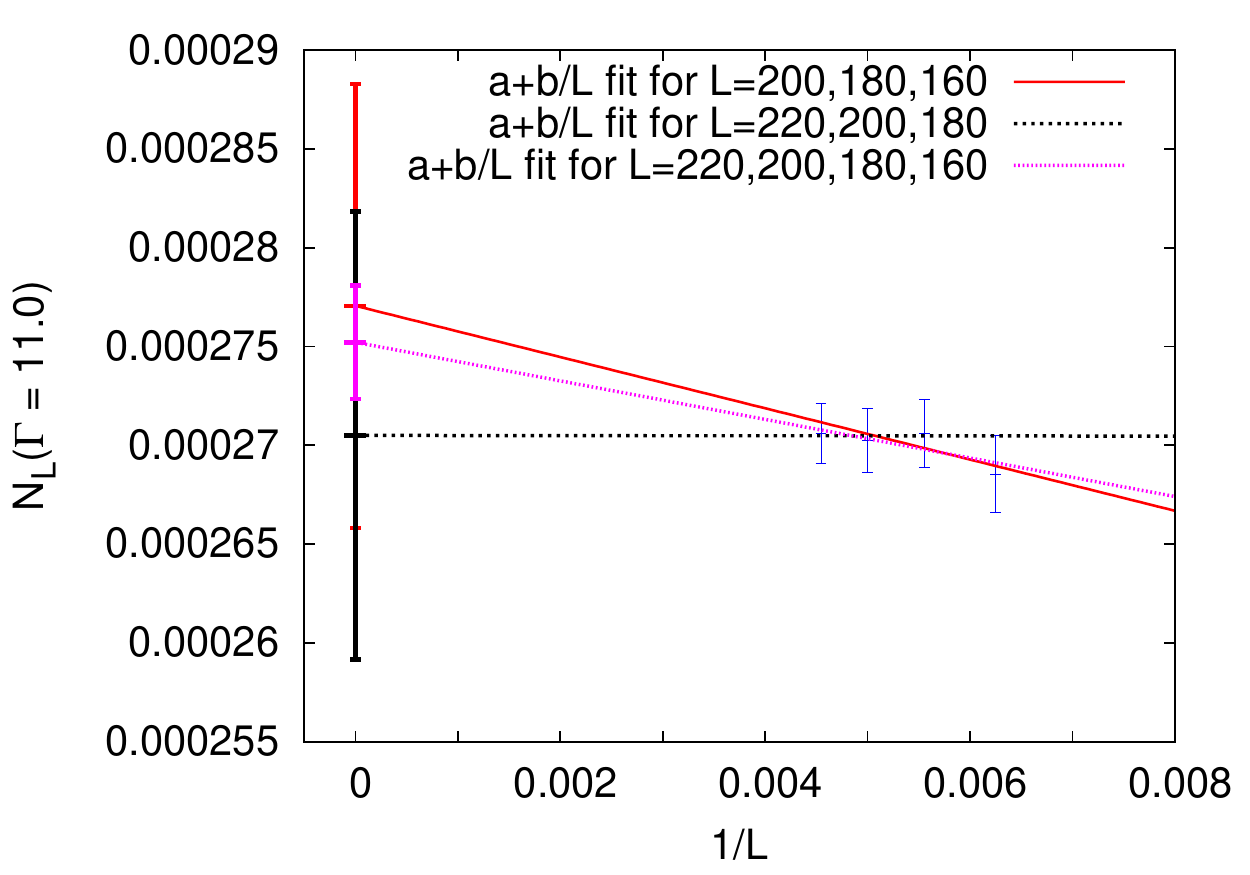}~
\includegraphics[width=0.3\columnwidth]{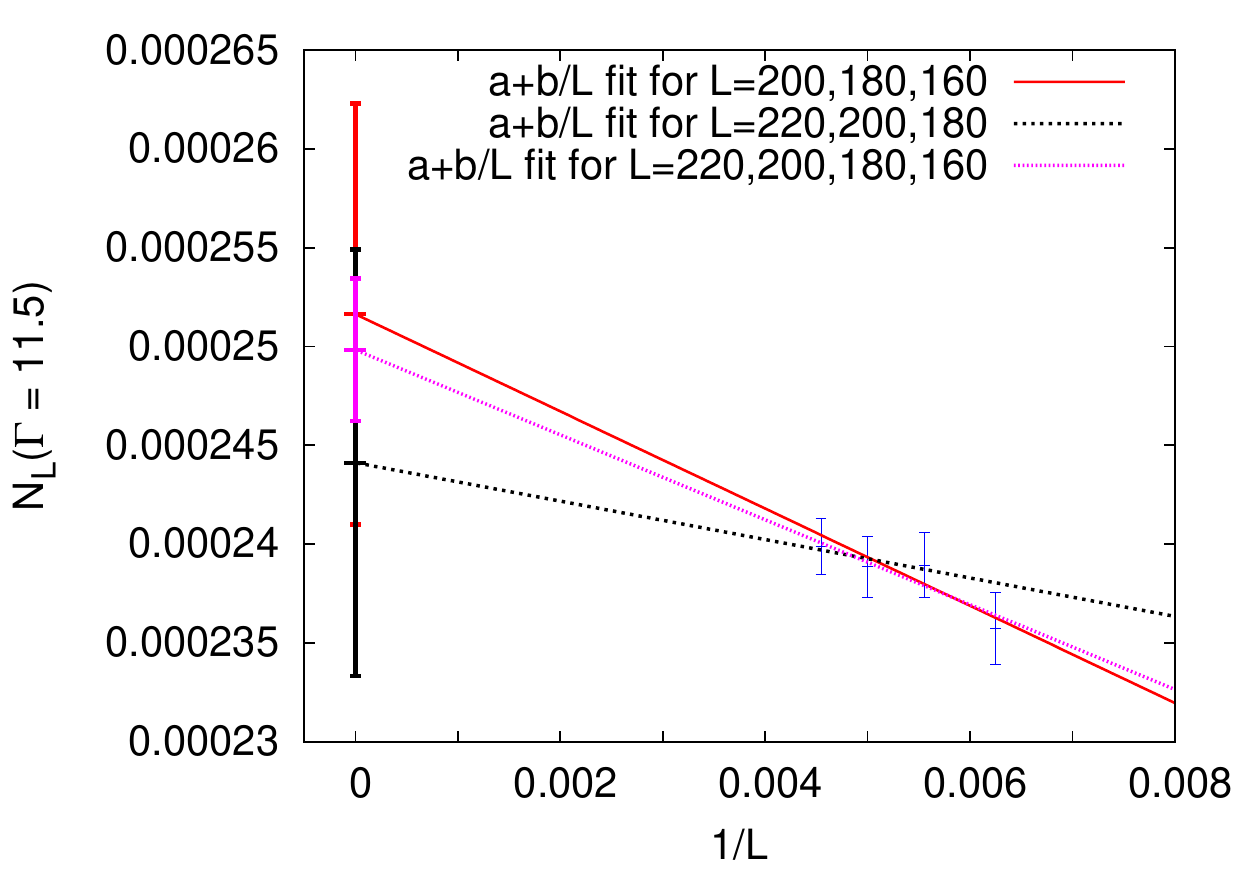}}
\centerline{\includegraphics[width=0.3\columnwidth]{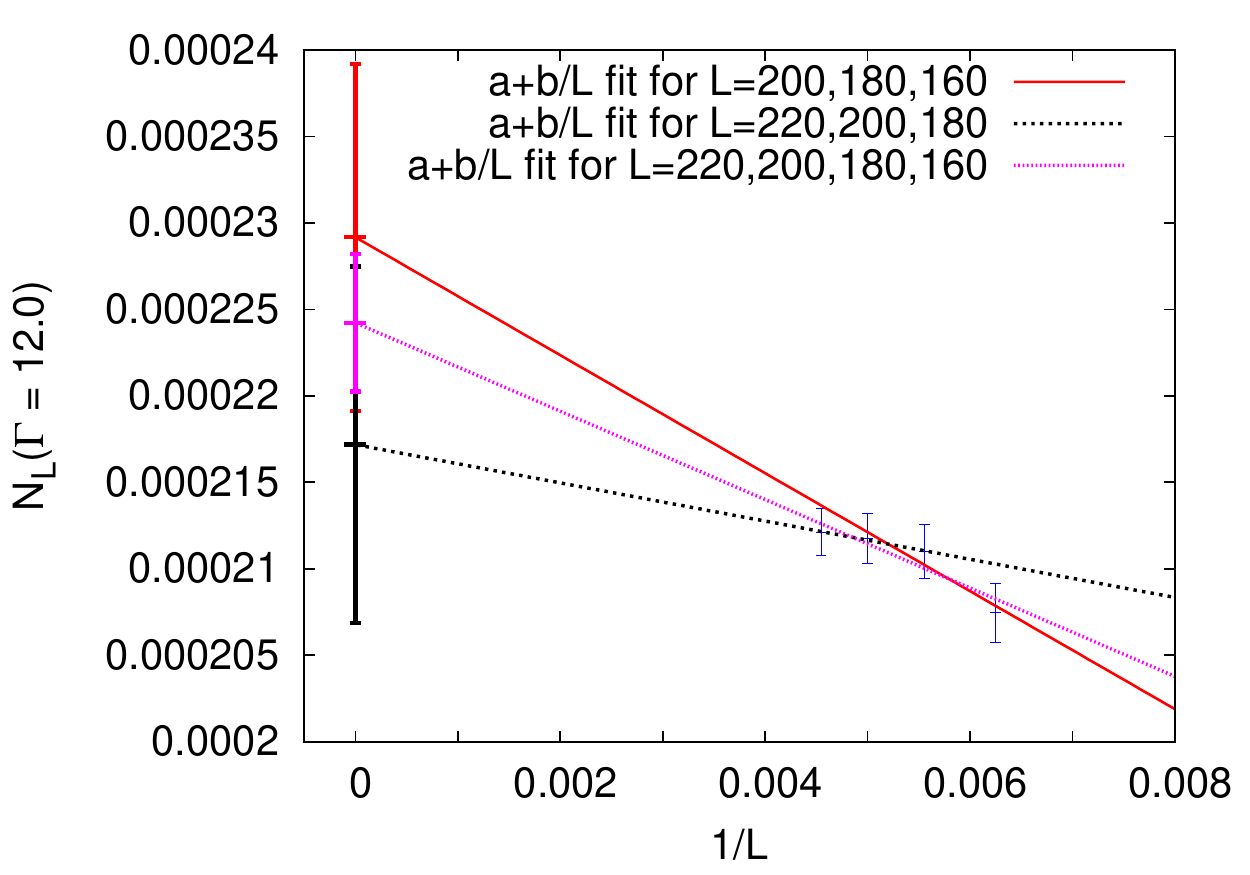}~
\includegraphics[width=0.3\columnwidth]{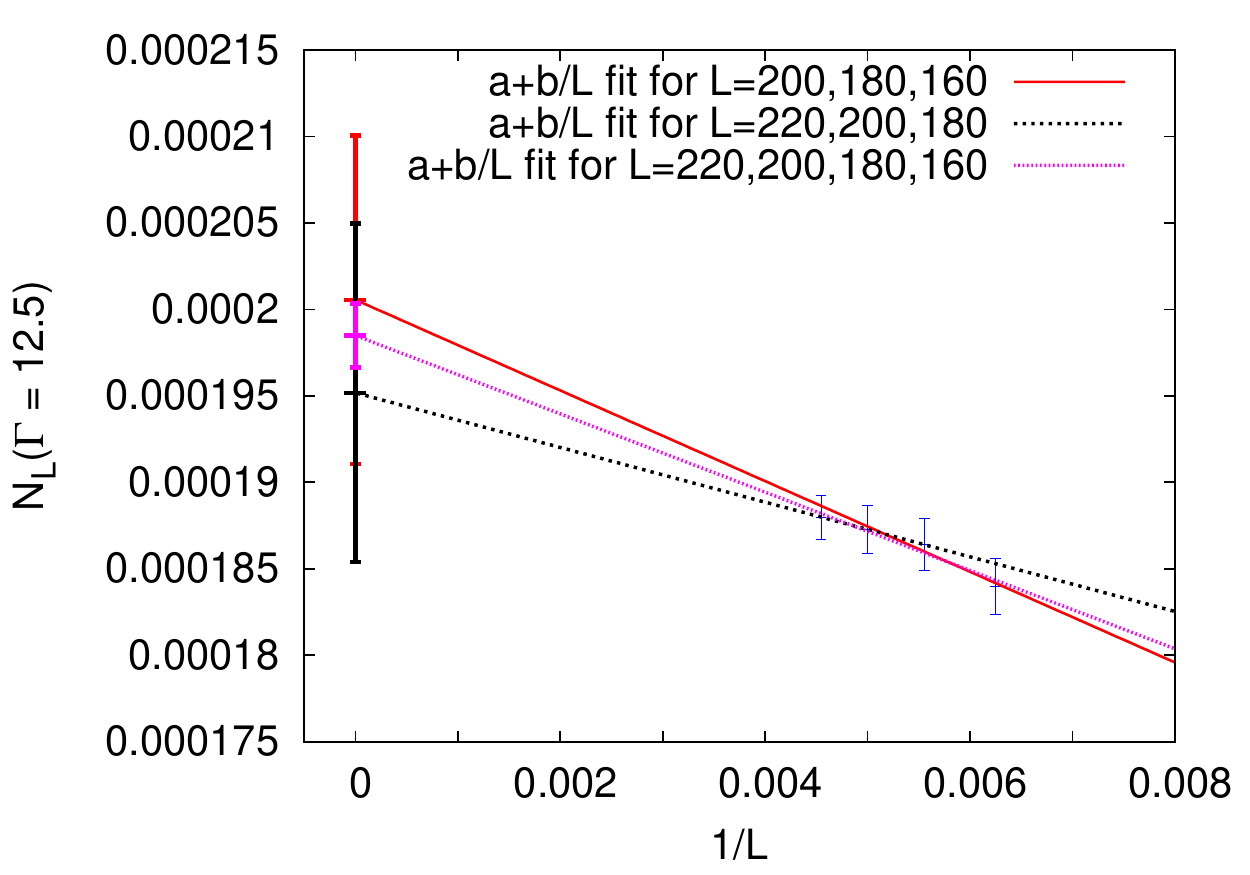}~
\includegraphics[width=0.3\columnwidth]{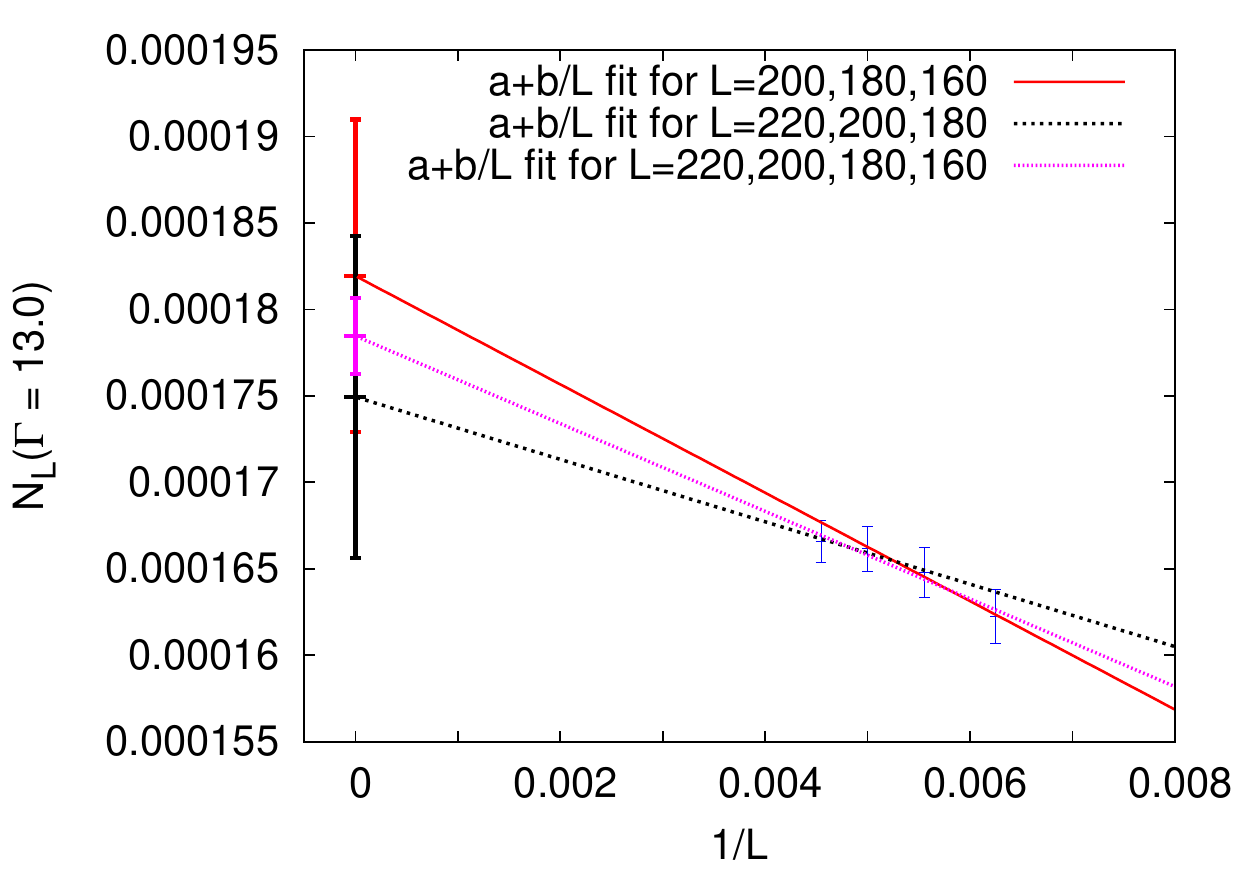}}
\centerline{\includegraphics[width=0.3\columnwidth]{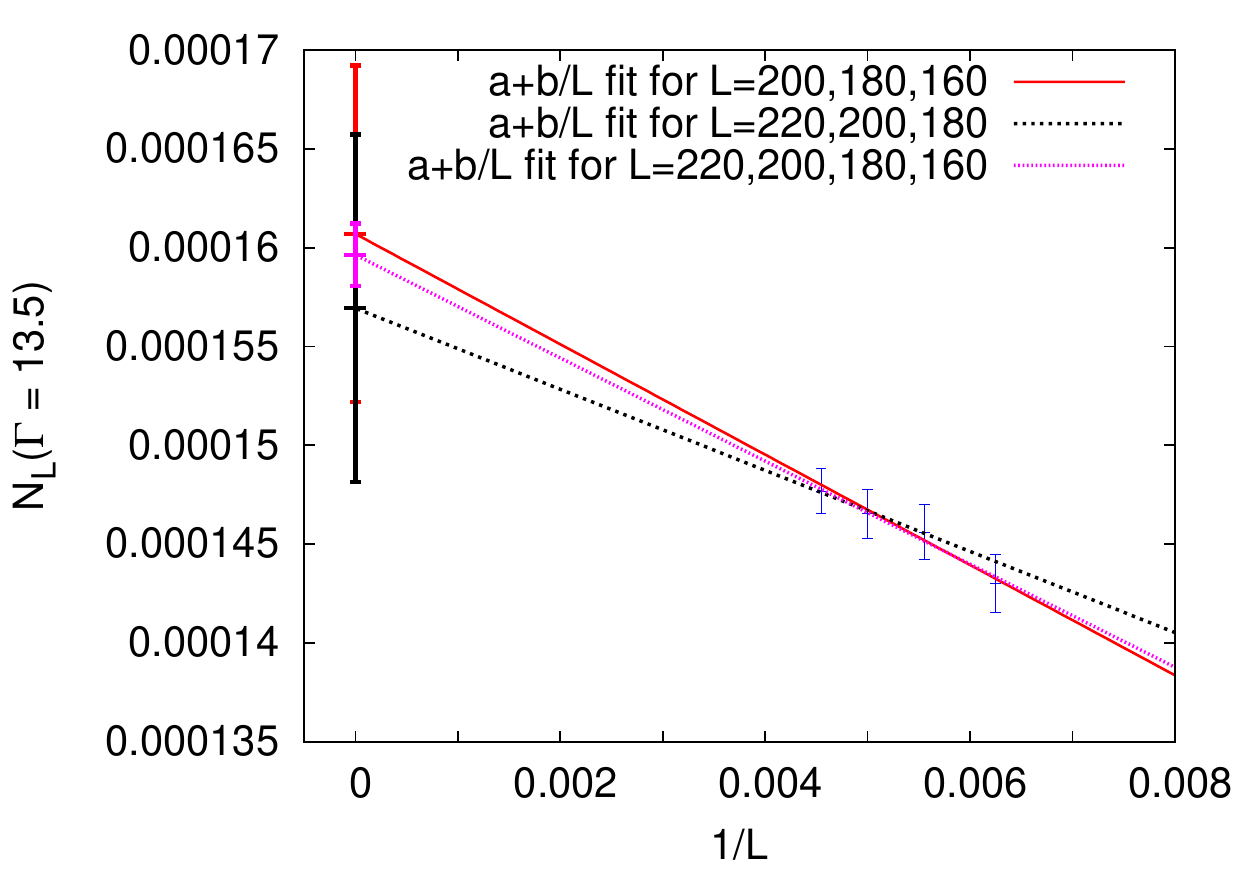}~
\includegraphics[width=0.3\columnwidth]{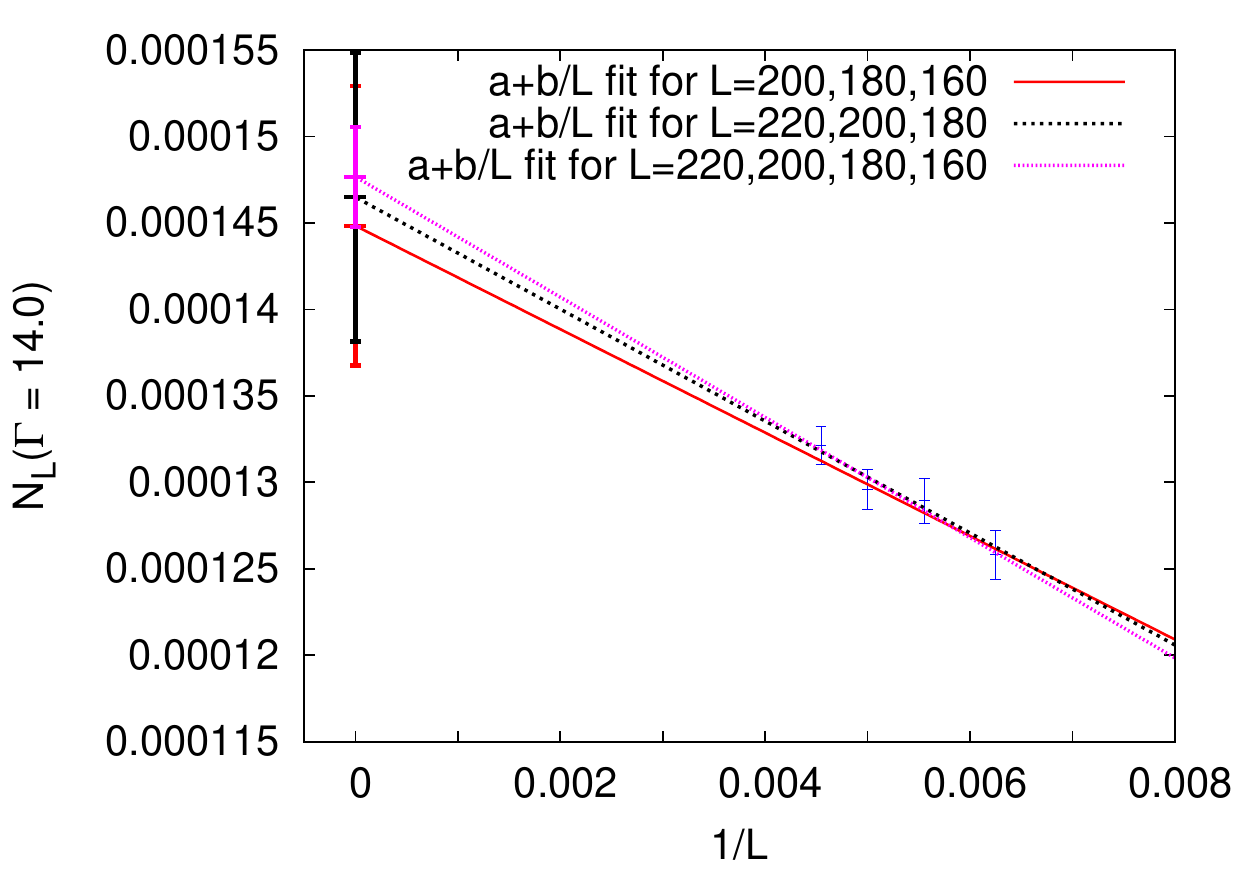}~
\includegraphics[width=0.3\columnwidth]{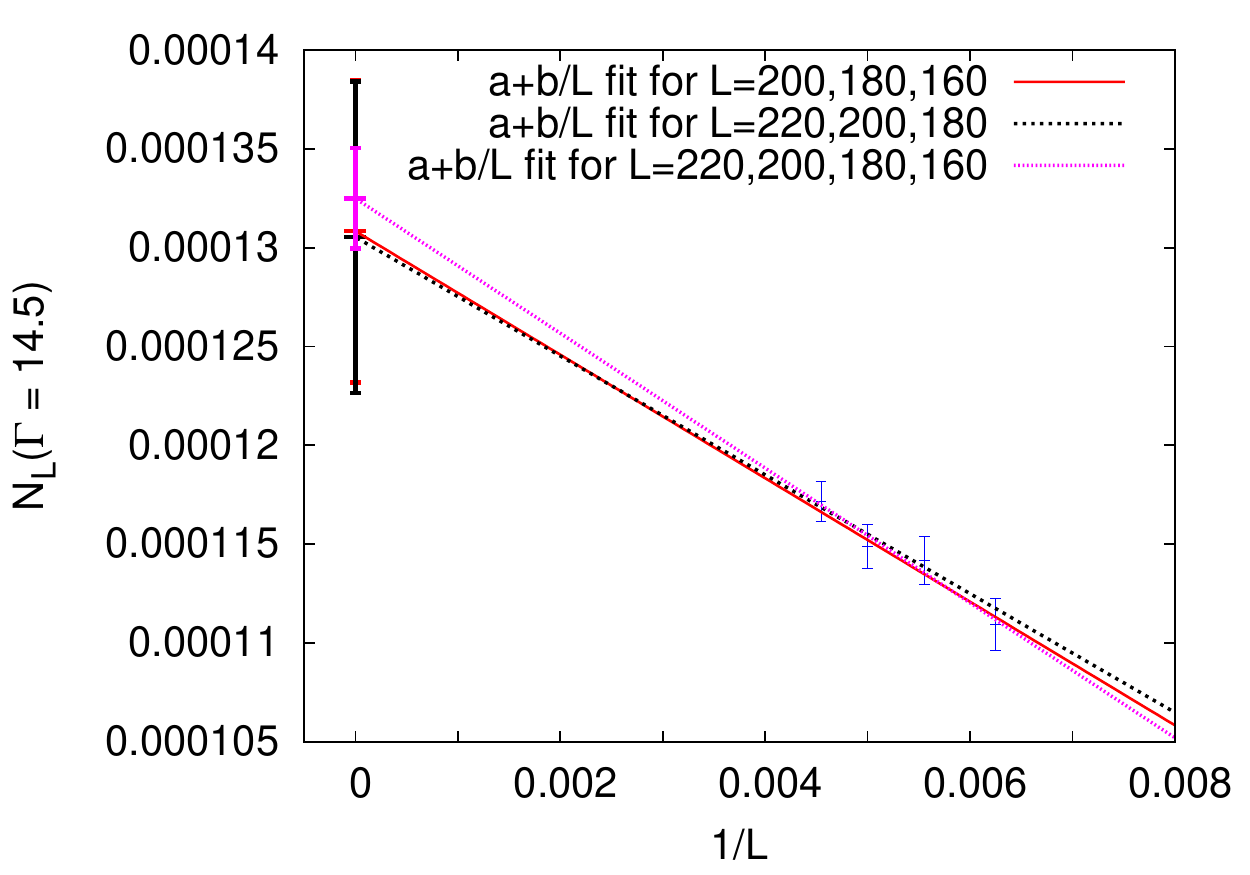}}
\centerline{\includegraphics[width=0.3\columnwidth]{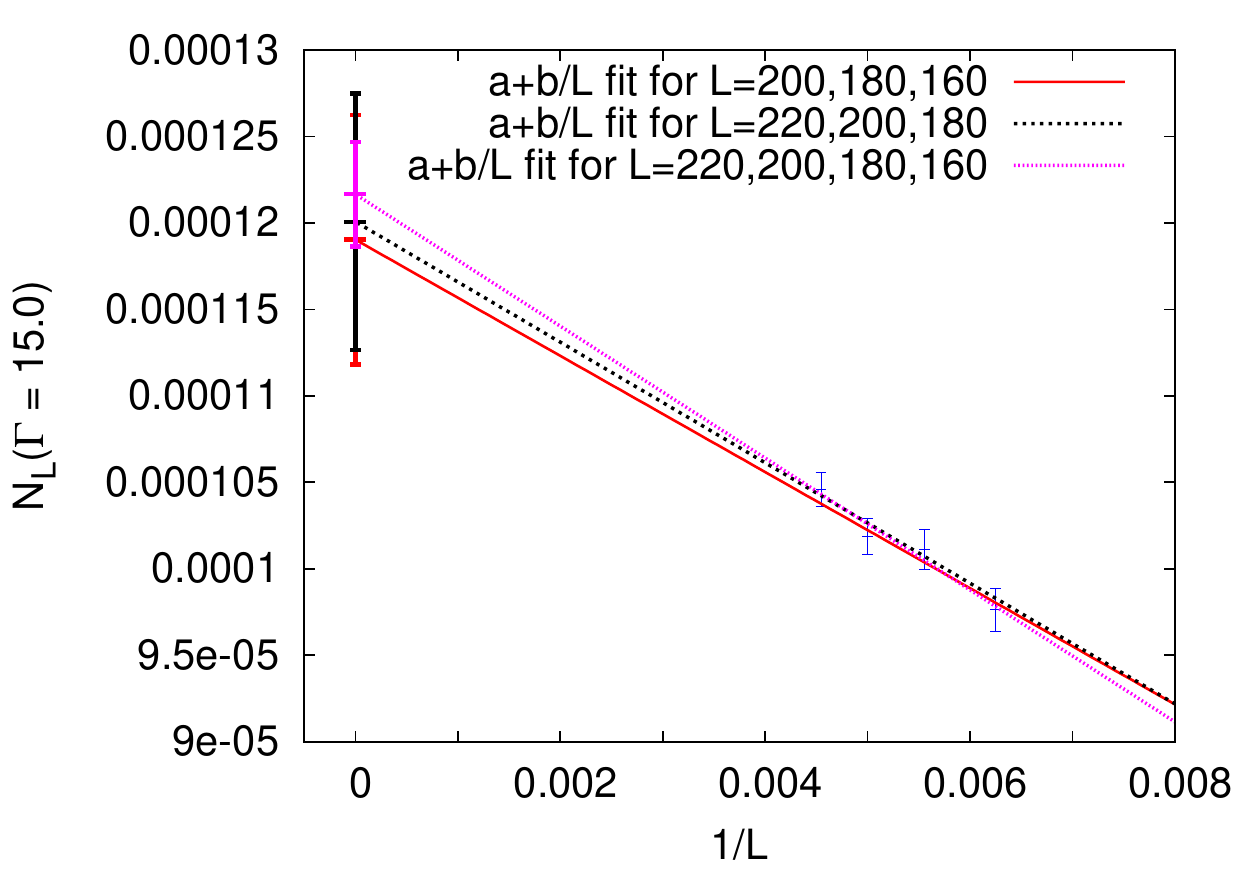}~
\includegraphics[width=0.3\columnwidth]{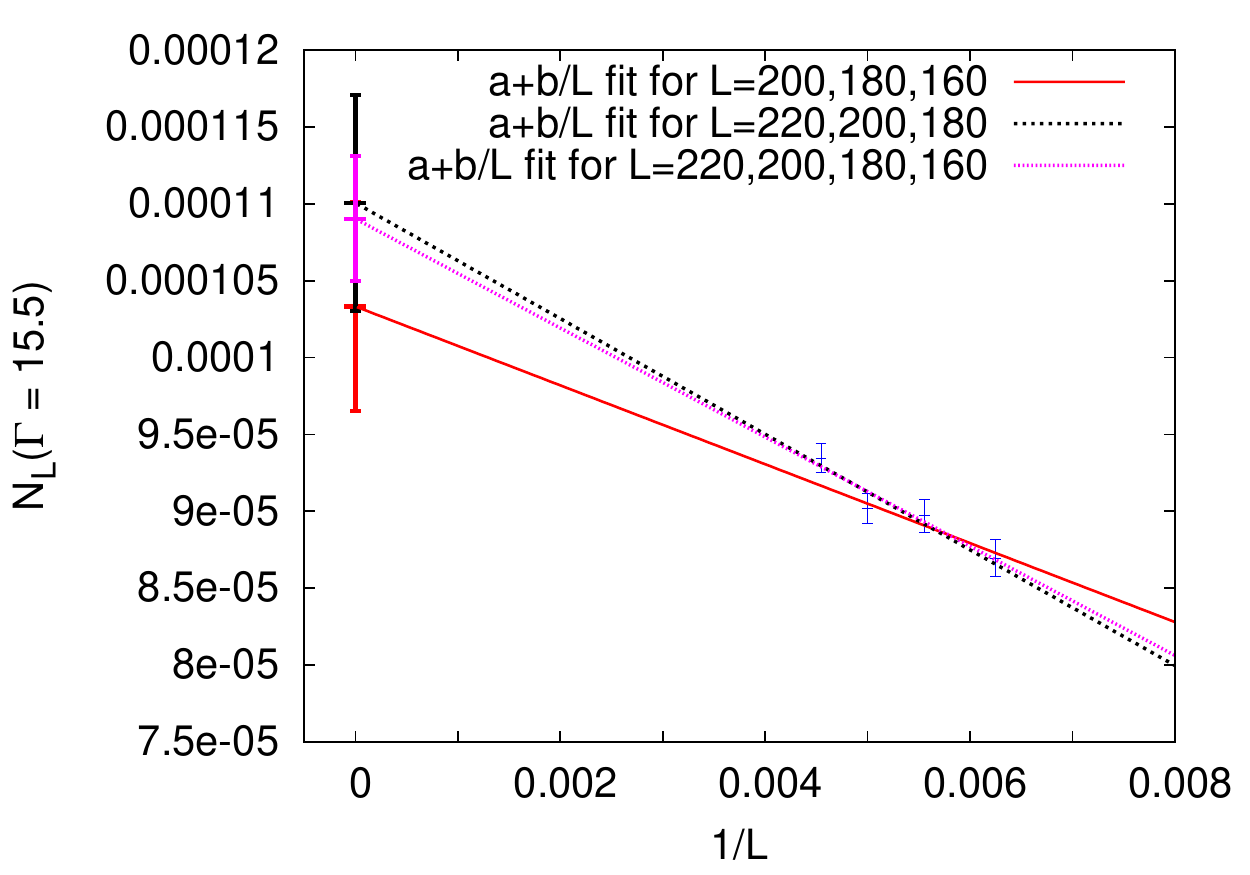}~
\includegraphics[width=0.3\columnwidth]{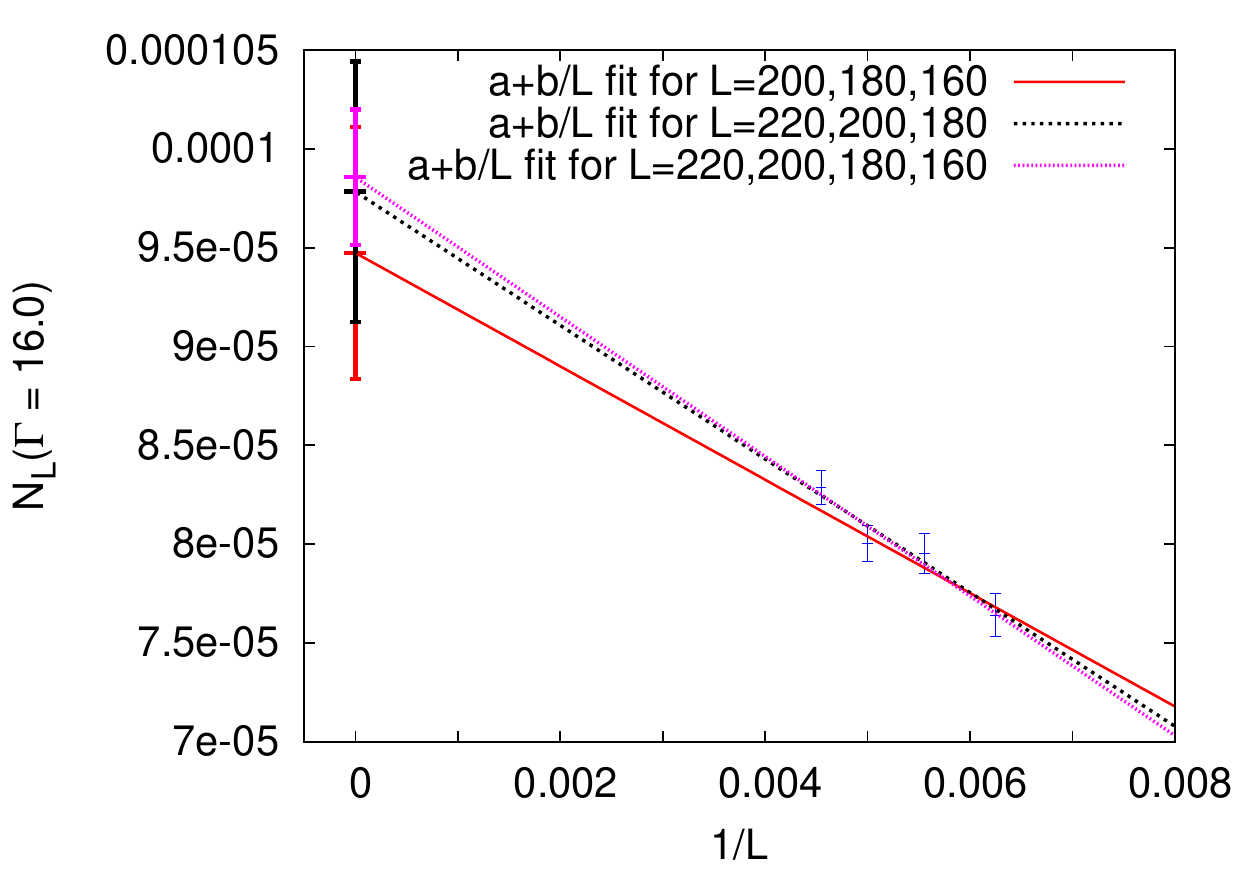}}
\centerline{\includegraphics[width=0.3\columnwidth]{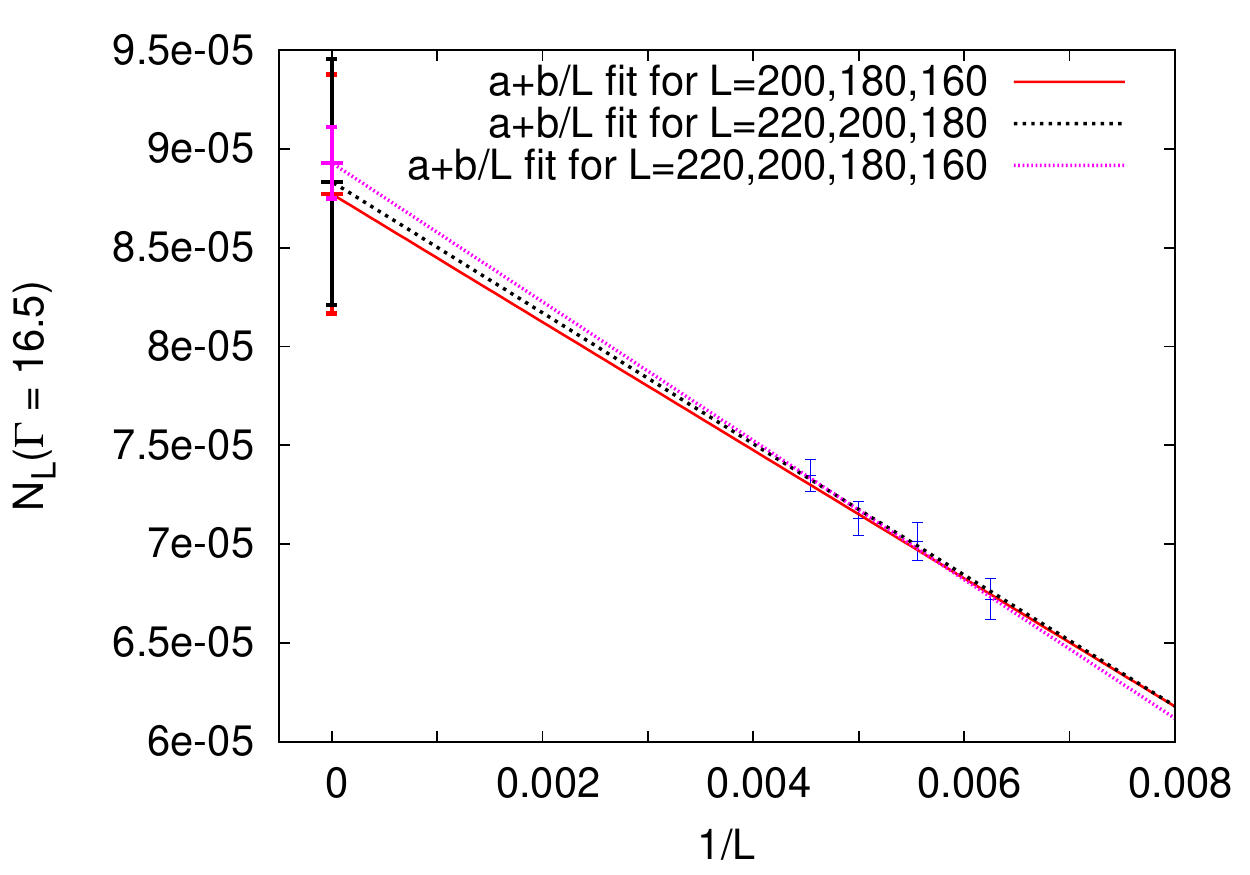}~
\includegraphics[width=0.3\columnwidth]{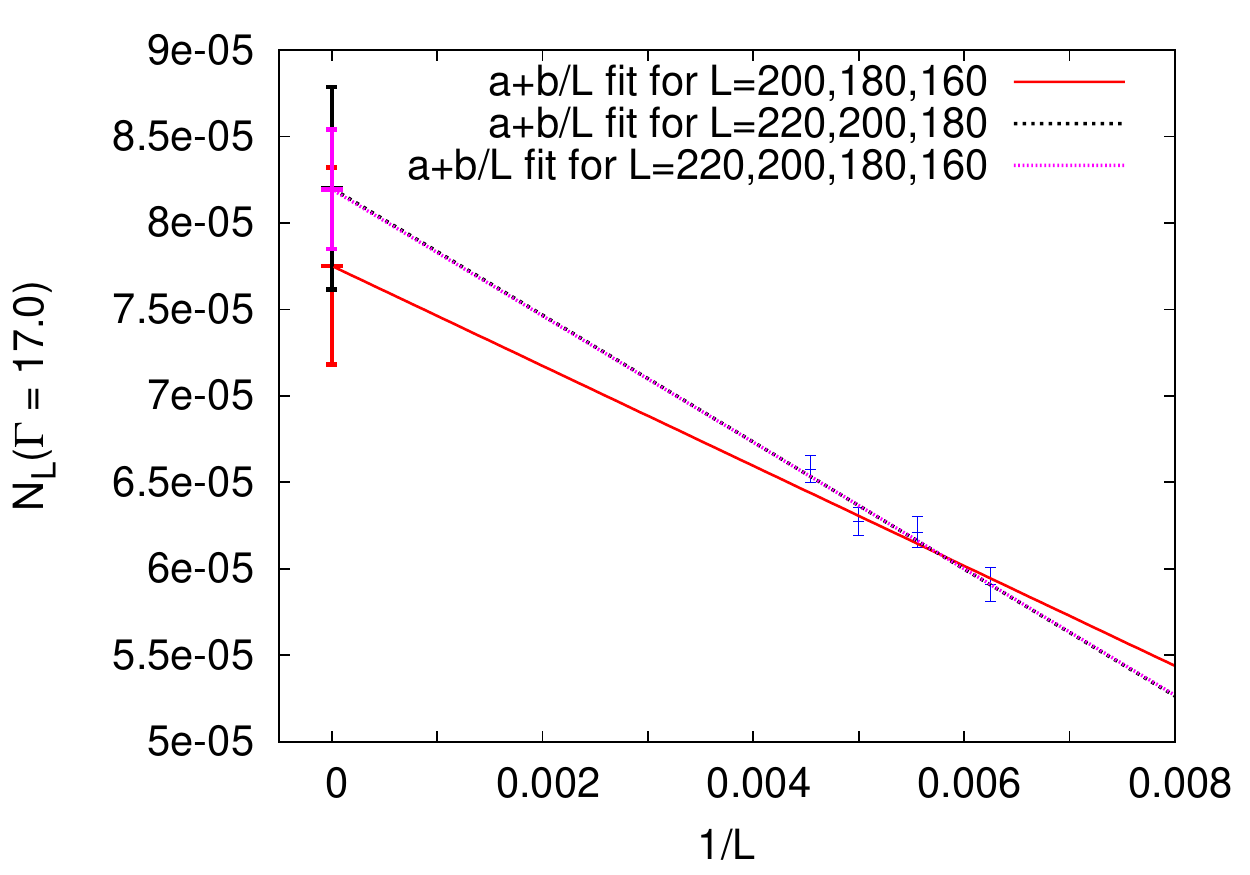}~
\includegraphics[width=0.3\columnwidth]{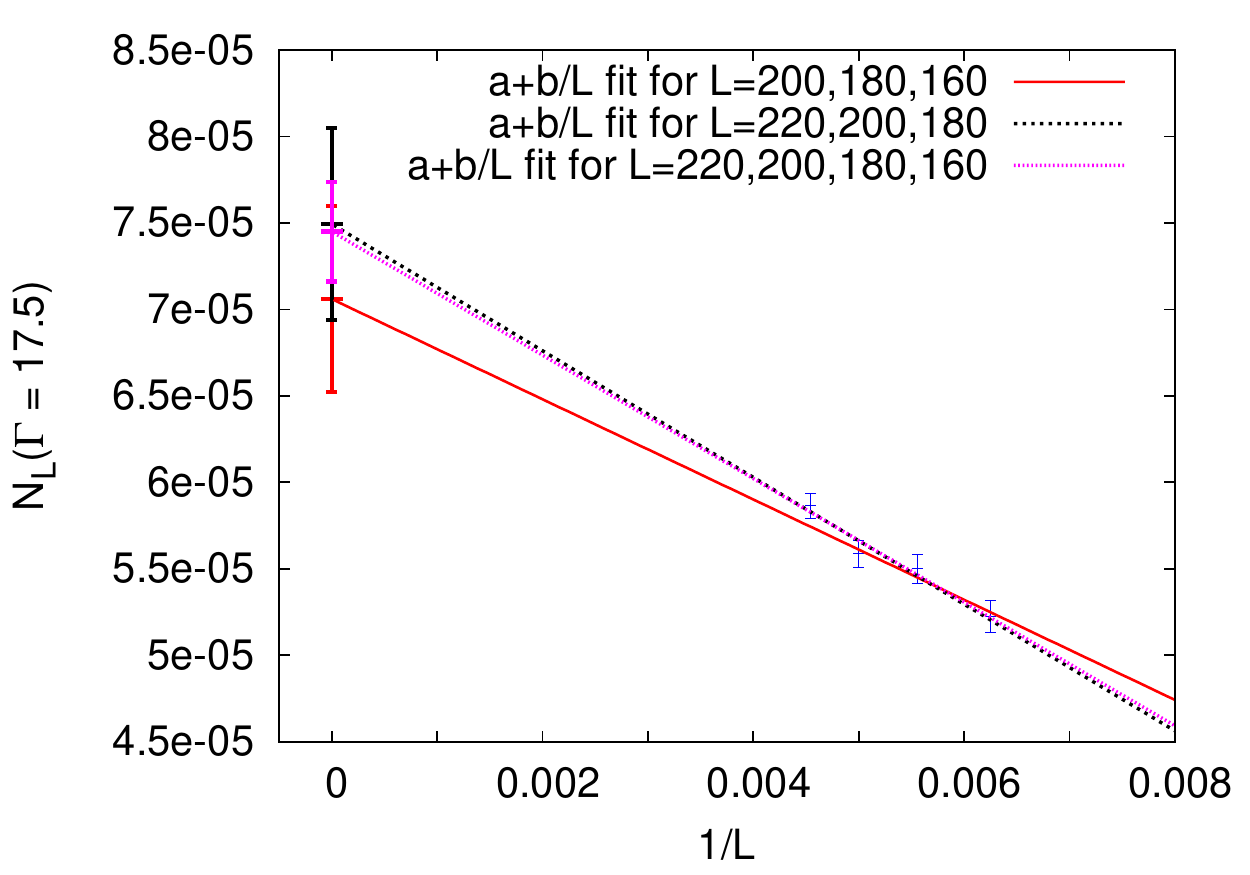}}
\centerline{\includegraphics[width=0.3\columnwidth]{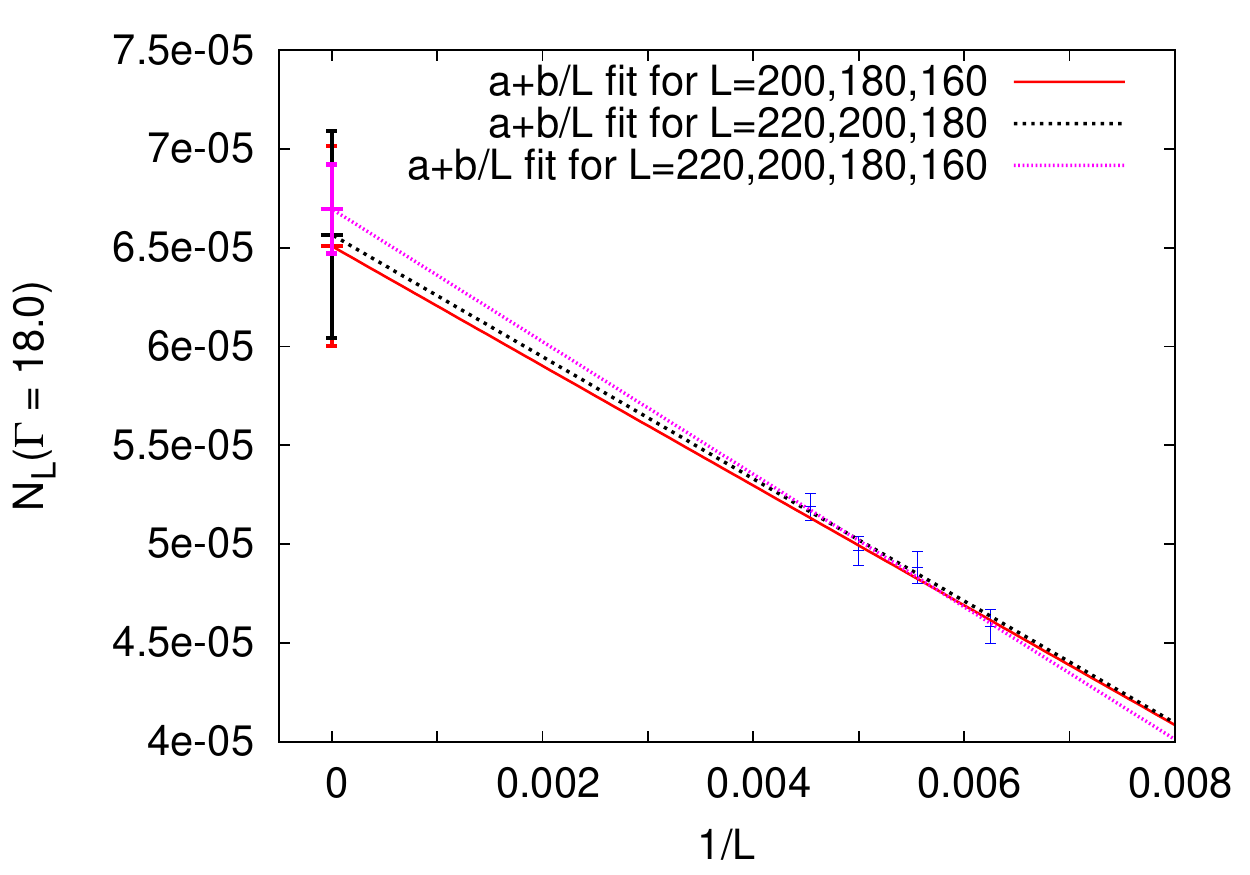}~
\includegraphics[width=0.3\columnwidth]{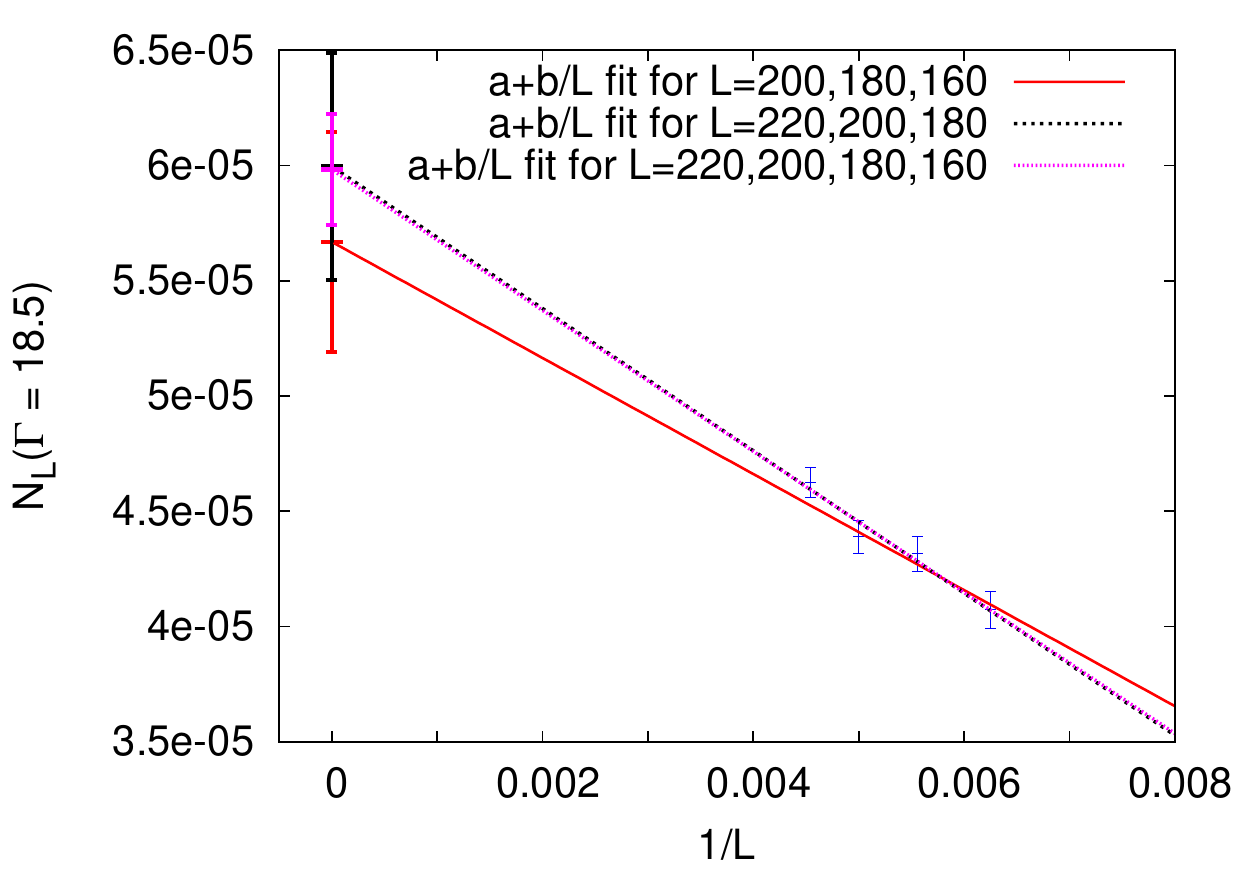}~
\includegraphics[width=0.3\columnwidth]{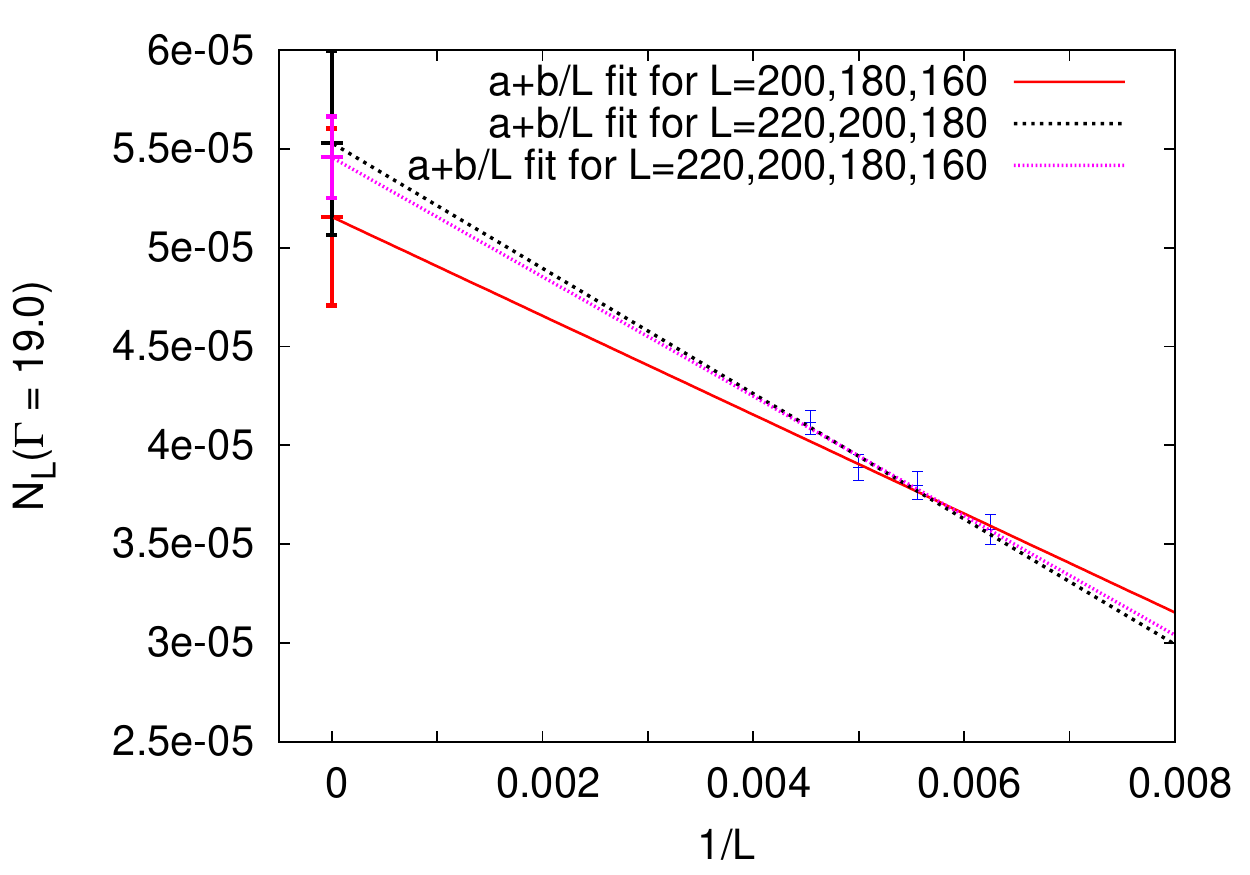}}
\end{figure}
\begin{figure}
\centerline{\includegraphics[width=0.3\columnwidth]{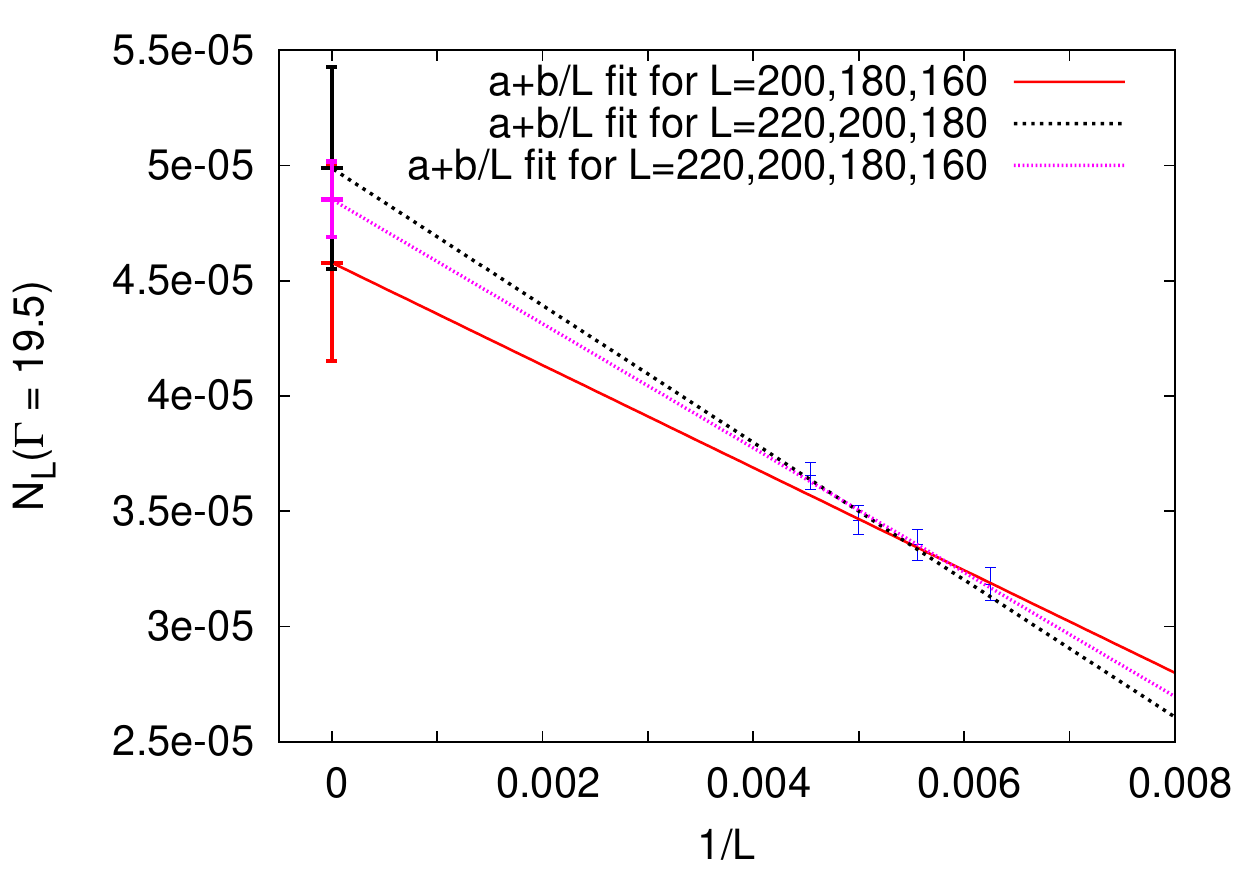}~
\includegraphics[width=0.3\columnwidth]{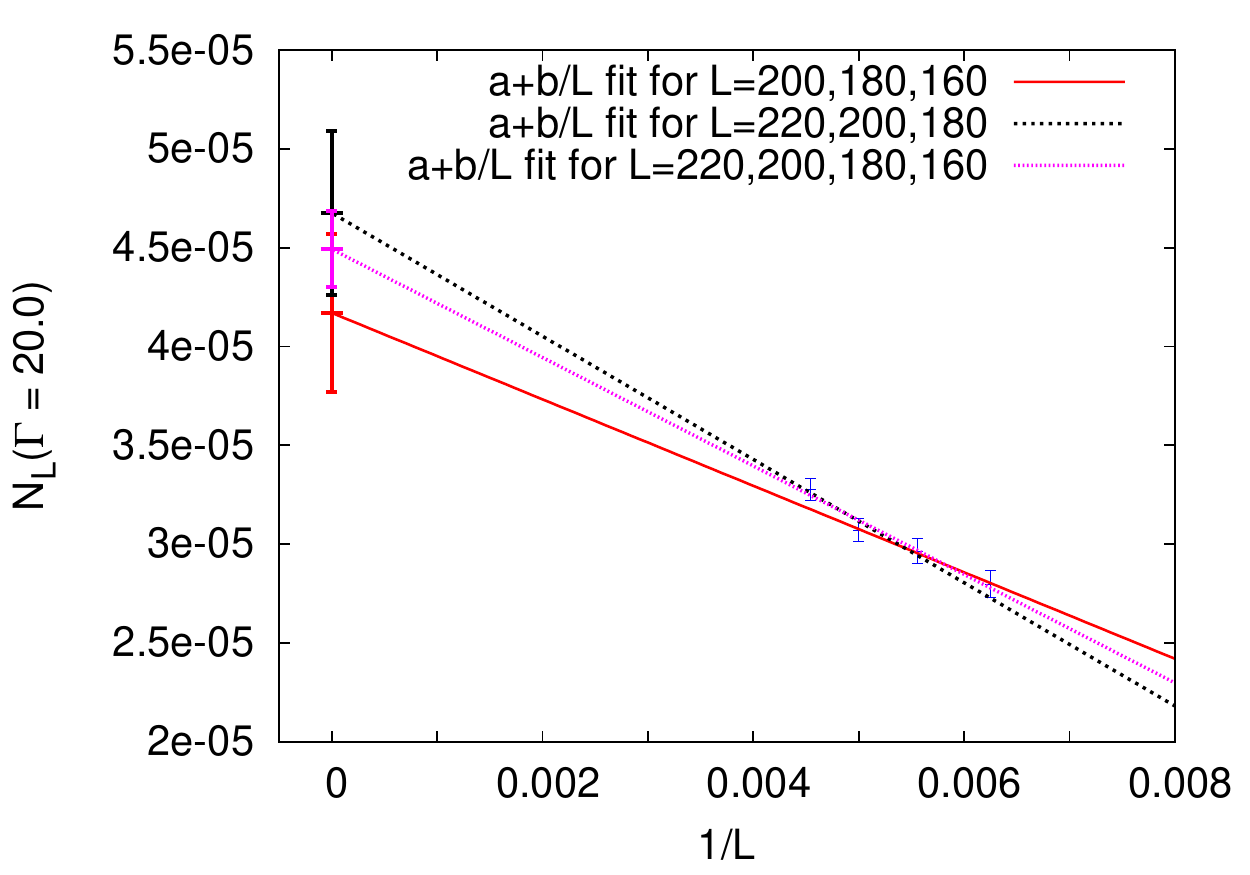}}
\caption{
Comparison of thermodynamic limit of $N_L(\Gamma)$ at $n_v= 0.0625$, taken with and without data at a larger size. At $n_v = 0.0625$, the estimated value of $\Gamma_c$ is $\Gamma_c \approx 7$.  Note that some correlation in the relative ordering of the three extrapolated values is expected over short intervals of $\Gamma$ since these data are correlated (obtained from the same set of random samples). However, over the range of $\Gamma$ from $\Gamma=7$ to $\Gamma=18$, we already see that there is no consistent ordering of the three extrapolated values, {\em i.e.}, the red points are not always higher than the black points or vice-versa. Additionally, the three different extrapolations fall within (or lie at the edge of)
each other's error bars. Also, the results quoted in the main text (values of $\Gamma_c$
and $y$ in fits to $N_{\rm 1D}$ and the quality of the different fits) do not change significantly if our analysis is performed on the thermodynamic limit $N(\Gamma)$ obtained by including data at the larger size. All this, taken together, provides compelling evidence that
our extrapolations to the thermodynamic limit are reliable. 
}
\label{SuppFig20}
\end{figure}

%}


\begin{thebibliography}{999}
\bibitem{Lee_Ramakrishnan}P.~A.~Lee and T.~V.~Ramakrishnan, Rev. Mod. Phys. {\bf 57}, 287 (1985).

\bibitem{Altland_Simons_Zirnbauer} A.~Altland, B.~D.~Simons, and M.~R.~Zirnbauer, Phys. Rep. {\bf 359}, 283 (2002).

\bibitem{Evers_Mirlin} F.~Evers and A.~D.~Mirlin,  Rev. Mod. Phys. {\bf 80}, 1355 (2008).



\bibitem{Gade} R.~Gade, Nucl. Phys. B {\bf 398}, 499 (1993).

\bibitem{Gade_Wegner} R.~Gade and F.~Wegner, Nucl. Phys. B {\bf 360}, 213 (1991).

\bibitem{Motrunich_Damle_Huse} O.~I.~Motrunich, K.~Damle, and D.~A.~Huse, 
Phys. Rev. B {\bf 65}, 064206 (2002).


\bibitem{Araujo_Terrones_Dresselhaus} P.~T.~Araujo, M.~Terrones, M.~S.~Dresselhaus,
Materials Today {\bf 15}, 98 (2012).

\bibitem{Forte_etal} G.~Forte, A.~Grassi, G.~M.~Lombardo, A.~La~Magna, G.~G.~N.~Angilella, R.~Pucci, R.~Vilardi, Phys. Lett. A {\bf 372}, 6168 (2008).

\bibitem{Pereira_dosSantos_CastroNeto} V.~M.~Pereira, J.~M.~B.~Lopes dos Santos, and A.~H.~Castro Neto, Phys. Rev.  B {\bf 77}, 115109 (2008).

\bibitem{Pereira_Guinea_dosSantos_Peres_CastroNeto} V.~M.~Pereira, F.~Guinea, J.~M.~B.~Lopes dos Santos, N.~M.~R.~Peres, and A.~H.~Castro Neto, Phys. Rev. Lett. {\bf 96}, 036801 (2006).

\bibitem{Wehling_Yuan_Lichtenstein_Geim_Katsnelson}
T.~O.~Wehling, S.~Yuan, A.~I.~Lichtenstein, A.~K.~Geim, and M.~I.~Katsnelson,
Phys. Rev. Lett. {\bf 105}, 056802 (2010).

\bibitem{Dyson} F.~J.~Dyson, Phys. Rev. {\bf 92}, 1331 (1953). 
\bibitem{Theodorou_Cohen} G.~Theodorou and M.~H.~Cohen, Phys. Rev. B {\bf 13}, 4597 (1976).

\bibitem{Eggarter_Riedenger}T.~P.~Eggarter and R.~Riedinger, Phys. Rev. B {\bf 18}, 569 (1978).
\bibitem{Motrunich_Damle_Huse_PRB0} O.~Motrunich, K. Damle, and D. A. Huse, Phys. Rev. B {\bf 63}, 134424 (2001).
\bibitem{Motrunich_Damle_Huse_PRB1} O.~Motrunich, K.~Damle, and D.~A.~Huse
Phys. Rev. B {\bf 63}, 224204 (2001).

\bibitem{Gruzberg_Read_Vishweshwara} I.~A.~Gruzberg, N.~Read, and S.~Vishveshwara,
Phys. Rev. B {\bf 71}, 245124 (2005).

\bibitem{Brouwer_Furusaki_Gruzberg_Mudry} P.~W.~Brouwer, A.~Furusaki, I.~A.~Gruzberg, and C.~Mudry, Phys. Rev. Lett. {\bf 85}, 1064 (2000).

\bibitem{Brouwer_Mudry_Furusaki} P.~W.~Brouwer, C.~Mudry, and A.~Furusaki,
Phys. Rev. Lett. {\bf 84}, 2913 (2000).

\bibitem{Titov_Brouwer_Furusaki_Mudry} M.~Titov, P.~W.~Brouwer, A.~Furusaki, and C.~Mudry, Phys. Rev. B {\bf 63}, 235318 (2001).

\bibitem{Mudry_Ryu_Furusaki} C.~Mudry, S.~Ryu, A.~Furusaki,
Phys. Rev.  B {\bf 67}, 064202 (2003).

\bibitem{Willans_Chalker_Moessner_PRB} A.~J.~Willans, J.~T.~Chalker, and R.~Moessner, Phys. Rev. B {\bf 84}, 115146 (2011)


\bibitem{Hafner_etal} V.~Hafner, J.~Schindler, N.~Weik, T.~Mayer, S.~Balakrishnan, R.~Narayanan, S.~Bera, and F.~Evers, Phys. Rev. Lett. {\bf 113}, 186802 (2014).

\bibitem{Ostrovsky_etal} P.~M.~Ostrovsky, I.~V.~Protopopov, E.~J.~Konig, I.~V.~Gornyi, A.~D.~Mirlin, and M.~A.~Skvortsov, Phys. Rev. Lett. {\bf 113}, 186803 (2014).

\bibitem{Lieb_Loss} E.~H.~Lieb and M.~Loss, Duke Math. J. {\bf 71}, 337 (1993).

\bibitem{Ryu_Hatsugai} S.~Ryu and Y.~Hatsugai, 
Phys. Rev. Lett. {\bf 89}, 077002 (2002).

\bibitem{Brey_Fertig} L.~Brey and H.~A.~Fertig, Phys. Rev. B {\bf 73}, 235411 (2006).

\bibitem{Brouwer_Racine_Furusaki_Hatsugai_Morita_Mudry}
P.~W.~Brouwer, E.~Racine, A.~Furusaki, Y.~Hatsugai, Y.~Morita, and C.~Mudry, Phys. Rev. B {\bf 66}, 014204 (2002).

\bibitem{ALGOL} https://en.wikipedia.org/wiki/ALGOL

\bibitem{Martin_Wilkinson} R.~S.~Martin and J.~H.~Wilkinson in {\em Handbook for Automatic Computation, Vol. II: Linear Algebra}, J.~H.~Wilkinson and C.~Reinsch (eds.), Springer-Verlag (Berlin, 1971).

\bibitem{Sanyal_Thesis} S.~Sanyal, Ph.D thesis, Tata Institute of Fundamental Research, Mumbai, 2014, http://theory.tifr.res.in/Research/Thesis/

\bibitem{GMP} https://en.wikipedia.org/wiki/ GNU\_Multiple\_Precision\_Arithmetic\_Library

\bibitem{LAPACK} https://en.wikipedia.org/wiki/LAPACK

%\bibitem{Supplemental} See Supplemental Material at http://link.aps.org/supplemental/xx/PhysRevLett.yy for additional numerical and analytical evidence in support of our conclusions.
\end{thebibliography}
\end{document}